\documentclass[10pt]{article}

\usepackage[english]{babel}
\usepackage{latexsym,amsmath,amssymb,amsbsy,amstext,amscd,amsfonts}

\usepackage{graphics,graphicx}
\usepackage{color}
\usepackage{tabularx}
\usepackage{multirow}
\usepackage{booktabs}
\usepackage{epstopdf}
\usepackage{natbib}
\usepackage{ulem}
\usepackage{float}
\usepackage [latin1]{inputenc}

\date{}

\title{On the pressure decline analysis for hydraulic fractures in elasto-plastic materials}

\author{M. Wrobel$^{(1)}\footnote{Corresponding author: wrobel.michal@ucy.ac.cy}$ , P. Papanastasiou$^{(1)}$ and M. Dutko$^{(2,3)}$  
\\
{\it $^{(1)}$\! Department of Civil and Environmental Engineering, University of Cyprus, }
\\ {\it 1678 Nicosia, Cyprus}
\\
{\it $^{(2)}$Rockfield Software Limited,}
\\ {\it Swansea, SA1 8AS, UK}
\\
{\it $^{(3)}$Department of Mathematics, Aberystwyth University,}
\\ {\it Ceredigion SY23 3BZ, Wales, UK}
 }

\begin{document}

\maketitle

\begin{abstract}
In this paper the problem of a plane strain hydraulic fracture in elasto-plastic material is analyzed. The analysis is based on Finite Element Method computations of crack propagation and closure to simulate the Mini-Frac calibration test. The basic trends of fracture evolution are identified for both, the propagation and the closure stage in elasto-plastic materials and compared with the results obtained for purely elastic deformation of rock. Pressure decline analysis is conducted to identify the closure stress and leak-off coefficient. The results show that the standard techniques of pressure decline analysis can be adopted also for the case where the fractured material deforms inelastically.

\end{abstract}

\providecommand{\keywords}[1]
{
  \small	
  \textbf{\textit{Keywords:}} #1
}


\section{Introduction}
\label{Introduction}

Hydraulic fracturing (HF) is a process of propagation of an internally pressurized crack. The underlying physical mechanism  manifests itself in many natural phenomena such as the subglacial drainage of water or magmatic intrusions in the earth crust. In technology, hydraulic fracturing can play an undesired role, jeopardizing the stability of hydraulic structures (e. g. dams) or CO$_2$ sequestration systems \citep{Papanastasiou_CO2, Li_2017}. However, the most prominent example of technological application of HF is fracking - a technique used to intensify extraction from the low-permeability hydrocarbon reservoirs.

Due to high magnitudes of the confining stress at the depths at which the fracking treatment is executed the fractured rock formation can exhibit plastic properties. In the last  decades the fracking technology has been frequently employed in the poorly consolidated formations for sand control. For the fractures produced in such weak formations the plastic deformation effects become especially important. It has been shown in many studies \citep{Papanastasiou_1997,Papanastasiou_algorithm,Sarris_2013,Wrobel_plasticity} that inelastic mechanism of solid deformation substantially affects the hydraulic fracturing process. In particular, the fracture grows shorter and wider than its elastic counterpart. Moreover, greater fluid pressure (and consequently the power input) is needed to propagate the fracture.  The aforementioned trends are related to the brittleness index that combines the rock strength and the in-situ parameters \citep{Papanastasiou_2015} but also to the pumping regime \citep{Papanastasiou_1997,Papanastasiou_algorithm}. Furthermore, the process of fracture closure after the fluid pumping shut-off depends greatly on the constitutive behaviour of the fractured material. In the paper by \cite{Papanastasiou_2000} it was shown the closure mechanism of a hydraulic fracture differs depending on whether the pressure decline involves a stationary or a previously propagated crack. In the former instance a hinge-like closure pattern holds, while in the latter case a zipping mechanism prevails. Proper identification and understanding of this mechanism is crucial in analyzing the problem of proppant placement and stability.

In technological applications the process of hydraulic fracturing is also routinely used to estimate the permeability and the in-situ stress of the rock formation \citep{Do_2020}. These estimations are carried out in the framework of an injection/falloff diagnostic test, the so-called Mini-Frac test. This technique includes breaking down the rock formation and creating a short fracture during the injection period with subsequent monitoring of the crack closure process in the falloff stage. The data recorded during the Mini-Frac test is then processed by means of dedicated techniques to obtain the values of reservoir permeability and the closure stress (minimum in-situ stress). Successful execution of the fracking treatments depends largely on proper identification of the aforementioned parameters. The Mini-Frac test is also employed in other engineering applications where the rock stress information is required either directly or as an input for numerical simulations. This includes for example  long and short-term stability of underground structures such as tunnels and shafts or design of rock support systems \citep{Ljunggren_2003}. Note that in the Mini-Frac tests performed in the weak formations the size of the plastic deformation zone is no longer small as compared to the fracture dimensions (the more so because the induced fracture is by definition short). Thus, all the peculiarities related to the propagation of the elasto-plastic hydraulic fractures,  that are mentioned in the previous paragraph, manifest themselves distinctly during the Mini-Frac treatments.

The data on the wellbore pressure, recorded during the falloff stage of the Mini-Frac test, is used in the pressure decline analysis. One of the main goals of this analysis is identification of the so-called shut-in pressure (or closure pressure). The shut-in pressure is defined as the lowest magnitude of wellbore pressure, during the pressure decay stage, required to hold the crack open against the far field horizontal stress \citep{Lee_1989}. As the hydraulic fracture propagates in the direction perpendicular to the minimum principal stress, the shut-in pressure is considered a direct estimator of the latter. There are several methods to determine the shut-in pressure from the pressure-time characteristics \citep{Hayashi_1989}. These may include: i) determination of the point at which the pressure graph starts to inflect, ii) determination of the point of intersection between tangents drawn to initial and later parts of the shut-in curve, iii) determination of the point of maximal curvature of pressure characteristics, or iv) determination of the point where the pressure diagram ultimately levels off. Additionally, these methods can be combined with logarithmic or semi-logarithmic pressure characteristic as well as dependence of the pressure decay rate on the wellbore pressure \citep{Choi_2003}. Clearly, there is no universal method to determine the shut-in pressure and different methods can produce different results from the analyzed data. Thus, estimation of the shut-in pressure requires proper selection and adjustment of the respective technique (depending on the particular analyzed data) and depends largely on the experience of the modeller \citep{Saeidi_2021}.

The permeability of the rock formation can be determined from the Mini-Frac results by applying the $g$-function analysis \citep{Nolte_1979,Nolte_1986,Nolte_1989}. This technique was developed for the Carter leak-off model on the assumption that the fracture surface evolves according to power law during the injection period and remains  constant (just as the fracture compliance) in the closure phase. For such conditions \cite{Nolte_1979} introduced an auxiliary dimensionless function, the so-called $g$ function, which is a monotonically increasing function of dimensionless shut-in time. Under an additional assumption, that the rock formation is described by the linear elasticity theory, one can prove that the fluid pressure after pumping shut-in behaves as a linear decreasing function of $g$ argument. The slope and the intercept of such an idealized pressure characteristics depend explicitly on the leak-off coefficients, provided that the crack geometry can be described by one of the three classical models: PKN, KGD or radial. The methodology of employing the $g$ function analysis assumes a straight line fitting of the measured pressure values in the decline stage. Next,  the leak-off coefficients are computed from the slope and intercept of the linear fitting function. Naturally, such strong assumptions are hardly met in the routine field practice, which leads to erroneous estimations of the rock permeability. Some of the issues arising due to the non-ideal conditions can be circumvented by relatively straightforward modifications of the $g$ function technique \citep{Valko,Economides}. Nevertheless, the application of the $g$ function analysis in the case of elasto-plastic model of rock deformation is still an open question.

In this paper we analyze the problem of a short hydraulic fracture propagating in the elasto-plastic material. Finite Element Method computations are employed in the framework of Elfen package \citep{Profit_2016,Profit_ARMA,Angus_2015} to simulate the Mini-Frac test.  The fluid leak-off to the rock formation is modeled by means of the Carter type relation with the leak-off coefficient depending on the fluid pressure. Two models of rock deformation are analyzed: i) the linear elastic model, ii) the elasto-plastic model based on the combination of  Mohr-Coulomb and Rankine yield criteria. The pressure decline analysis is performed with elements of the $g$ function technique.  This includes determination of the shut-in pressure and leak-off coefficient alongside identification of the crack closure patterns. Note that in our simulations, unlike in the field practice, the shut-in pressure is known a priori which enables the approximation error to be quantified. The accepted formulation of the problem violates at least two basic assumptions of the original $g$ function analysis, i. e.: i) plastic deformations of rock are present in one of the analyzed variants, ii) the fluid leak-off coefficient is pressure dependent. We verify to what degree these deviations from the idealized conditions affect the results of analysis.

The paper is structured as follows. In section \ref{problem_form} we introduce the mathematical formulation of analyzed problem and description of the computational algorithm implemented in the Elfen software. Section \ref{numerical_results} describes the numerical computations performed to simulate the Mini-Frac test. Here, in subsection \ref{frac_prop_s}, we show the results obtained for the crack propagation stage. The crack closure phase and the pressure decline analysis are addressed in subsection \ref{frac_clos_s}. The final conclusions are presented in section \ref{concl}.

\section{Mathematical formulation of the problem}
\label{problem_form}

Let us consider a problem of a plane strain hydraulic fracture propagating in elasto-plastic material. The basic geometry of the problem corresponds to the classical KGD model. Note that, despite its geometrical simplicity, the KGD model can be utilized to describe properly the basic physical mechanisms that govern the HF problem (see e.g. \cite{Wrobel_2015,Wrobel_2017,Wrobel_2018,Wrobel_2021,Wrobel_redirection,Perkowska_2016}).

 The general system of governing equations employed in the Elfen software describes \citep{Profit_2016,Profit_ARMA,Profit_2018}:
\begin{itemize}
\item{Equilibrium of the total stress in the rock formation and the external loads (rock deformation is governed by the effective stress).}
\item{Porous flow in the rock formation.}
\item{Fluid flow inside the fracture.}
\end{itemize}
In our analysis the component problem of porous flow in the rock formation is neglected. The phenomenon of fluid loss to the formation is reduced to the Crater leak-off model. In the general case the Elfen package employs Darcy law to describe the fluid transport within the solid phase.

The corresponding equation for the stress equilibrium reads:
\begin{equation}
\label{stress_eq}
\nabla\left(\mbox{\boldmath$\sigma$}' -\alpha \textbf{m}p_\text{s}\right) +\rho_\text{B}\textbf{g}=\textbf{0},
\end{equation}
where $\mbox{\boldmath$\sigma$}'$ is the effective stress tensor (the compressive stress is assumed negative), $\alpha$ is the Biot coefficient, $\textbf{m}$ is the identity tensor, $p_\text{s}$ stands for the pore fluid pressure, $\rho_\text{B}$ is the wet bulk density, while $\textbf{g}$ denotes the gravity vector.

The flow of Newtonian fluid inside the fracture is described by the relation:
\begin{equation}
\label{fluid_eq}
\nabla \left(\frac{k^\text{fr}}{\mu_\text{n}}\left(\nabla p_\text{n}-\rho_\text{fn}\textbf{g} \right) +q_\text{L}\right)=C^\text{fr}\frac{\text{d} p_\text{n}}{\text{d} t}+\alpha \Delta \dot w_\varepsilon,
\end{equation}
where $k^\text{fr}$ is the intrinsic permeability of the fractured region defined as:
\begin{equation}
\label{kfr_def}
k^\text{fr}=\frac{w^2}{12}.
\end{equation}
with $w$ being the crack opening. The remaining symbols from \eqref{fluid_eq} denote: $\mu_\text{n}$ - the viscosity of the fracturing fluid, $p_\text{n}$ - the fracturing fluid pressure, $\rho_\text{fn}$ - the density of the fracturing fluid, $\Delta \dot w_\varepsilon$ - the aperture strain rate, $C^\text{fr}$ - the compressibility (storage) coefficient computed in the following way:
\begin{equation}
\label{sfr_def}
C^\text{fr}=\frac{1}{w}\left(\frac{1}{K_\text{n}^\text{fr}} +\frac{w}{K_\text{f}^\text{fr}}\right),
\end{equation}
where $K_\text{n}^\text{fr}$ and $K_\text{f}^\text{fr}$ are the fracture normal stiffness and the bulk modulus of the fracturing fluid, correspondingly. 

The leak-off term is defined as:
\begin{equation}
\label{ql_def}
q_\text{L}=\frac{C_\text{L}}{\sqrt{t-t_\text{exp}}},
\end{equation}
with $t_\text{exp}$ being an inverse of the crack length function $L(t)$:
\[
t_\text{exp}(x)=L^{-1}(t).
\]
The leak-off coefficient, $C$, is related to the fluid pressure according to the following relation:
\begin{equation}
\label{C_def}
C_\text{L}=\frac{2 \hat C_I \hat C_{II}}{\hat C_I+\sqrt{\hat C_I^2+4\hat C_{II}^2}}, \quad \hat C_I=\sqrt{\frac{k_\text{f}\phi_\text{f}(p_\text{n}-p_\text{r})}{2\mu_\text{f}}}, \quad \hat C_{II}=\sqrt{\frac{k_\text{r}c_\text{T}\phi_\text{T}}{\mu_\text{r}}}(p_\text{n}-p_\text{r}),
\end{equation}
where $k_\text{f}$ is the filtrate permeability, $\phi_\text{f}$ is the porosity of the filtrate zone, $\mu_\text{f}$ is the effective viscosity of the fracturing fluid filtrate, $k_\text{r}$ is the reservoir permeability, $c_\text{T}$ is the reservoir compressibility, $\phi_\text{r}$ is the reservoir porosity, $\mu_\text{r}$ denotes the viscosity of the reservoir fluid, while  $p_\text{n}$ and $p_\text{r}$ stand for the fracking fluid pressure and  reservoir pore pressure, respectively.

\begin{figure}[H]
\begin{center}
\includegraphics[scale=0.35]{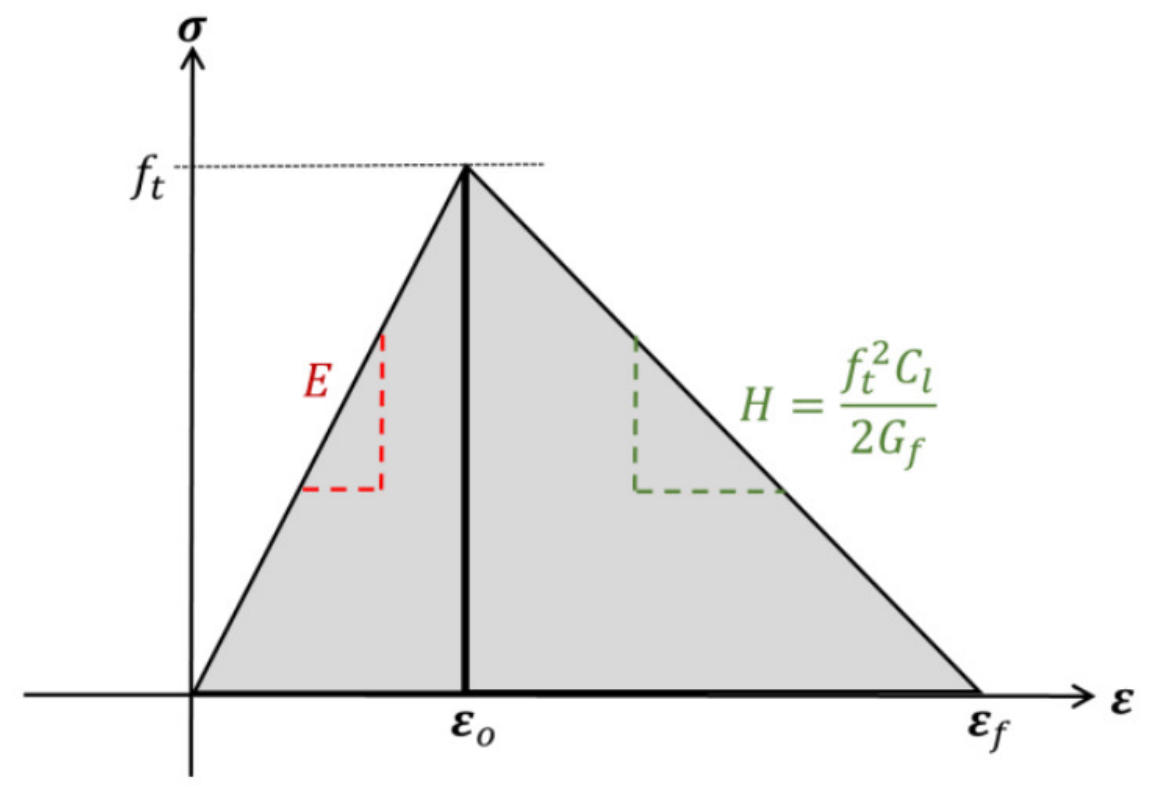}
\caption{Constitutive behaviour of the material in the cohesive zone  \citep{Profit_2016}. }
\label{strain_soft}
\end{center}
\end{figure}

The boundary conditions include the zero opening opening condition at the fracture tip and the influx boundary condition at the crack inlet. The crack propagation condition is based on the cohesive zone model \citep{Bazant_1998,Profit_2016}. The constitutive behaviour of material in the cohesive zone is depicted in Figure \ref{strain_soft}. The pre-yield characteristics is determined by the Young's modulus, $E$, and the Poisson's ratio, $\nu$. After the tensile strength, $f_t$, is reached at the  yield strain, $\varepsilon_0$, a linear approximation of the softening behaviour is employed. The softening slope, $H$, depends on the  uniaxial tensile strength, $f_t$, the strain energy release rate, $G_f$, and the element characteristic length, $C_l$. At the failure strain, $\varepsilon_f$, the closing stress reaches zero and the fracture propagates.

The above system of equations is supplemented by a pertinent constitutive model for the rock deformation. Respective models employed in our studies are described in section \ref{numerical_results}.

\begin{figure}[htb!]
\begin{center}
\includegraphics[scale=0.30]{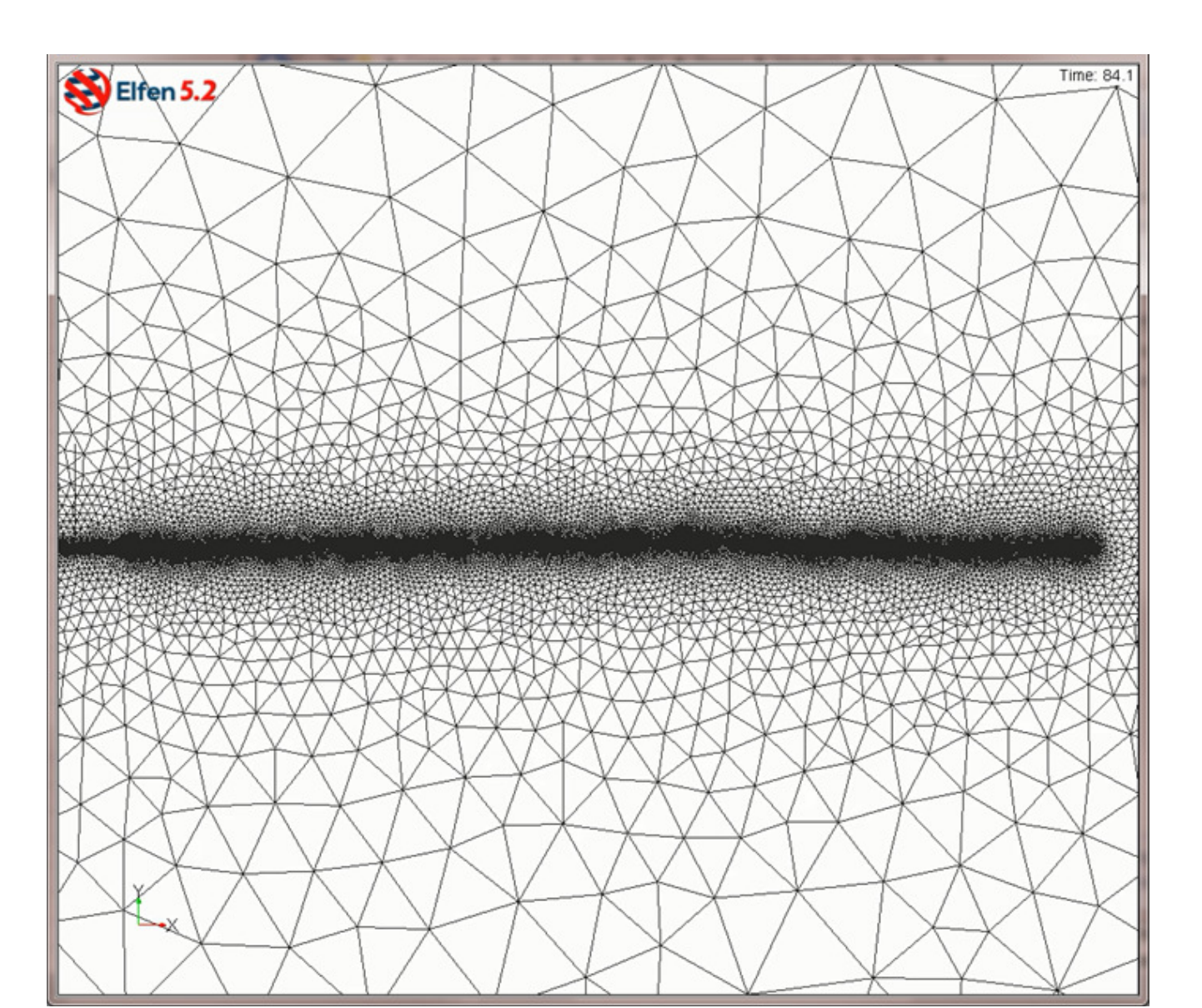}
\caption{Mesh of finite elements.}
\label{mesh}
\end{center}
\end{figure}

The system of governing PDEs \eqref{stress_eq}--\eqref{fluid_eq} is discretized by means of the finite element method. For a detailed description of the discretization technique and the resulting algebraic equations the reader is referred to the publication by \cite{Profit_2016}.
The mesh of finite elements is composed of triangular solid elements for rock formation for both stress and fluid flow analyses and 2-noded variable width line element for fluid flow. A dedicated remeshing mechanism is used based on a local algorithm that only updates mesh around the fracture tip region; hence the need for computationally expensive global remeshing is avoided.  The use of local remeshing also contributes to a significant reduction of parameter dispersion due to mapping between changing meshes. The employed mesh pattern is displayed in Figure \ref{mesh}. High density of discretization can be observed along the fracture path.

The solution is sought in a process of a staggered iterative solving the solid and fluid mechanics equations, with an explicit component solver employed for the stress and fracture width equations and implicit solver used for the fluid pressure computation (for details see \cite{Profit_2016}).

\section{Numerical results}
\label{numerical_results}

\begin{figure}[H]
\begin{center}
\includegraphics[scale=0.30]{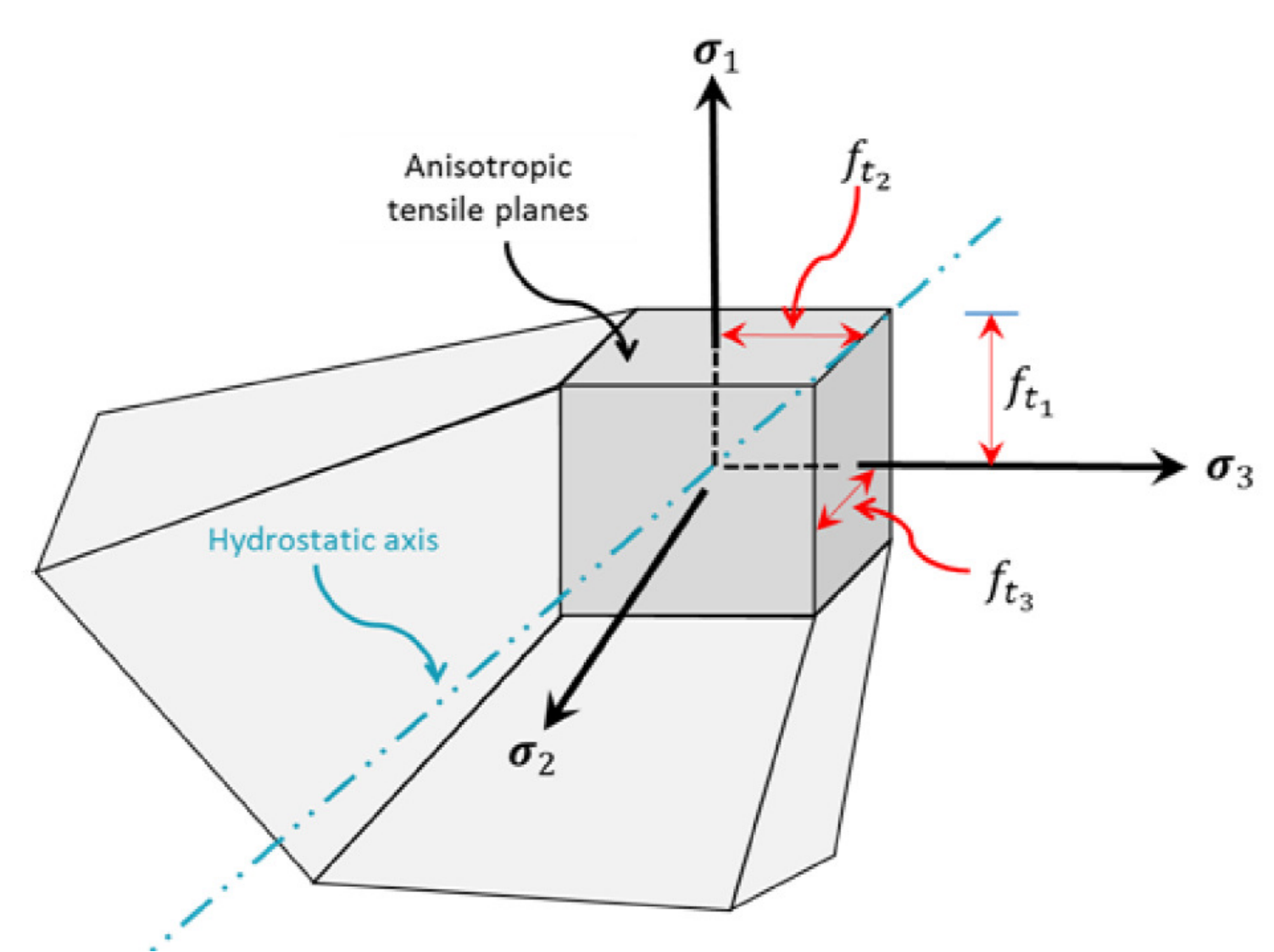}
\caption{A combined Mohr-Coulomb/Rankine yield surface according to \cite{Profit_2016}. The values of $f_{t1}$, $f_{t2}$, $f_{t3}$  define the respective tensile strengths.}
\label{yield_surf}
\end{center}
\end{figure}

The analyzed computational example involves simulation of the hydraulic fracture propagation under a constant influx and subsequent fracture closure after the pumping is stopped. Two constitutive models of the solid material are considered:
\begin{itemize}
\item{The linear elastic model of rock deformation. Note that, even though the bulk of the fractured material does not undergo  plastic deformation in this case, the plasticity tensile effects are still embedded in the cohesive zone model employed for the crack propagation condition.}
\item{The elasto-plastic model of rock. In this variant of the problem the linear elasticity holds up to the point of material yield. The plastic yield is governed by a combination Mohr-Coulomb and Rankine criteria in a way described by \cite{Profit_2016}. The yield surface in the space of principal stresses is depicted schematically in Figure  \ref{yield_surf}. While in tension, the material yields according to the Rankine theory:
\[
\sigma_i=f_{t_i},
\] 
where $\sigma_i$ is the maximal principal stress and $f_{t_i}$ stands for the corresponding tensile strength of the material. Otherwise the classical Mohr-Coulumb criterion holds:
\[
|\sigma_i-\sigma_j|=2c\cos(\phi)+(\sigma_i+\sigma_j)\sin(\phi),
\]
with $c$ being material cohesion, $\phi$ denoting the angle of friction and $\sigma_i$ and $\sigma_j$ standing for the maximal and minimal principal stresses, respectively. In this study we assume isotropic  properties of the material.}
\end{itemize}

In the following we will consider three computational examples that were selected to validate the proposed methodology of pressure decline analysis. In the first variant of the HF problem, only the elastic model of the fractured material is assumed. In the second variant, the elasto-plastic mechanism of rock deformation is accounted for. The plastic properties of the material are set here to such values that, under the applied hydraulic loading, substantial plastic deformation takes place (the size of the plastic deformation zone is of the order of the fracture length). The aforementioned examples are intended to describe two limiting cases: i) the case of elastic deformations, ii) the case of large plastic deformations. As will be shown later, the proposed methodology is applicable even if the plastic deformations of the fractured material are substantial. Thus, the methodology is expected to be practicable for smaller extents of plastic yield as well. In order to substantiate this claim we  provide computational results for the third variant of the problem. Here, the the elasto-plastic constitutive model of material is used again. However, the respective material constants are taken in a way to reduce the extent of plastic deformations with respect to the second variant.

From now on, the following nomenclature will be employed for the respective variants of the problem:
\begin{itemize}
\item{The first variant of the problem will be referred to as `elastic case' or `elastic fracture'. In the corresponding figures the subscript `$elast$' will be used to denote the components of solution.}
\item{The second variant of the problem will be called a `large yield case'. The subscript `$plast(1)$' will be employed for this variant.}
\item{For the third variant of the problem a name `small yield case' will be used. The corresponding subscript is `$plast(2)$'.}
\end{itemize}
Adjectives `large/small' in the above classification should be considered in a conventional meaning rather than in a literal (strict in terms of absolute values) sense. Additionally, the second and third variants will also be referred to as elasto-plastic cases.

\begin{table}[]
\begin{center}
\begin{tabular}{|lll|}
\hline
\multicolumn{3}{|c|}{Elastic constants}                                                                                                \\ \hline
\multicolumn{1}{|l|}{Young's modulus}                       & \multicolumn{2}{l|}{$E_\text{unloading}=16.2$ GPa}                       \\ \hline
\multicolumn{1}{|l|}{Poisson ratio}                         & \multicolumn{2}{l|}{$\nu=0.3$}                                           \\ \hline \hline
\multicolumn{3}{|c|}{Plastic constants}                                                                                                \\ \hline
\multicolumn{1}{|l|}{Initial uniaxial compressive strength} & \multicolumn{2}{l|}{$\sigma_\text{c}^0=4$ MPa}                           \\ \hline
\multicolumn{1}{|l|}{Loading modulus}                       & \multicolumn{2}{l|}{$E_\text{loading}=1.785$ GPa}                        \\ \hline
\multicolumn{1}{|l|}{Cohesion}                              & \multicolumn{1}{l|}{$c=1.33$ MPa}               & $c=2$ MPa              \\ \hline
\multicolumn{1}{|l|}{Angle of friction}                     & \multicolumn{2}{l|}{$\phi=35^o$}                                         \\ \hline
\multicolumn{1}{|l|}{Angle of dilation}                     & \multicolumn{2}{l|}{$\rho=35^o$}                                         \\ \hline \hline
\multicolumn{3}{|c|}{In-situ effective stress}                                                                                         \\ \hline
\multicolumn{1}{|l|}{Vertical stress}                       & \multicolumn{2}{l|}{$\sigma_1=14$ MPa}                                   \\ \hline
\multicolumn{1}{|l|}{Minimal horizontal stress}             & \multicolumn{2}{l|}{$\sigma_2=3.7$ MPa}                                  \\ \hline
\multicolumn{1}{|l|}{Maximal horizontal stress}             & \multicolumn{2}{l|}{$\sigma_3=9$ MPa}                                    \\ \hline \hline
\multicolumn{3}{|c|}{Fluid flow parameters}                                                                                            \\ \hline
\multicolumn{1}{|l|}{Fluid viscosity}                       & \multicolumn{2}{l|}{$\mu_\text{n}=0.001$ Pa$\cdot$s}                     \\ \hline
\multicolumn{1}{|l|}{Inlfux}                                & \multicolumn{2}{l|}{$q_0=5 \cdot 10^{-4}$ $\frac{\text{m}^2}{\text{s}}$} \\ \hline
\end{tabular}
\caption{The material and HF process parameters used in the numerical simulations. In the second variant of the problem (the large yield case) the cohesion value of $c=1.33$ MPa is employed, while in the third one (the small yield case) $c=2$ MPa is imposed.}
\label{param_table}
\end{center}
\end{table}

The computations are carried out for the values of material and HF process parameters collated in Table \ref{param_table}. Respective plastic constants for the second and third variants of the problem are the same except for the material cohesion ($c=1.33$ MPa for the second variant and $c=2$ MPa for the third one). The pumping regime assumes a constant influx magnitude for $t \in [0,80]$ s and zero influx for $t>80$ s (the crack closure stage). 
During the stage of crack propagation the leak-off coefficient $C_\text{L}$ from equation \eqref{ql_def} is set to zero (impermeable solid). For the closure stage we assume a non-zero fluid loss to formation. The following simplified representation of the coefficients $\hat C_I$ and $\hat C_{II}$ from \eqref{C_def} is adopted:
\begin{equation}
\label{C_def_simp}
 \hat C_I=2 \cdot 10^{-7}\sqrt{p_\text{n}-p_\text{res}}, \quad \hat C_{II}=2 \cdot 10^{-7}(p_\text{n}-p_\text{res}).
\end{equation}
The above assumption was accepted in order to simplify the analysis. Considering the aim of our studies it does not detract from the generality of the analysis and conclusions. Additionally, the computational cost is reduced. 

The imposed initial conditions include an immobile zero-opening fracture of the half-length $L_0=0.5$ m. The following analysis is divided into two stages: i) the crack propagation stage, ii) the crack closure stage where the pressure decline analysis is conducted.

\begin{figure}[H]
\begin{center}
\includegraphics[scale=0.30]{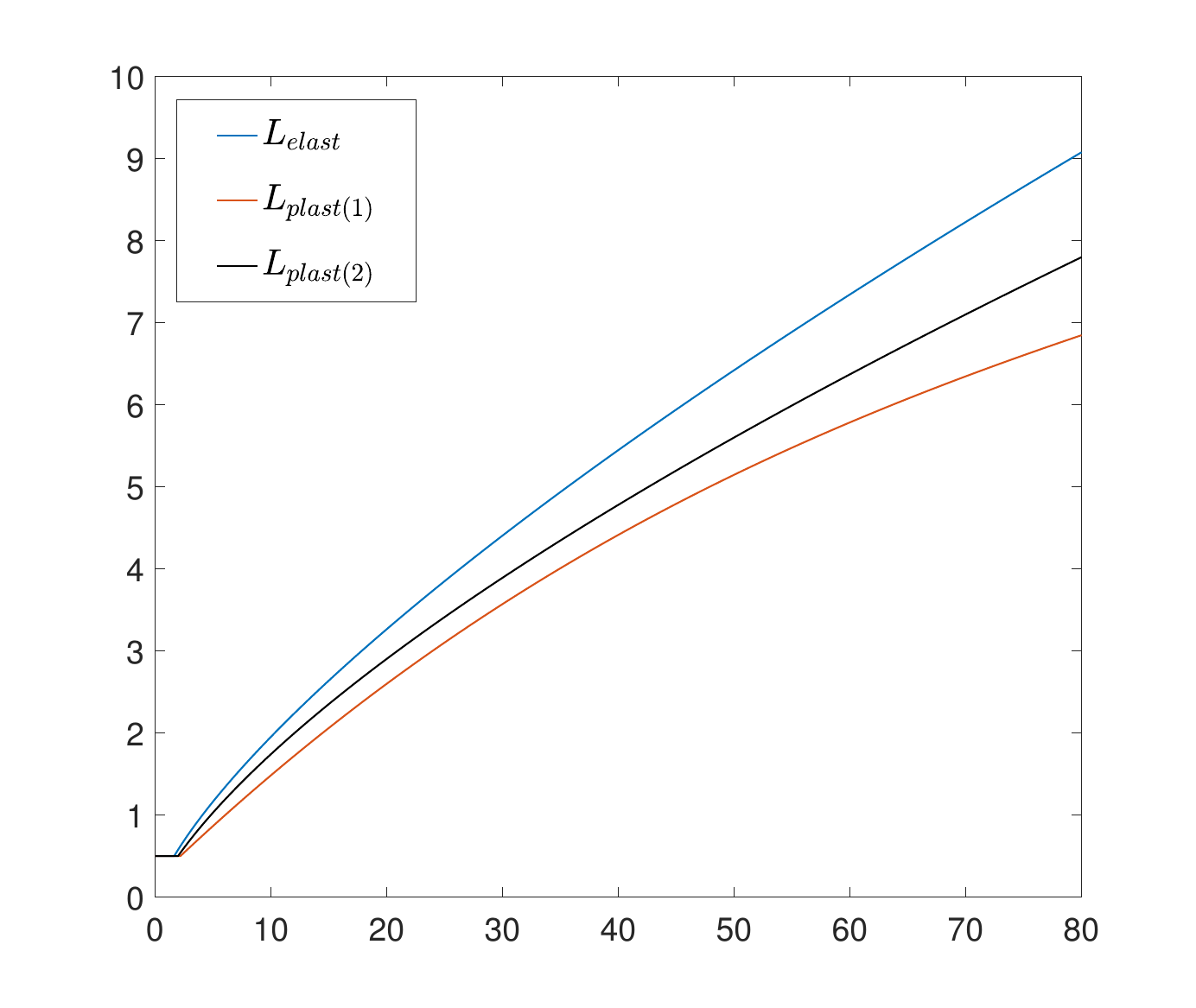}
\hspace{0mm}
\includegraphics[scale=0.30]{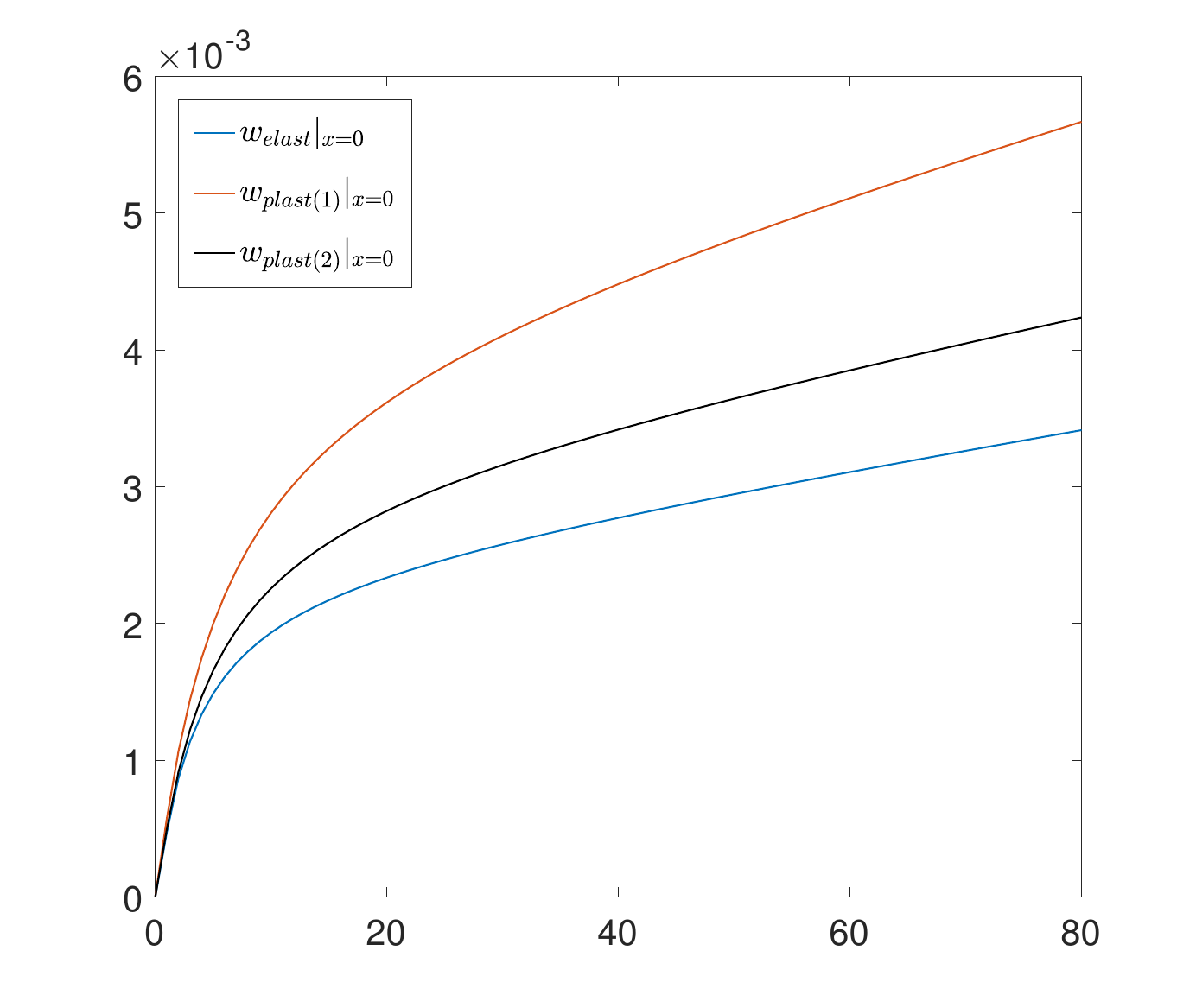}
\put(-310,0){$t$ [s]}
\put(-105,0){$t$ [s]}
\put(-440,155){$\textbf{a)}$}
\put(-215,155){$\textbf{b)}$}
\put(-410,75){\rotatebox{90}{$L(t)$ [m]}}
\put(-200,70){\rotatebox{90}{$w(0,t)$ [m]}}
\begin{center}
\includegraphics[scale=0.30]{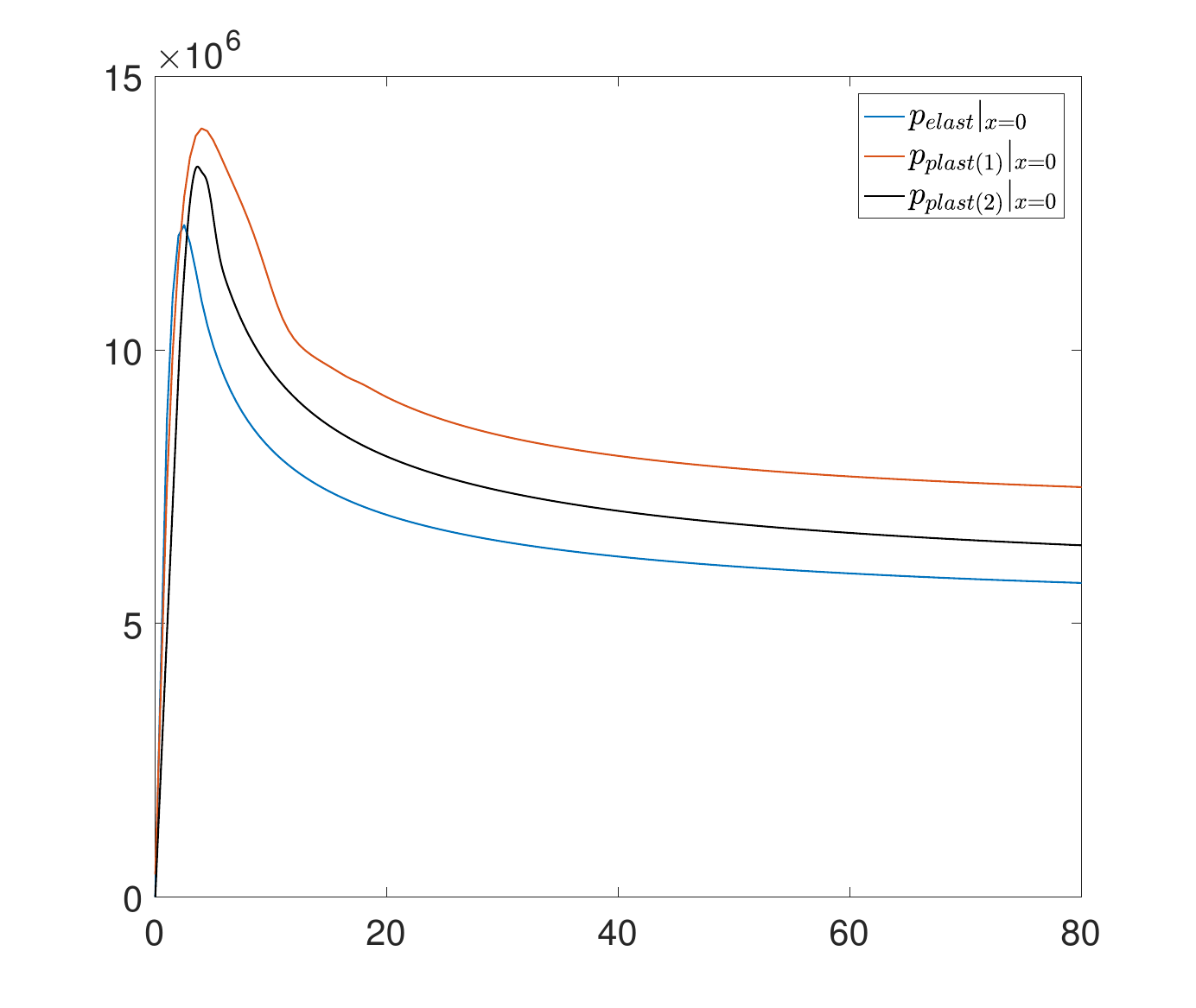}
\put(-210,155){$\textbf{c)}$}
\put(-200,70){\rotatebox{90}{$p_\text{w}$ [Pa]}}
\put(-105,0){$t$ [s]}
\end{center}
\caption{Fracture evolution during the propagation stage: a) the crack half-length, $L(t)$ [m], b) the crack opening at the fracture mouth, $w(0,t)$ [m], c) the wellbore pressure, $p_\text{w}$ [Pa]. }
\label{L_w_p_prop}
\end{center}
\end{figure}
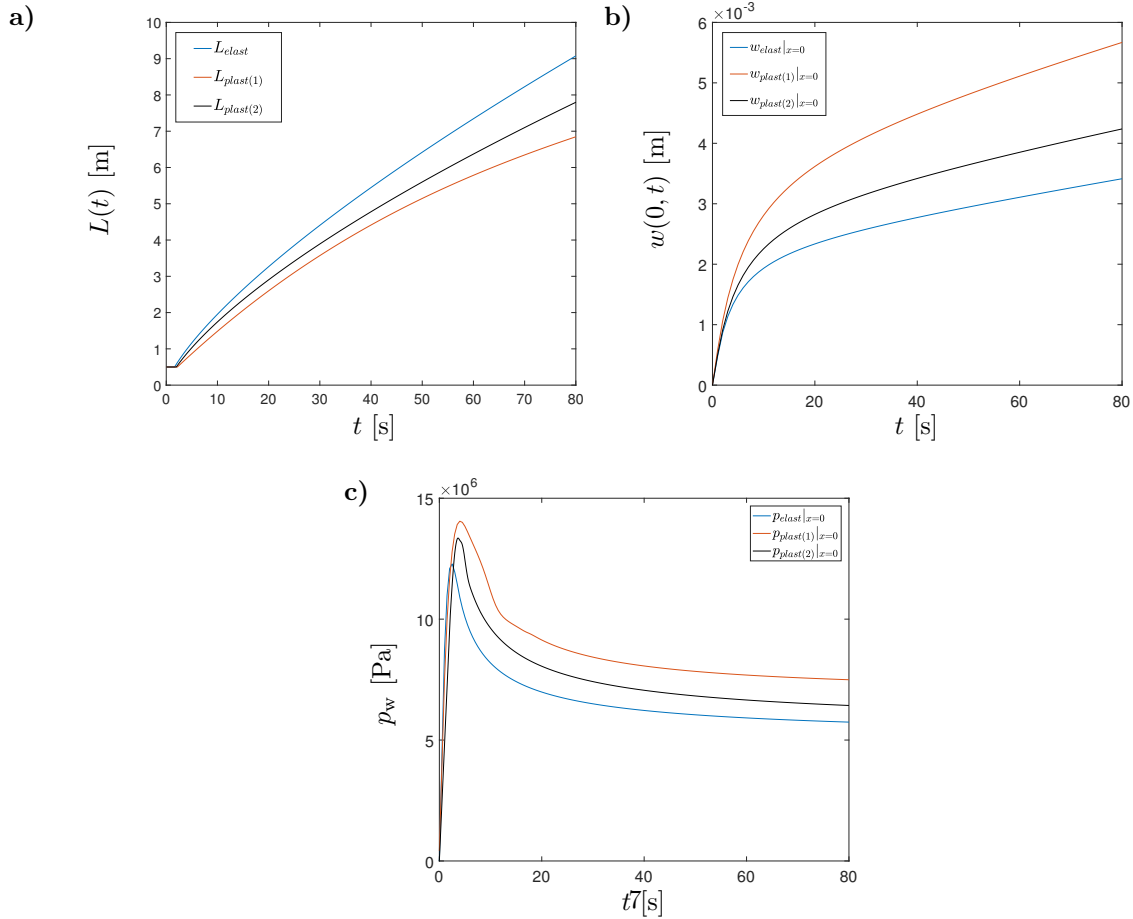

\subsection{Fracture propagation}
\label{frac_prop_s}

\begin{figure}[H]
\begin{center}
\includegraphics[scale=0.30]{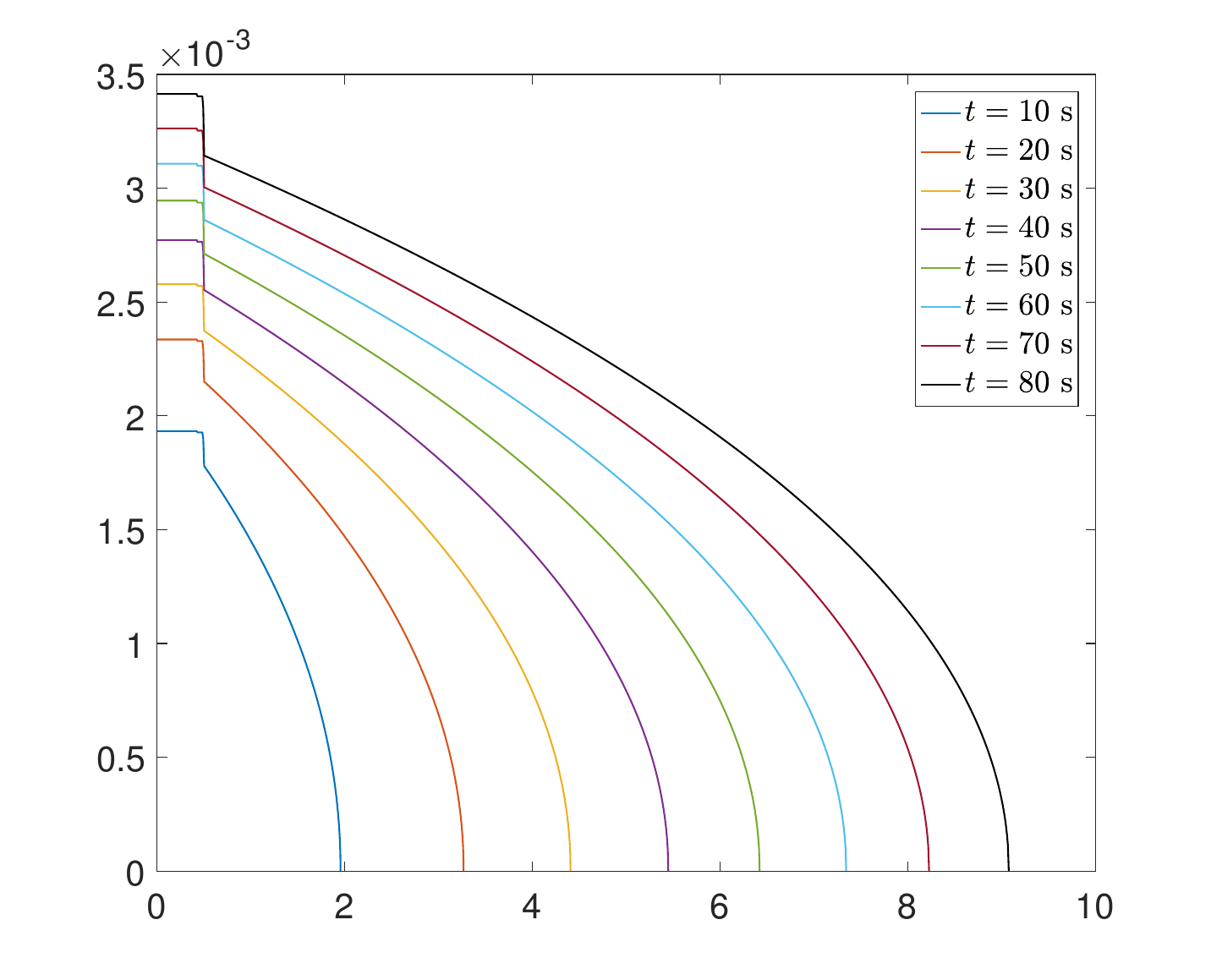}
\hspace{0mm}
\includegraphics[scale=0.30]{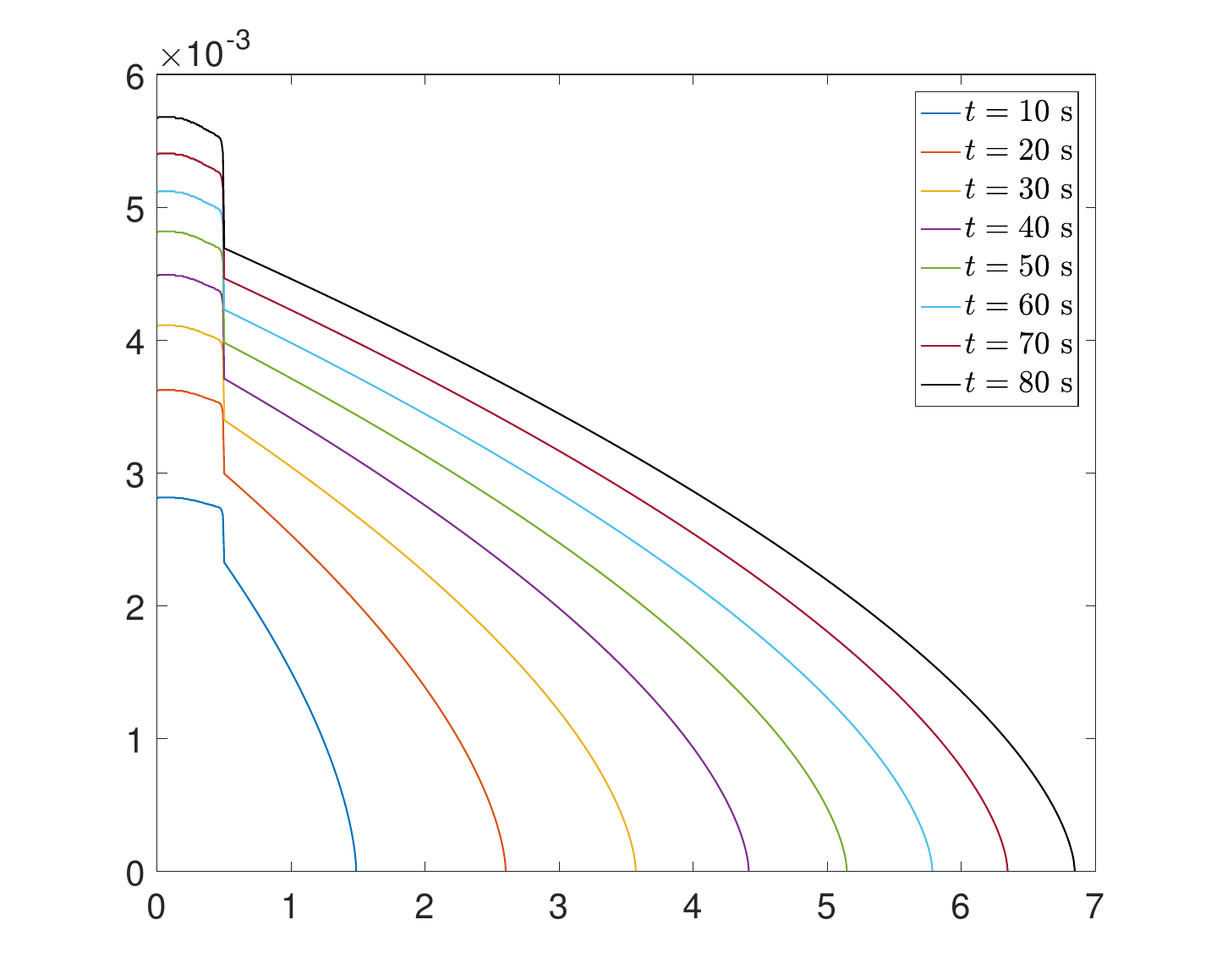}
\put(-325,0){$x$ [m]}
\put(-112,0){$x$ [m]}
\put(-440,155){$\textbf{a)}$}
\put(-215,155){$\textbf{b)}$}
\put(-420,75){\rotatebox{90}{$w(x)$ [m]}}
\put(-205,75){\rotatebox{90}{$w(x)$ [m]}}
\begin{center}
\includegraphics[scale=0.30]{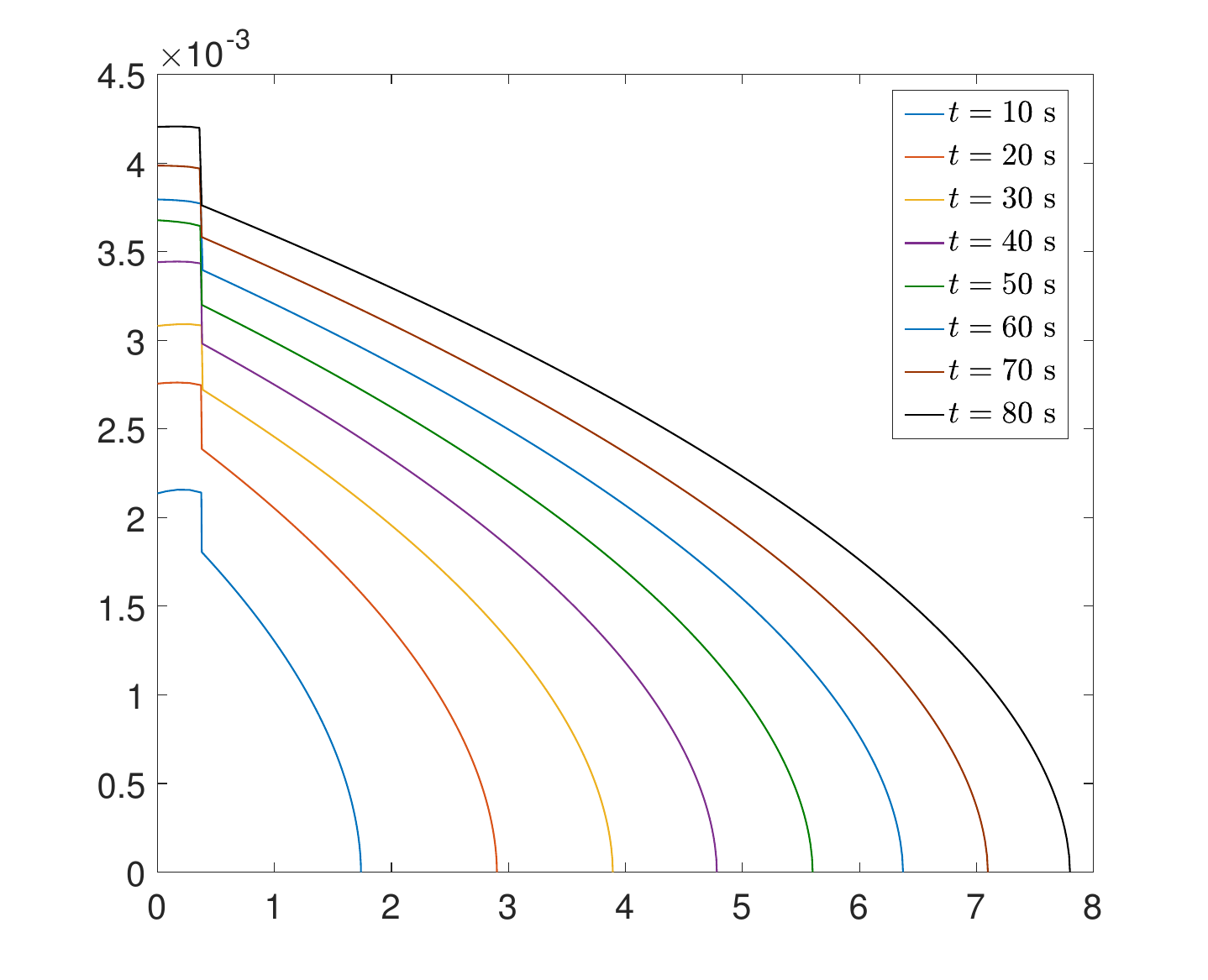}
\put(-210,155){$\textbf{c)}$}
\put(-210,70){\rotatebox{90}{$w(x)$ [m]}}
\put(-112,0){$x$ [m]}
\end{center}
\caption{Crack opening evolution, $w$ [m], during the propagation stage for: a) the elastic case, b) the large yield case ($c=1.33$ MPa), c) the small yield case ($c=2$ MPa). }
\label{w_profile_prop}
\end{center}
\end{figure}
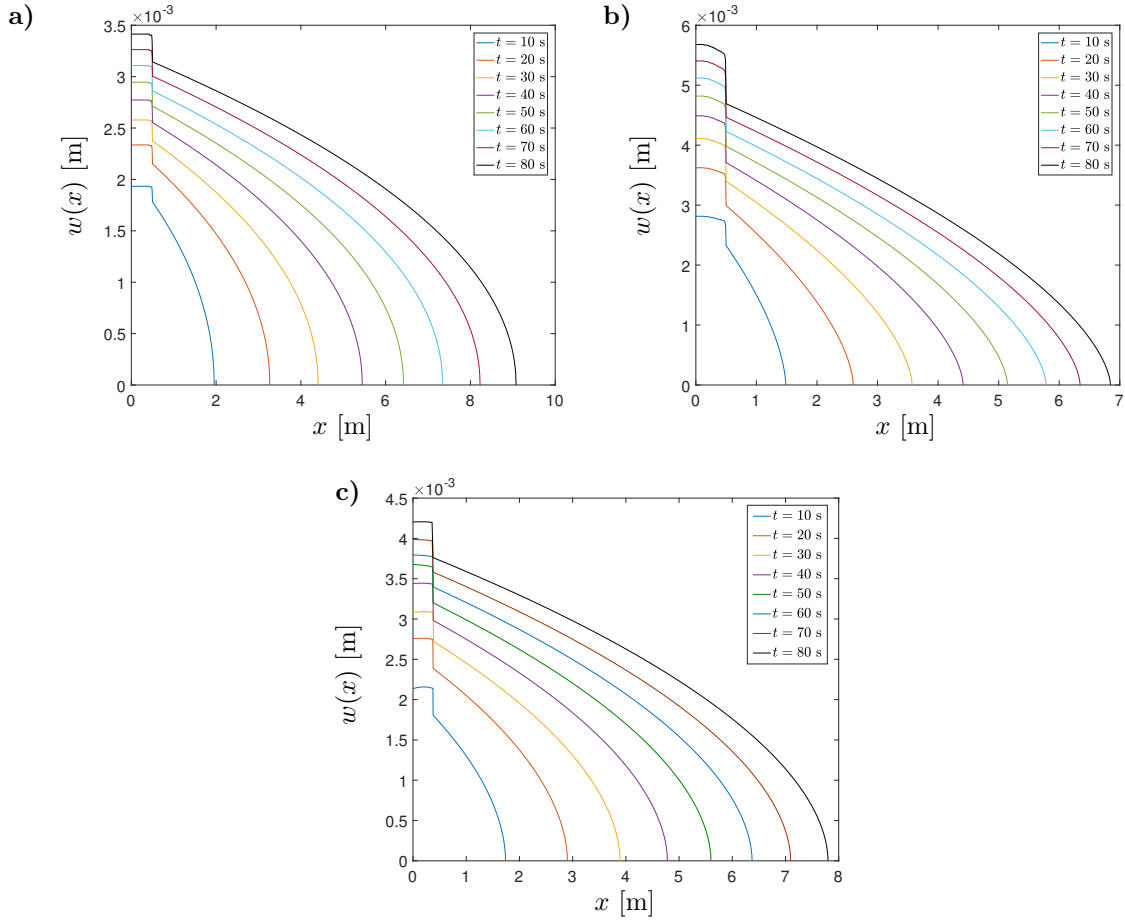

\begin{figure}[H]
\begin{center}
\includegraphics[scale=0.30]{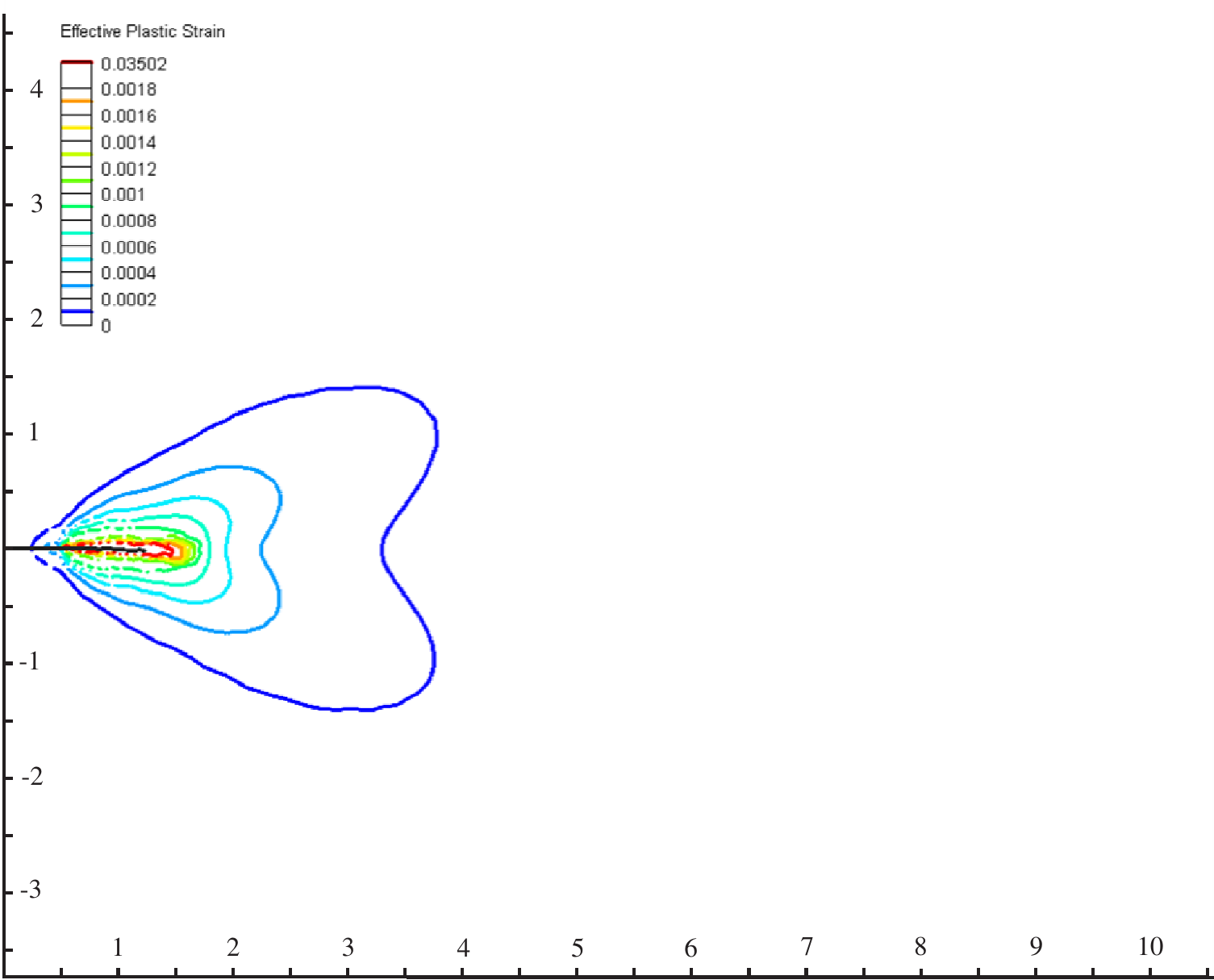}
\includegraphics[scale=0.30]{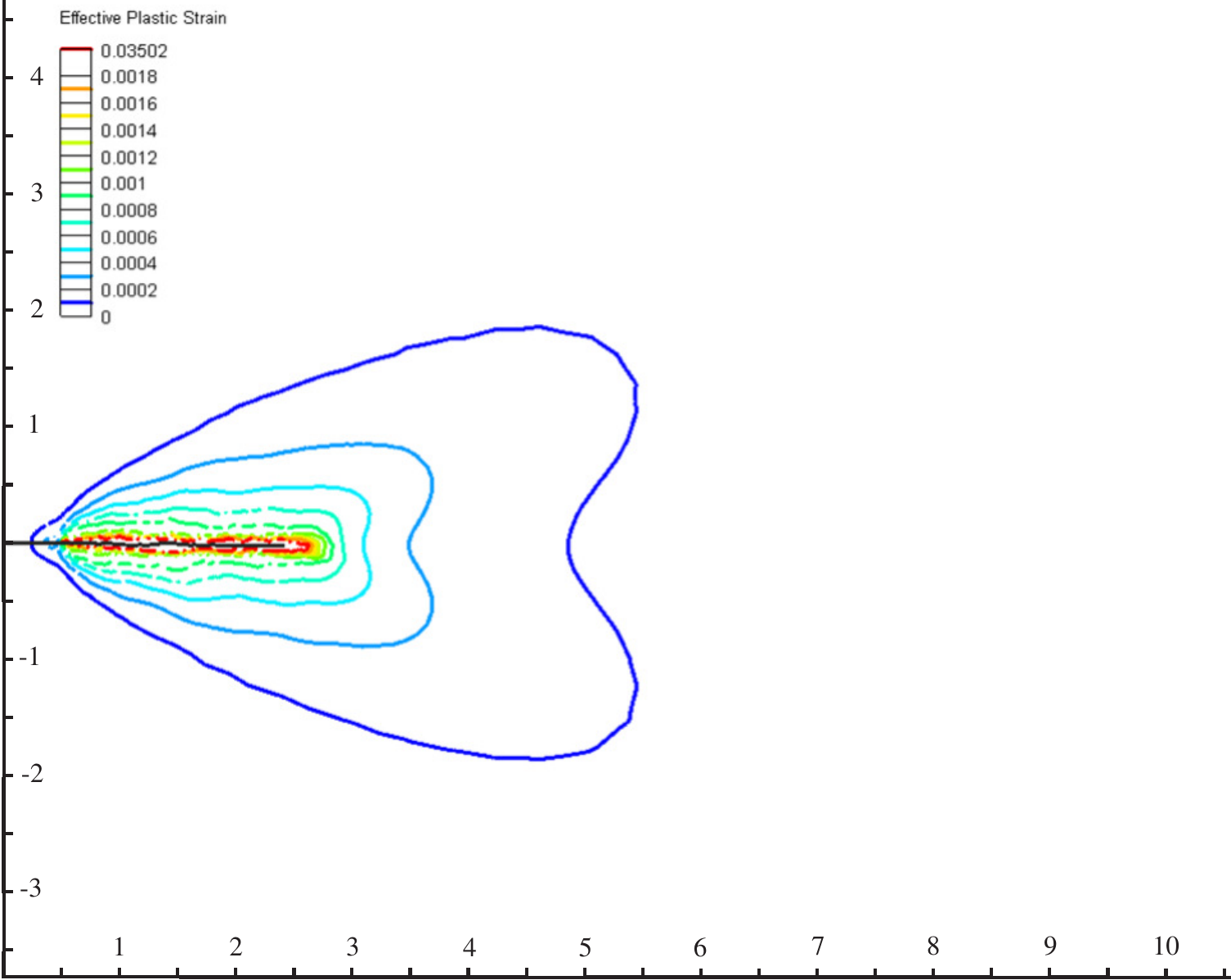}
\put(-270,155){$t=10$ s}
\put(-45,155){$t=20$ s}
\begin{center}
\includegraphics[scale=0.30]{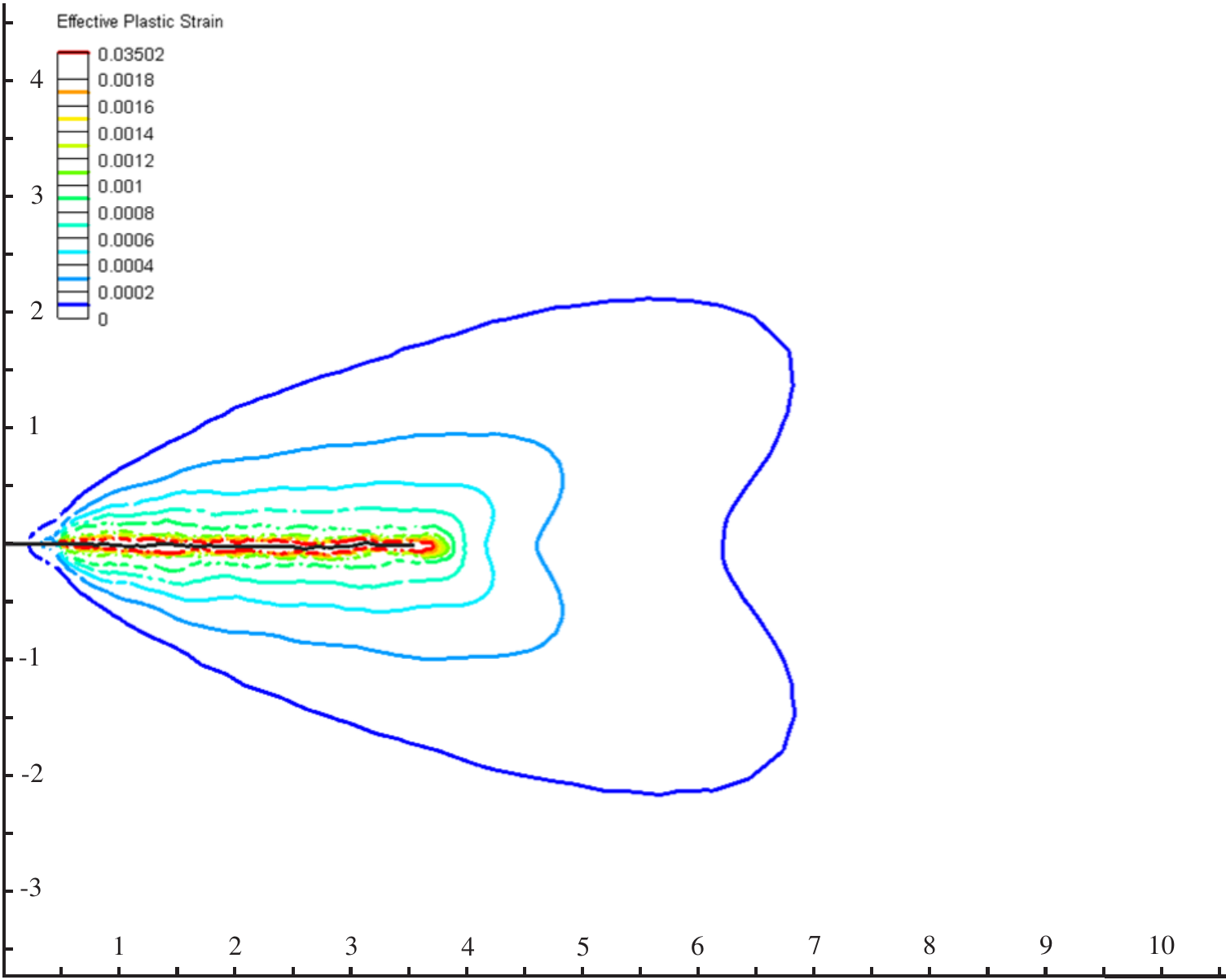}
\includegraphics[scale=0.30]{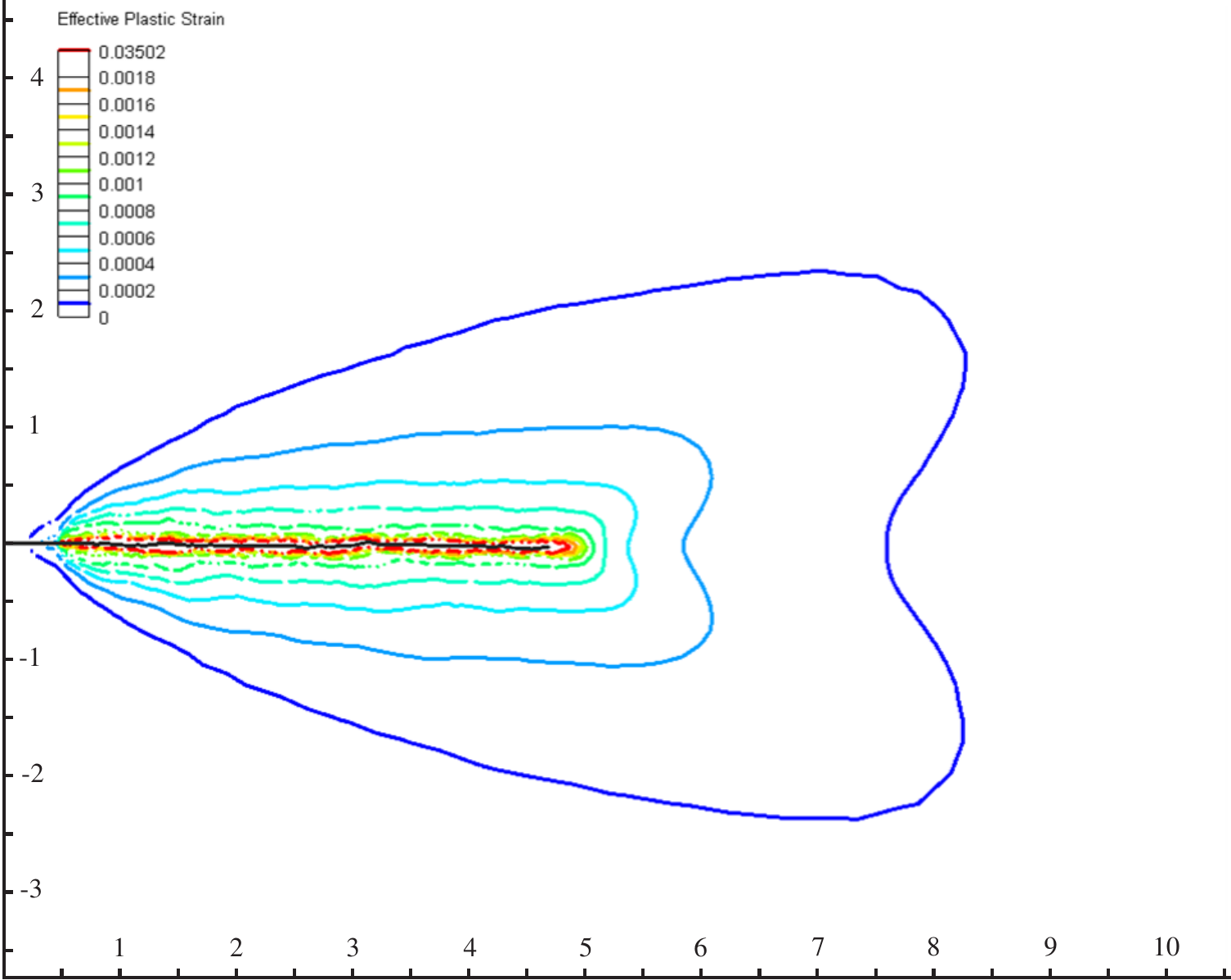}
\put(-270,155){$t=30$ s}
\put(-45,155){$t=40$ s}
\end{center}
\includegraphics[scale=0.30]{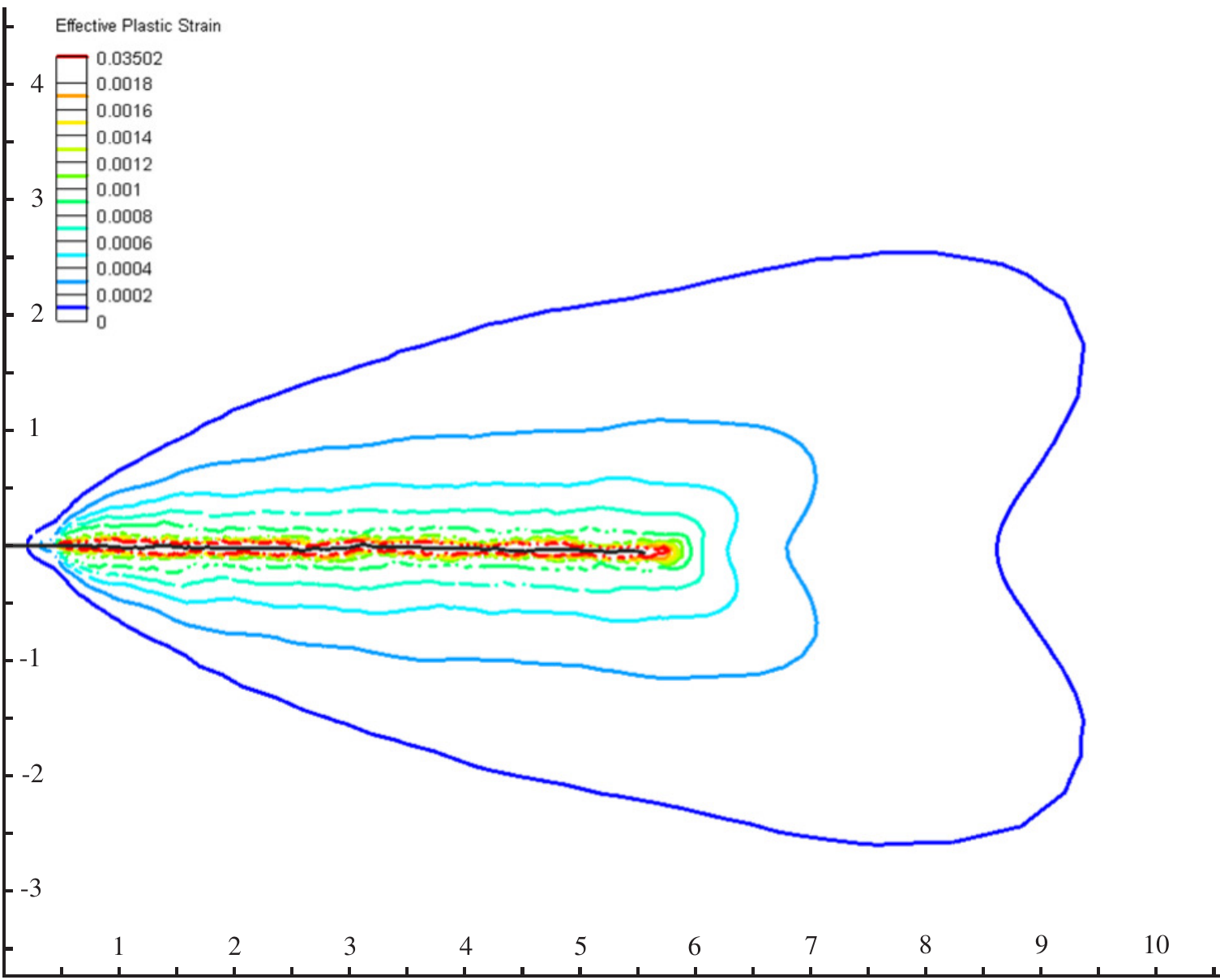}
\includegraphics[scale=0.30]{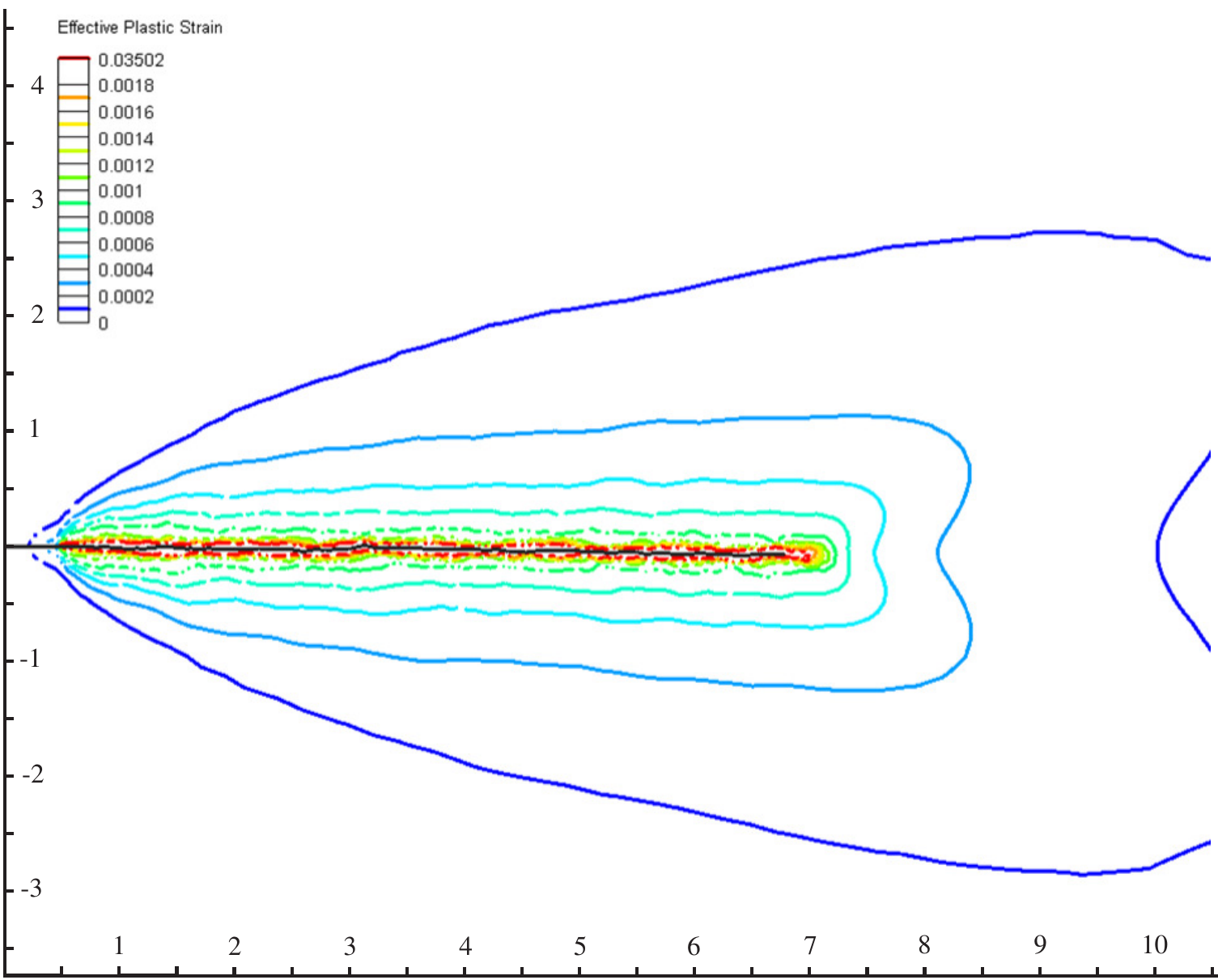}
\put(-270,155){$t=60$ s}
\put(-45,155){$t=80$ s}
\caption{Evolution of the plastic deformation zone for the elasto-plastic fracture depicted by the accumulated effective plastic strain. In each of the subplots the most external isoline refers to the value $10^{-4}$. The spatial scale is in meters.}
\label{plastic_zone}
\end{center}
\end{figure}

\begin{figure}[H]
\begin{center}
\includegraphics[scale=0.30]{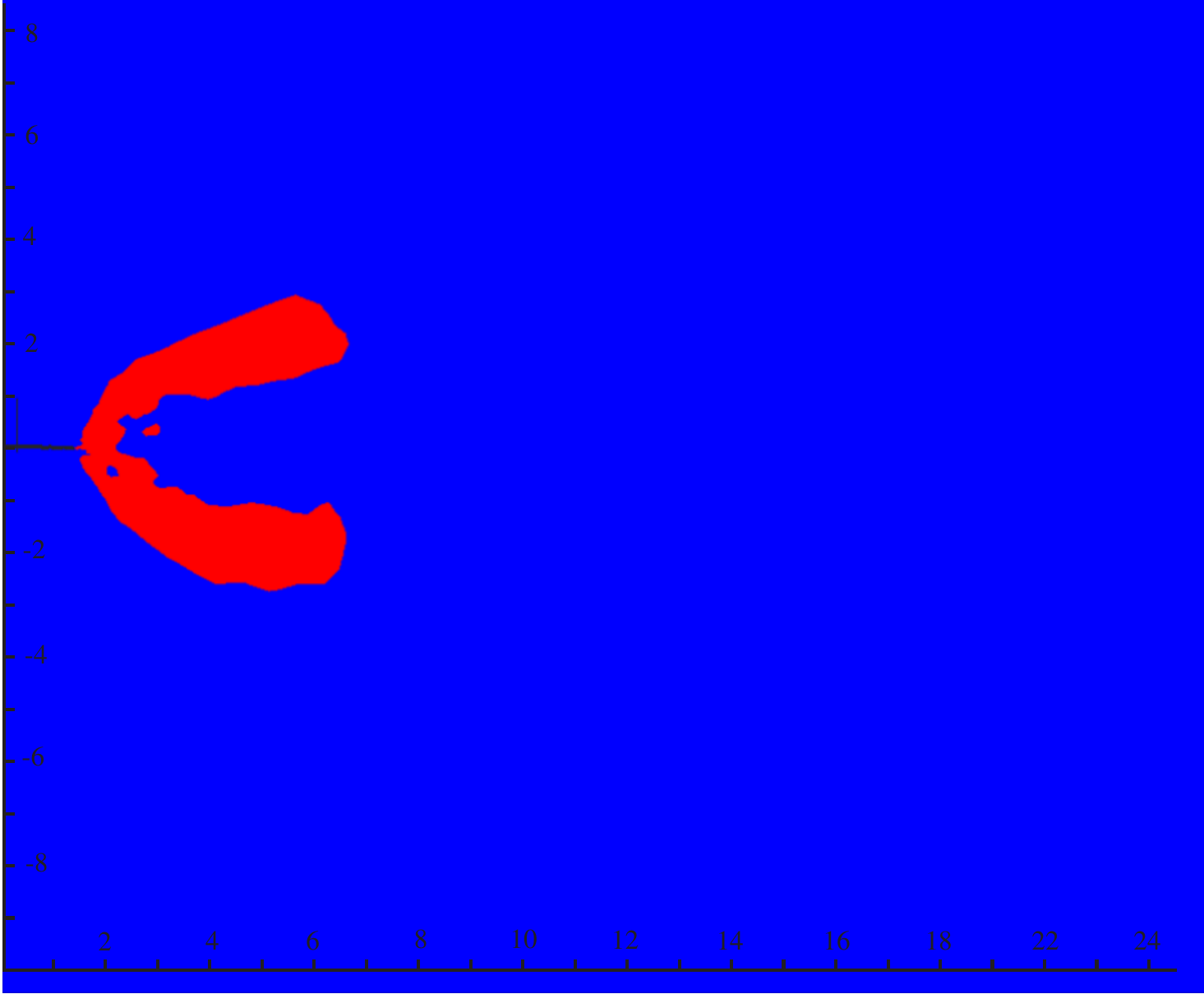}
\includegraphics[scale=0.30]{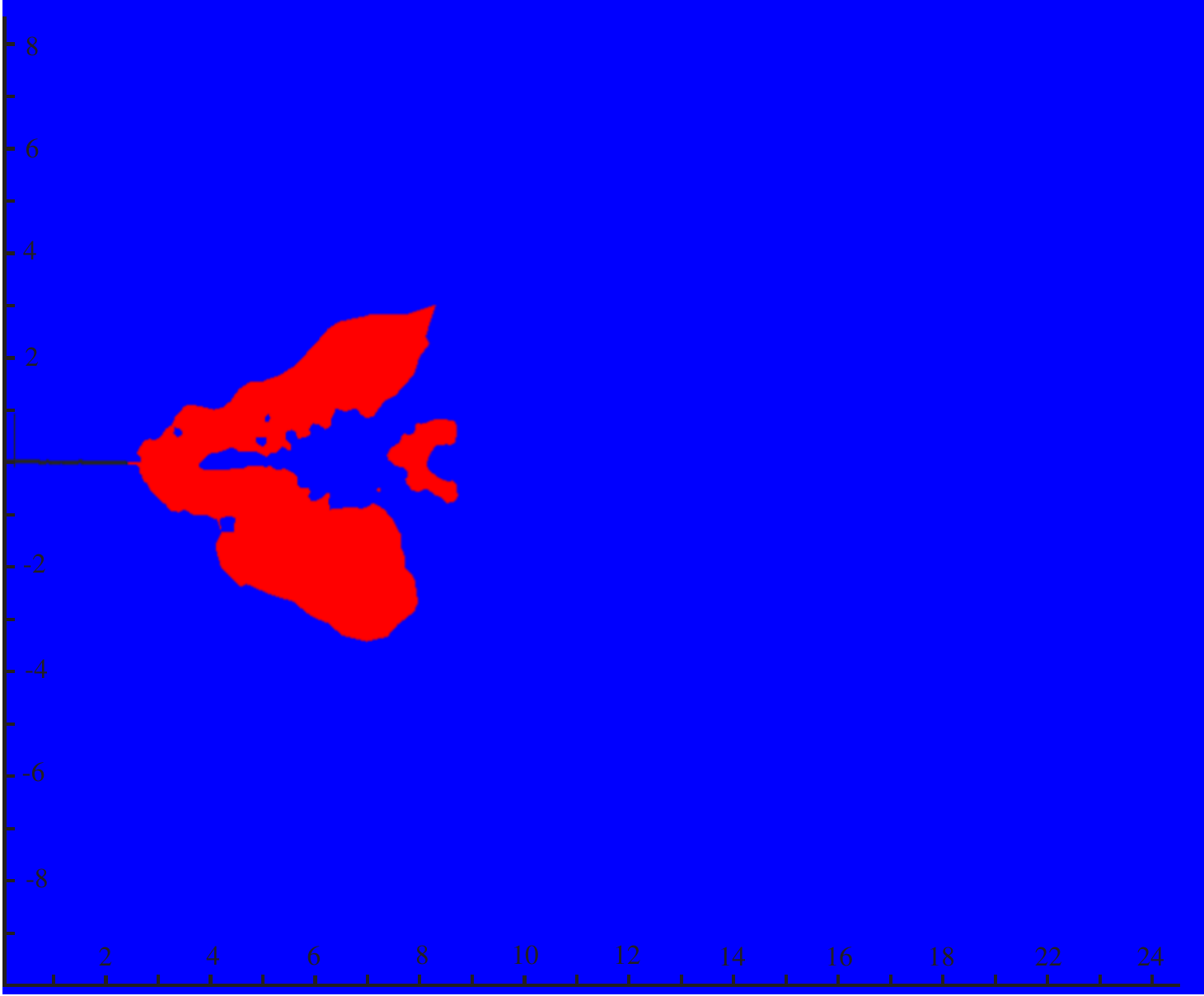}
\put(-270,155){$t=10$ s}
\put(-45,155){$t=20$ s}
\begin{center}
\includegraphics[scale=0.30]{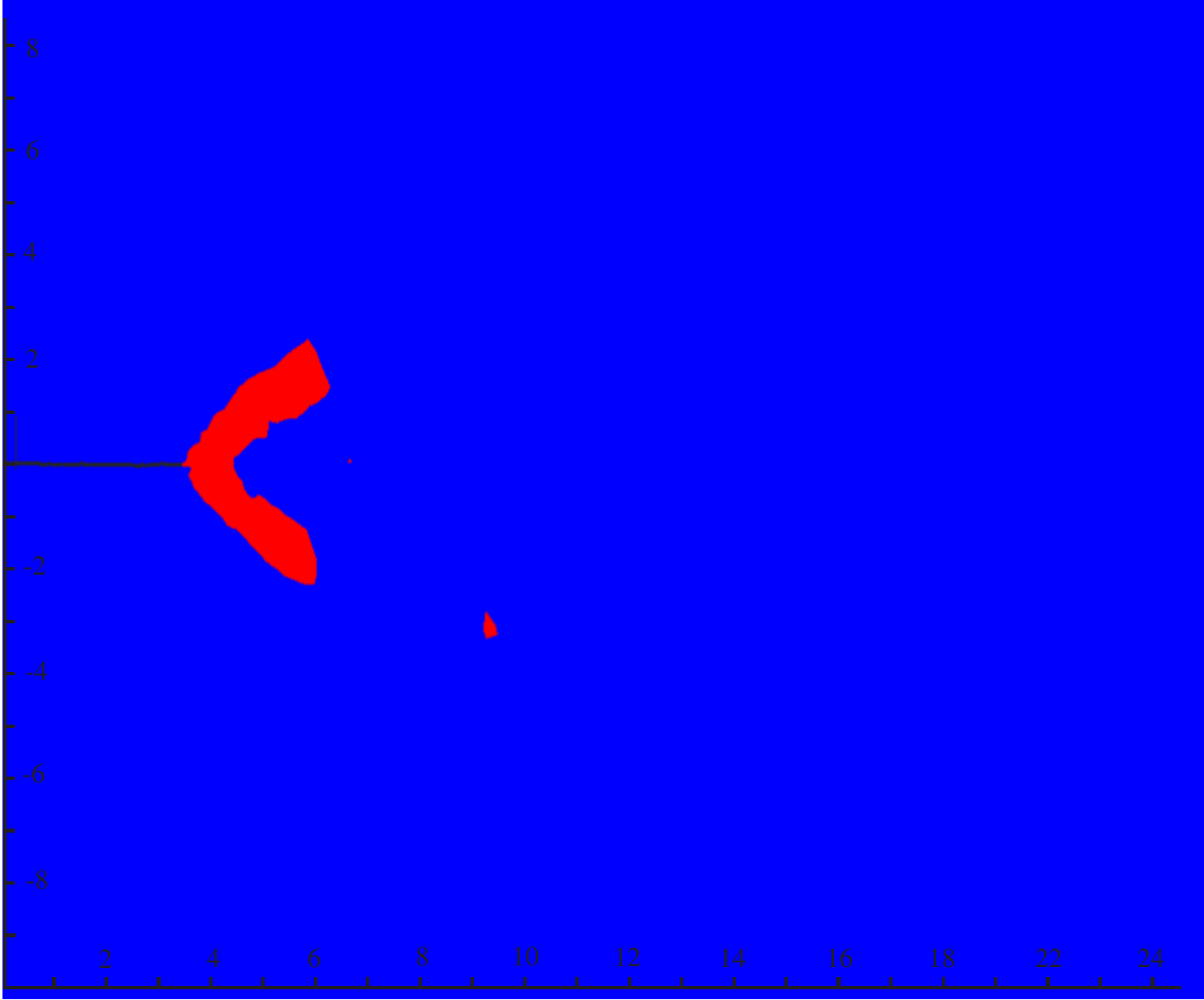}
\includegraphics[scale=0.30]{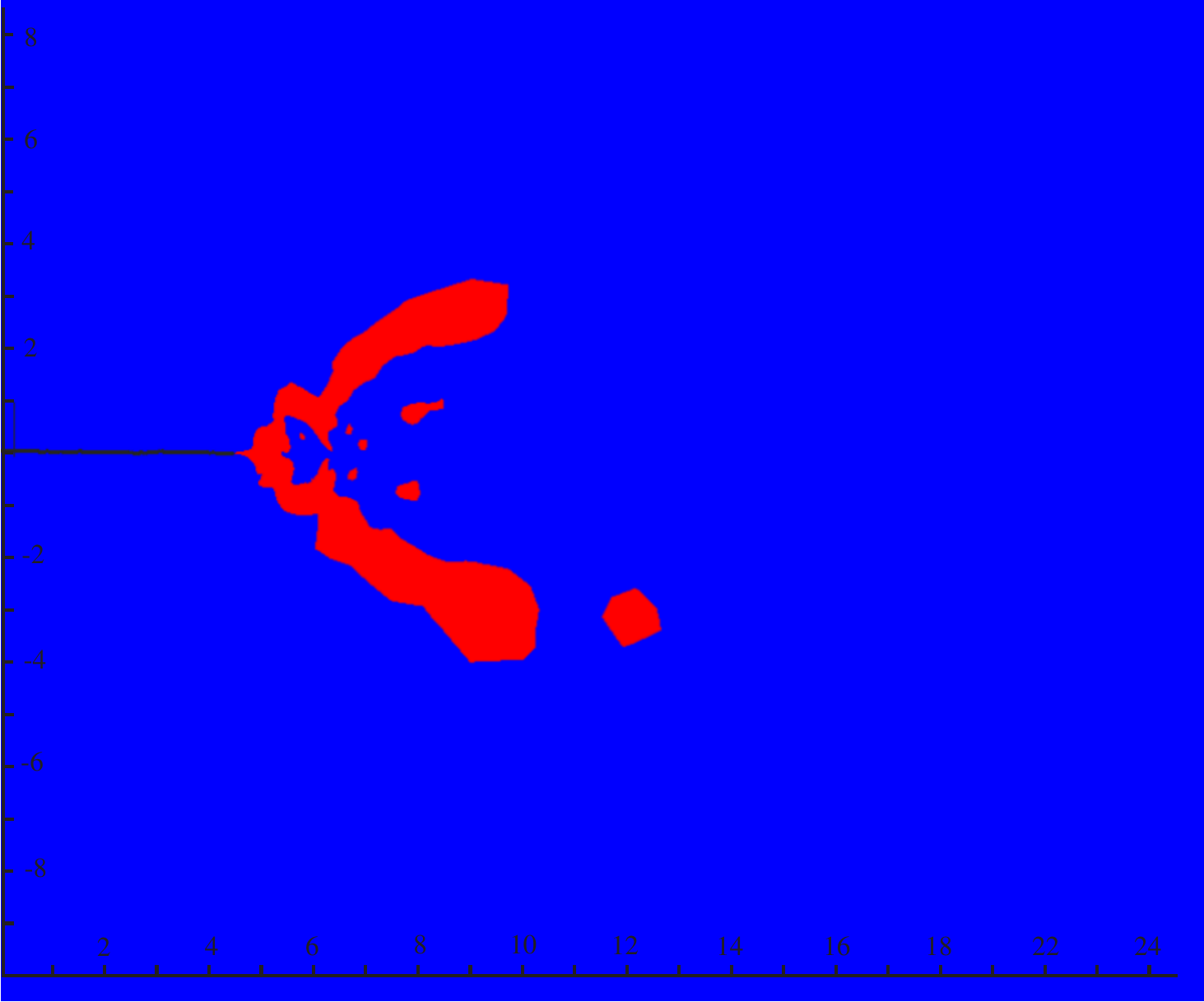}
\put(-270,155){$t=30$ s}
\put(-45,155){$t=40$ s}
\end{center}
\includegraphics[scale=0.30]{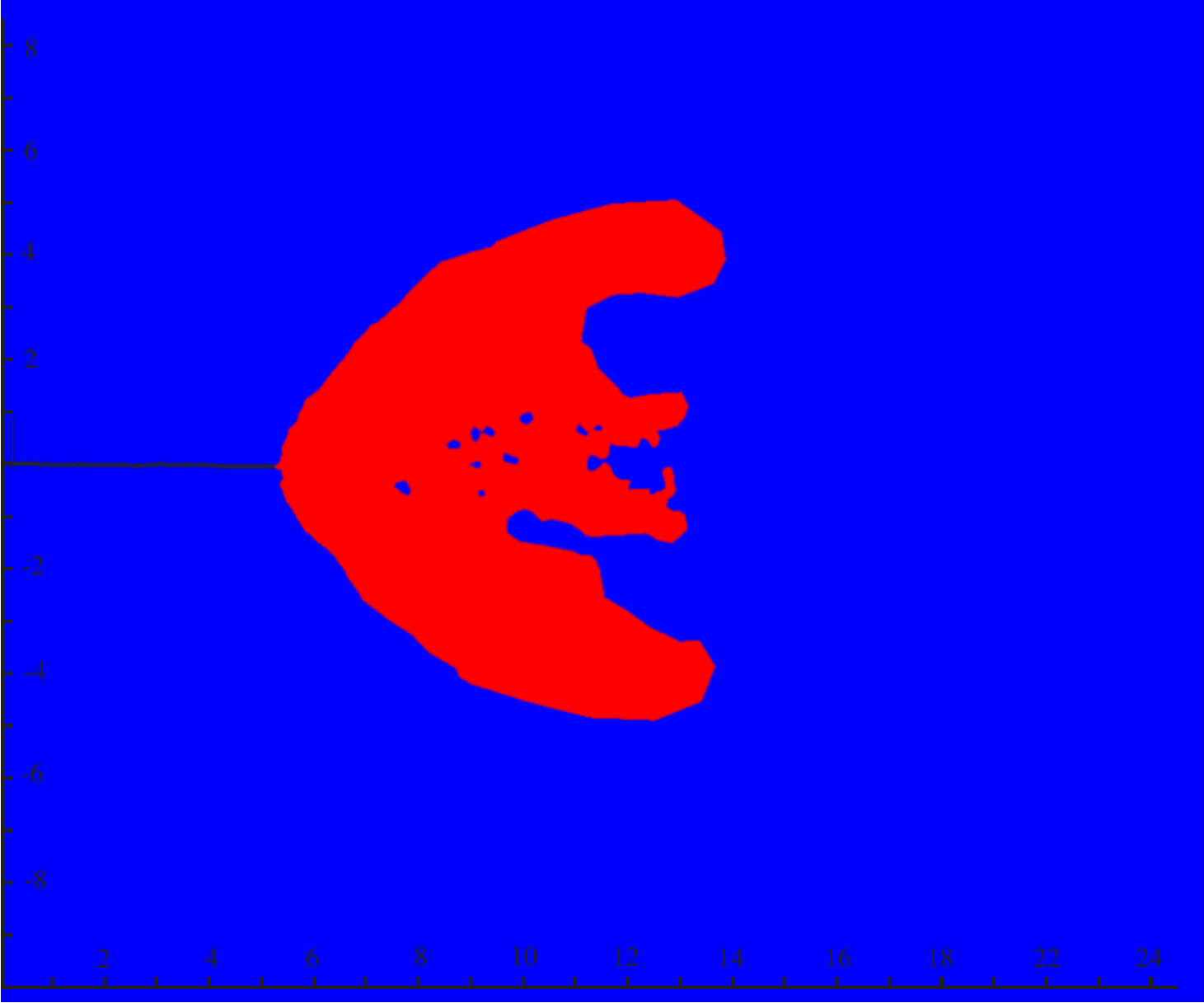}
\includegraphics[scale=0.30]{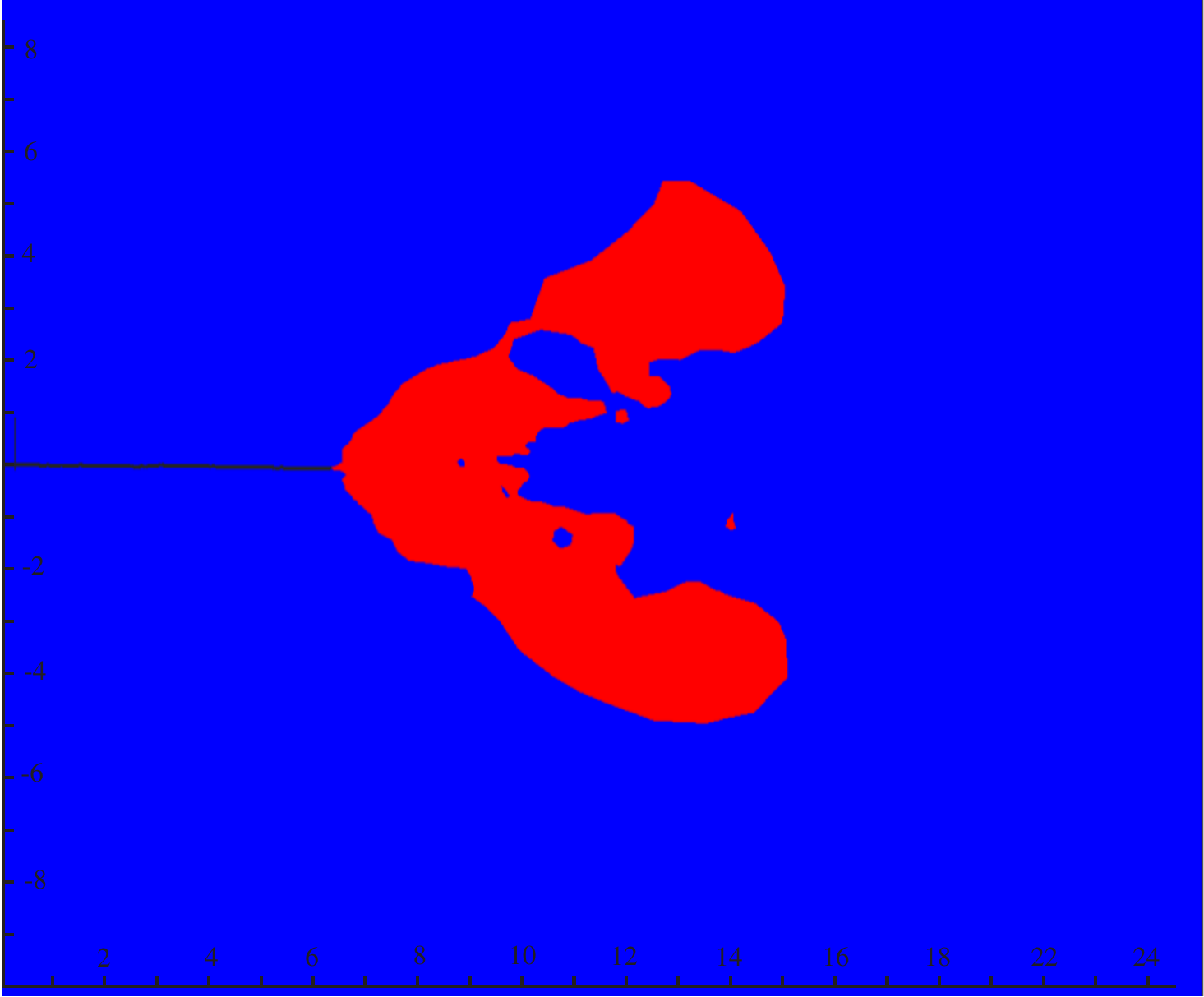}
\put(-270,155){$t=60$ s}
\put(-45,155){$t=80$ s}
\caption{Instantaneous plastic deformation for the elasto-plastic fracture at different moments of time depicted by the plastic flag. The red zones denote the elements that undergo plastic yield at various time instants. The spatial scale (in meters) is different from that shown in Figure \ref{plastic_zone}.}
\label{plastic_flag}
\end{center}
\end{figure}

The computational results for the crack propagation stage are depicted in Figures \ref{L_w_p_prop}--\ref{plastic_flag}. The general trends, already reported in many studies (see e. g. \cite{Papanastasiou_1997,Papanastasiou_algorithm,Papanastasiou_1999a,Papanastasiou_2000,Wrobel_plasticity}), that involve comparison of elastic and elasto-plastic fractures, hold also for the obtained results. The plastic deformation mechanism causes the hydraulic fracture to grow shorter and wider than its elastic counterpart. As can be seen in Figure \ref{L_w_p_prop}c) the wellbore pressure, $p_\text{w}$, is noticeably higher in the elasto-plastic cases. Consequently, the power input needed to propagate the fracture is expected to be larger as compared to the scenario where the elastic deformations prevail. Note also that in all cases the crack length remains  constant up to the moment when the breakdown pressure is achieved. This initial time interval is slightly longer for the elasto-plastic fractures.

In Figure \ref{w_profile_prop} we demonstrate the fracture profiles at different time instants. A footprint of the initial fracture is discernible in all analyzed cases. This effect originates from the form of initial condition (initial crack opening) which does not include plastic deformation. Thus, as the fracture advances, the plastic deformation zone develops starting from the locus of the initial crack length (due to the cohesive zone type crack propagation condition this phenomenon involves also the elastic fracture). As a result, 
the shape peculiarity is observed over the span of the initial crack length. Similar trend has already been reported in other studies on the elasto-plastic hydraulic fractures (see e. g. the papers of \cite{Papanastasiou_1997,Wrobel_plasticity}). Clearly, the corresponding effect become less pronounced with the fracture growth.

The plastic deformation of the fractured material can substantially affect the closure pattern of the hydraulic fracture \citep{Papanastasiou_2000,van_Dam}. For this reason, a proper identification of the extent of plastic yielding is important for the pressure decline analysis conducted in the next subsection. In Figure \ref{plastic_zone} we present the contour plots of the accumulated plastic strains (which represent the active plastic zones) obtained for the large yield case (the second variant of the HF problem) at different time instants. It can be seen that a substantial part of the rock adjacent to the crack surface undergoes plastic deformation. The plastic deformation zone grows with the fracture advance. As explained above, no plastic yield is observed over the span of the initial crack length. The characteristic shape of the plastic deformation zone results from the near tip deformations related to the stress concentration. In Figure \ref{plastic_flag} we present the plastic flag parameter obtained at the time instants that correspond to the respective subplots of plastic strains. The plastic flag parameter has a binary value (0 or 1) and informs which finite elements undergo plastic yield at a given moment of time (the plastic areas behind the advancing tip unload elastically). It shows that the inelastic deformations are concentrated along two strips originating from the crack tip and inclined symmetrically with respect to the fracture plane. The strips form a v-shaped zone. Similar deformation pattern for the Mohr-Coulomb yield criterion was obtained by \cite{Papanastasiou_1997}.

\begin{figure}[H]
\begin{center}
\includegraphics[scale=0.30]{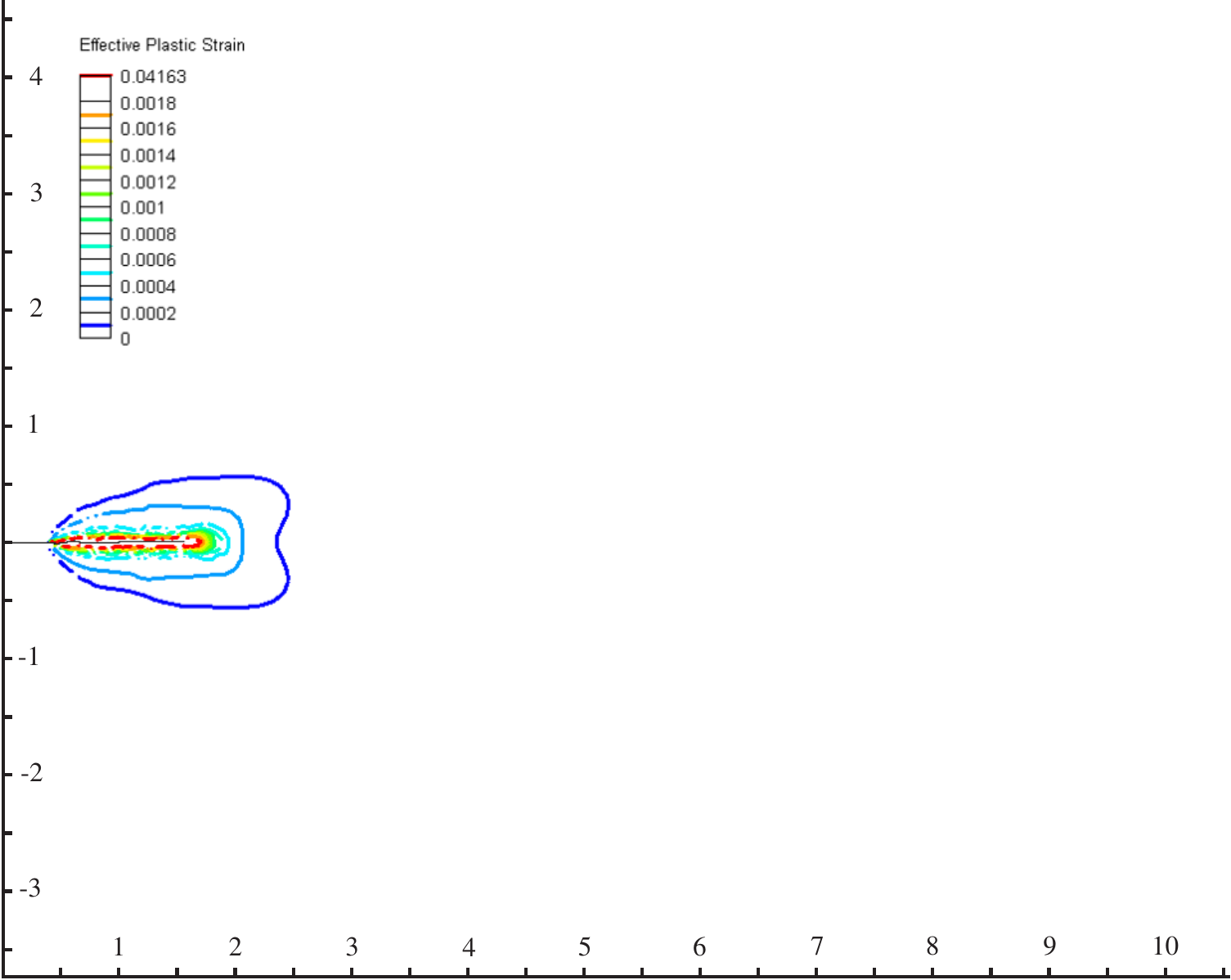}
\includegraphics[scale=0.30]{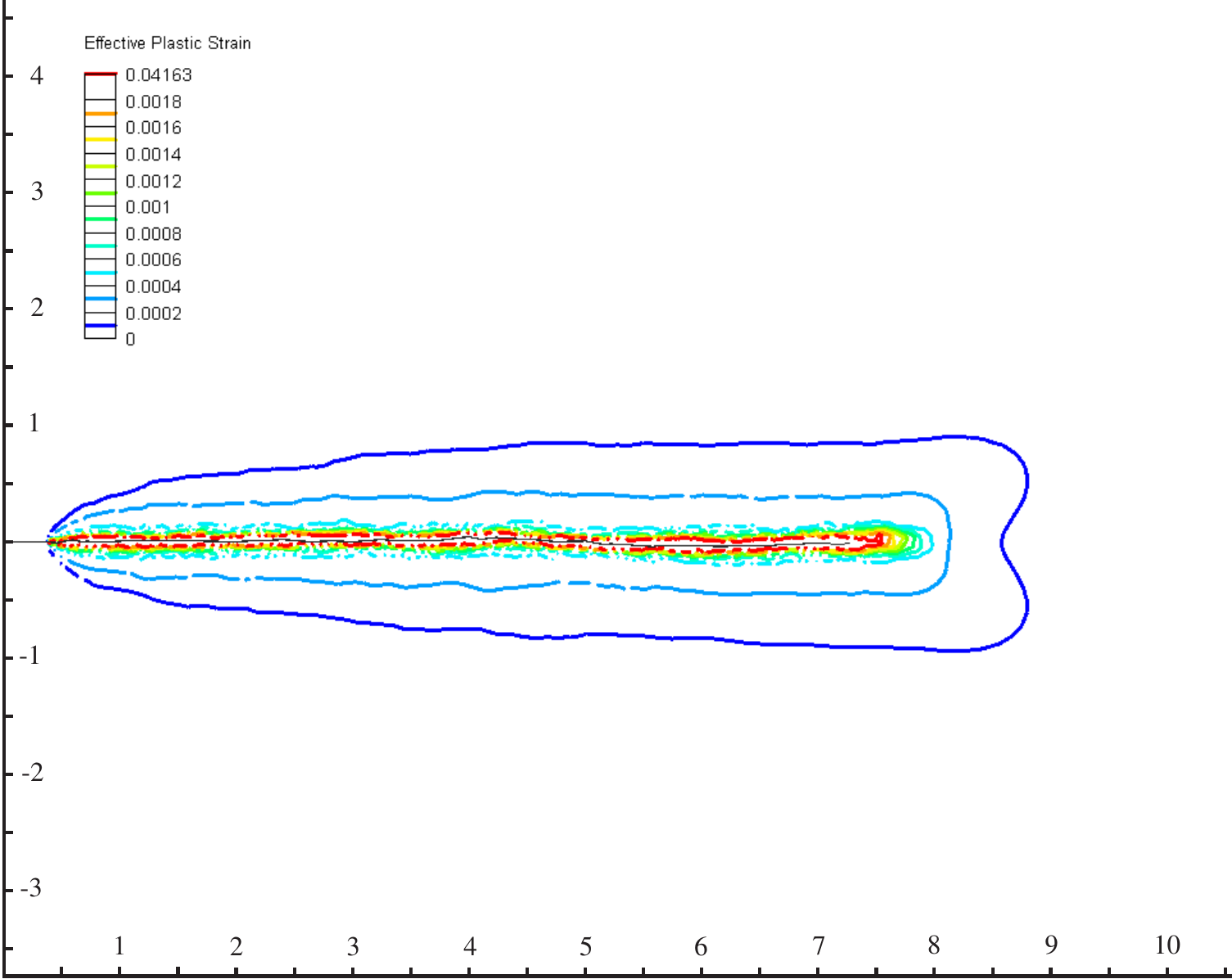}
\put(-270,155){$t=10$ s}
\put(-45,155){$t=80$ s}
\caption{Evolution of the plastic deformation zone for the small yield case depicted by the accumulated effective plastic strain. In each of the subplots the most external isoline refers to the value $10^{-4}$. The spatial scale is in meters.}
\label{plastic_zone_small_yield}
\end{center}
\end{figure}

\begin{figure}[htb!]
\begin{center}
\includegraphics[scale=0.32]{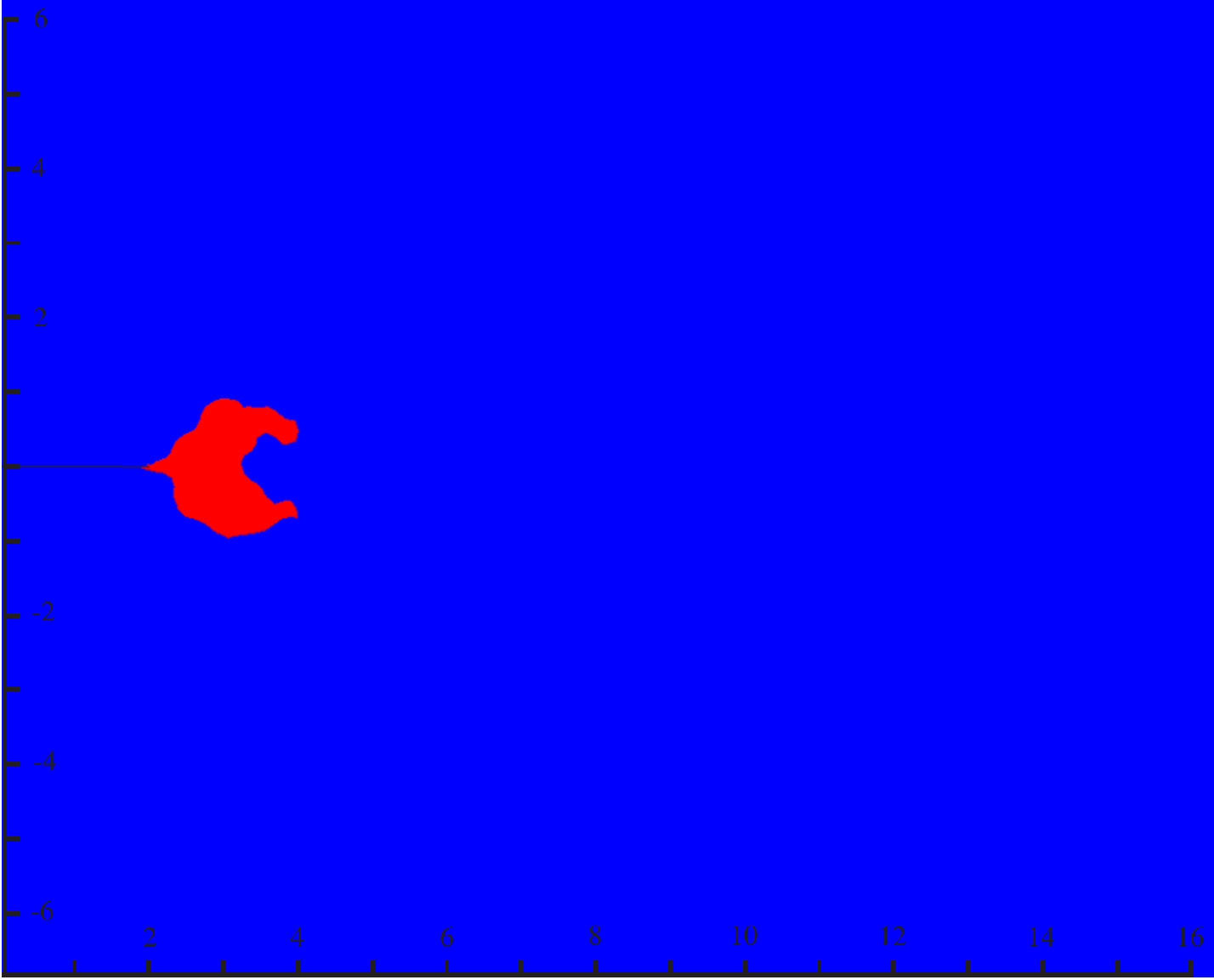}
\includegraphics[scale=0.32]{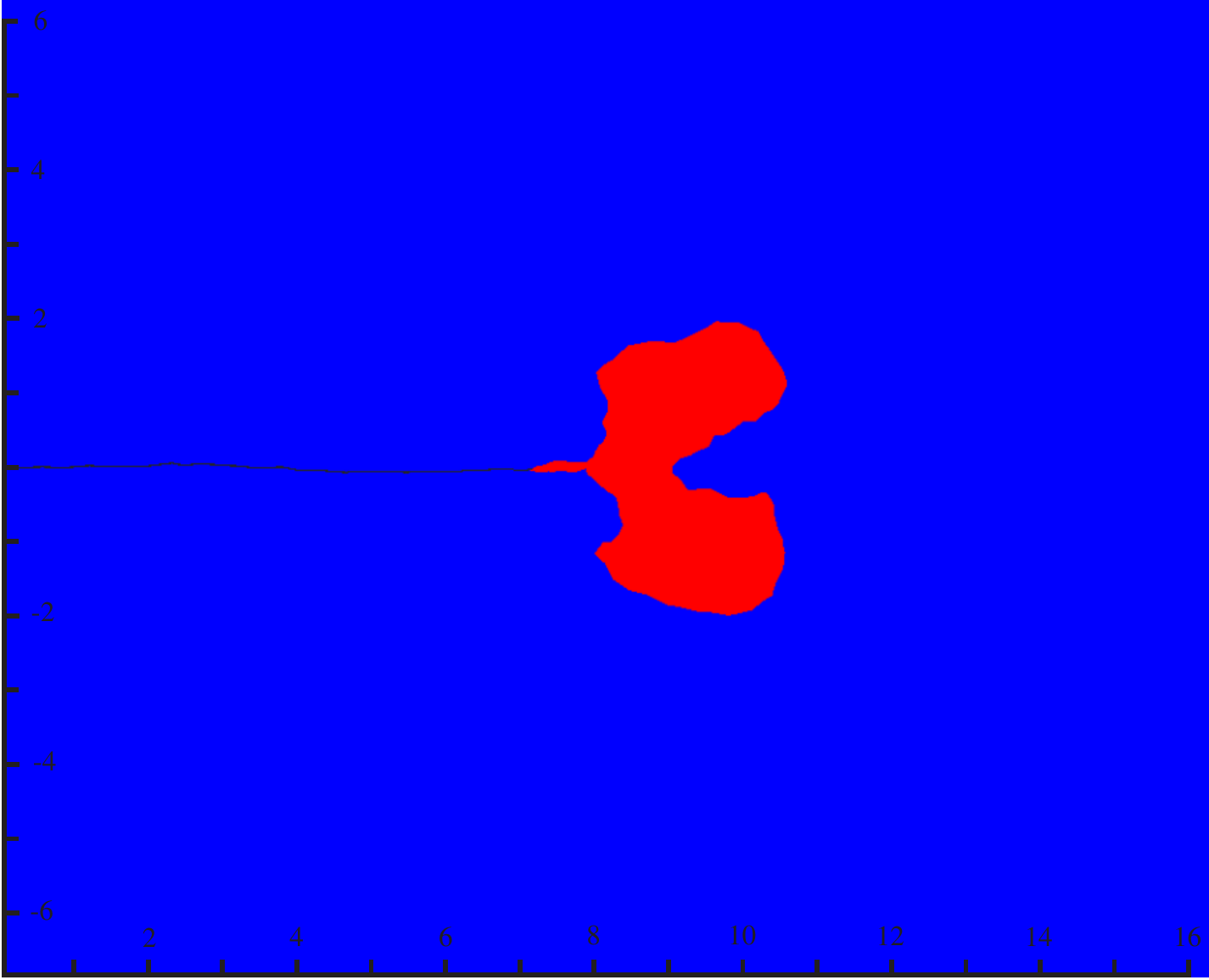}
\put(-270,155){$t=10$ s}
\put(-45,155){$t=80$ s}
\caption{Instantaneous plastic deformation for the small yield case at different moments of time depicted the plastic flag. The red zones denote the elements that undergo plastic yield at various time instants. The spatial scale (in meters) is different from that shown in Figure \ref{plastic_zone_small_yield}.}
\label{plastic_flag_small_yield}
\end{center}
\end{figure}

Respective graphs of the plastic yield indicators for the small yield case (the third variant of the problem - $c=2$ MPa) are shown in Figures \ref{plastic_zone_small_yield} - \ref{plastic_flag_small_yield} for two time instants: $t=10$ s and $t=80$ s. As the general trends discussed above hold also here, we do not depict the full evolution of the zone (i. e. at all time instants shown in Figures \ref{plastic_zone} - \ref{plastic_flag}) . Naturally, the sizes of the plastic deformation zones are smaller than before.

\begin{figure}[H]
\begin{center}
\includegraphics[scale=0.30]{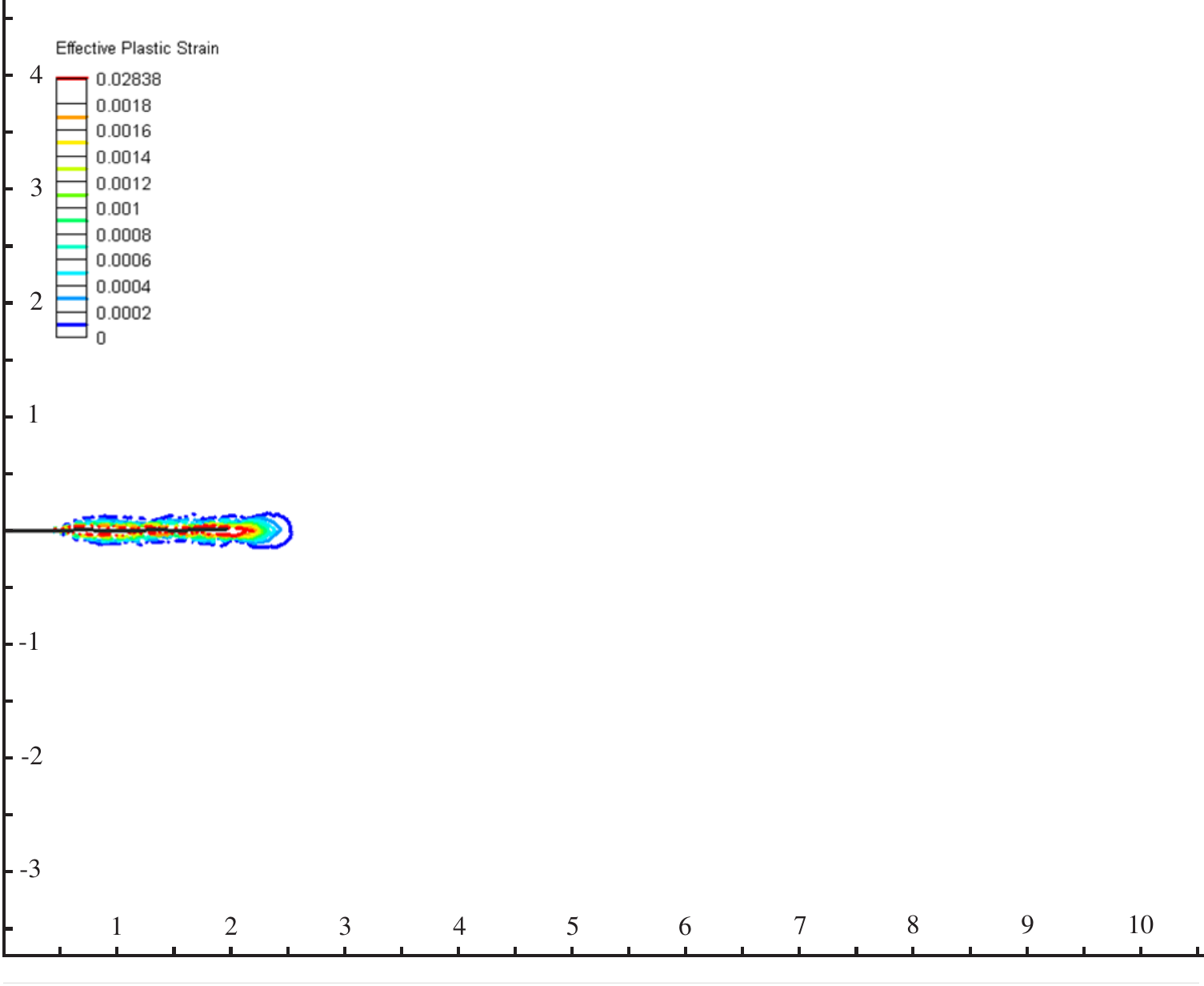}
\includegraphics[scale=0.30]{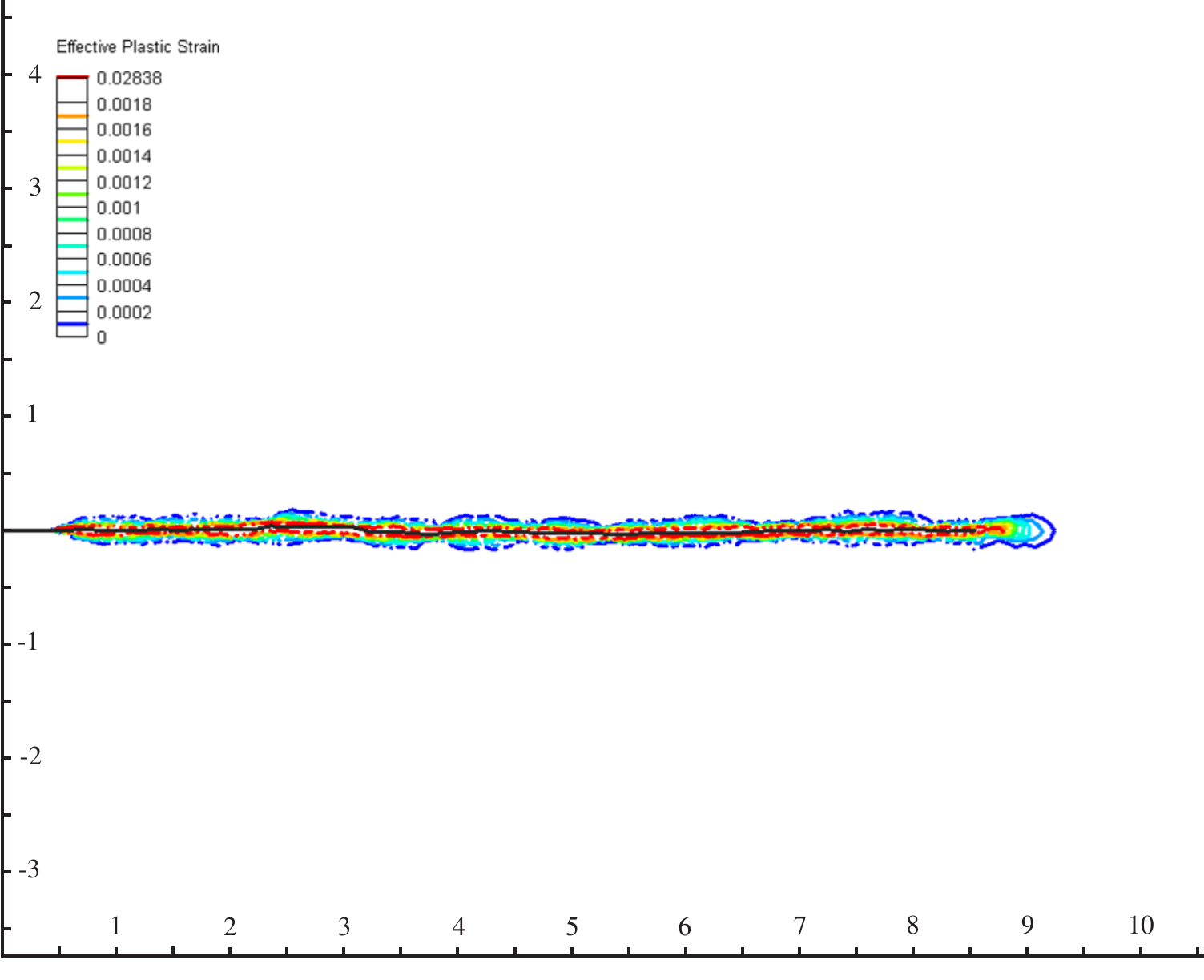}
\put(-270,155){$t=10$ s}
\put(-45,155){$t=80$ s}
\caption{Evolution of the plastic deformation zone for the elastic fracture depicted by the accumulated effective plastic strain. In each of the subplots the most external isoline refers to the value $10^{-4}$. The spatial scale is in meters.}
\label{plastic_zone_elast}
\end{center}
\end{figure}

\begin{figure}[htb!]
\begin{center}
\includegraphics[scale=0.30]{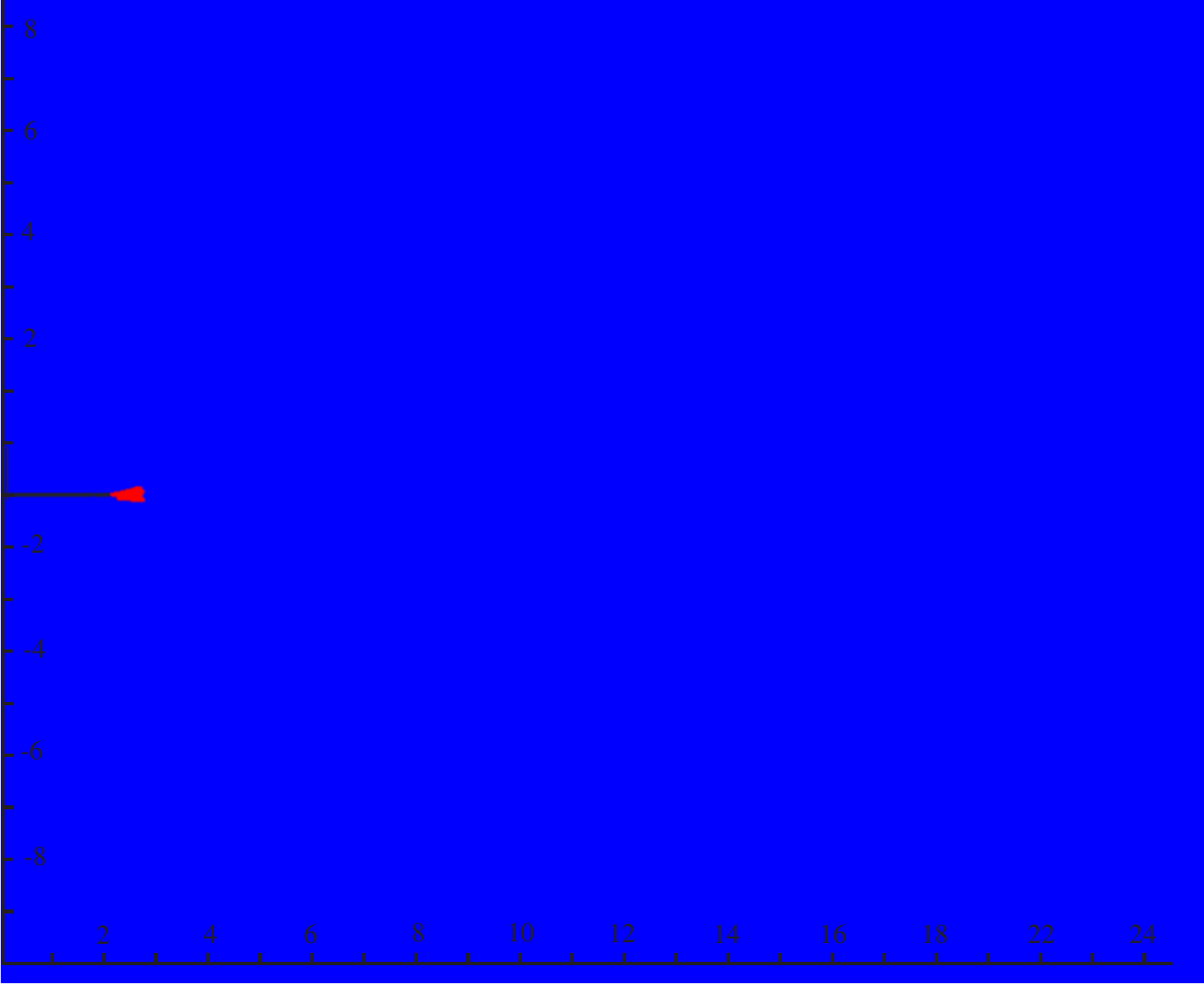}
\includegraphics[scale=0.30]{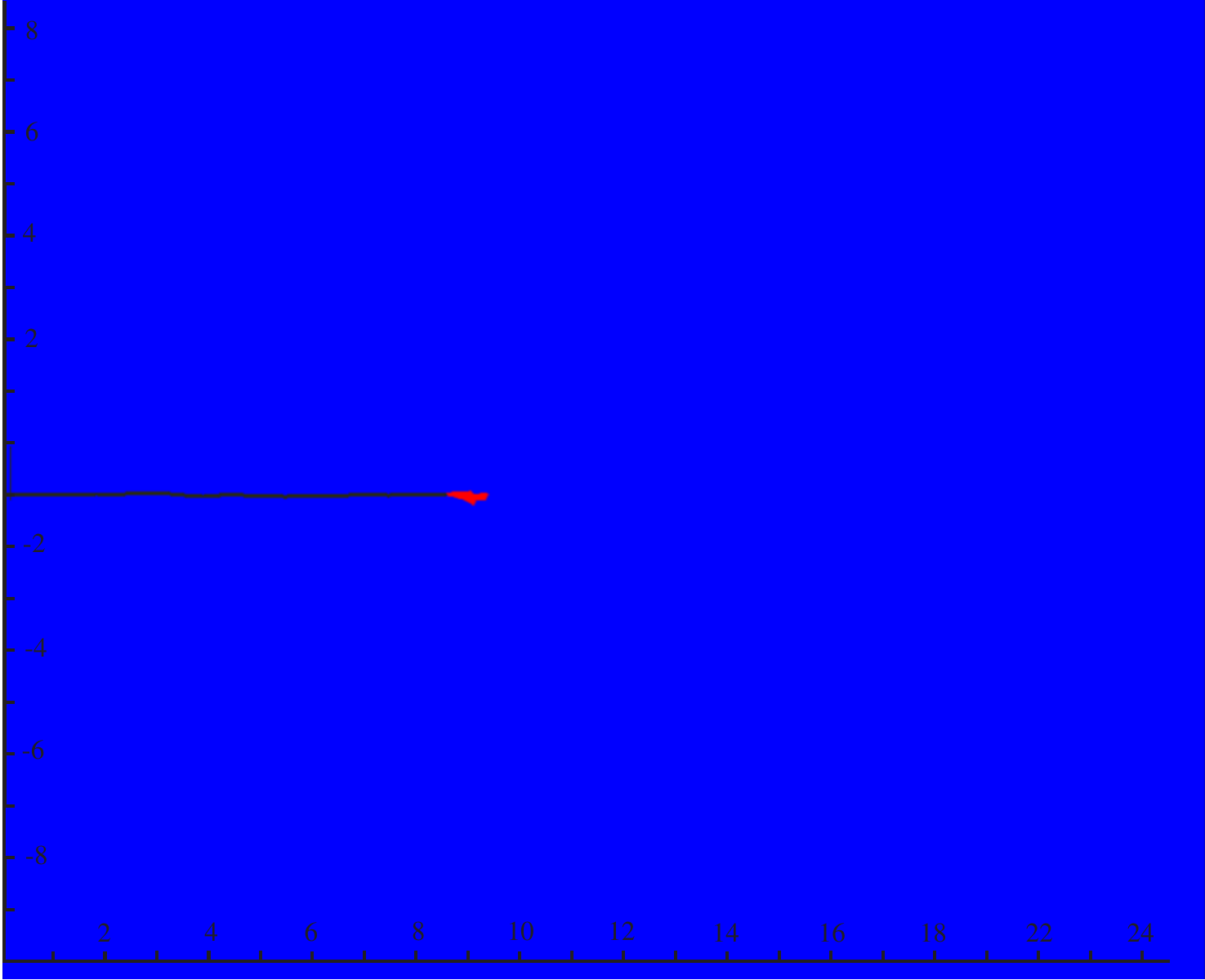}
\put(-270,155){$t=10$ s}
\put(-45,155){$t=80$ s}
\caption{Instantaneous plastic deformation for the elastic fracture at different moments of time depicted the plastic flag. The red zones denote the elements that undergo plastic yield at various time instants. The spatial scale (in meters) is different from that shown in Figure \ref{plastic_zone_elast}.}
\label{plastic_flag_elast}
\end{center}
\end{figure}

For the sake of comparison we show in Figures \ref{plastic_zone_elast} -- \ref{plastic_flag_elast} the accumulated plastic strains and the plastic flag parameter for the elastic fracture at two time instants ($t=10$ s and $t=80$ s). This time the plastic yield is only related to tensile deformation in the cohesive zone. Thus, the respective inelastic deformation area forms a thin strip around the crack surface. The unit value of the plastic flag is obtained only in the fracture plane across the length of the cohesive zone.

\subsection{Fracture closure - pressure decline analysis}
\label{frac_clos_s}

\begin{figure}[htb!]
\begin{center}
\includegraphics[scale=0.30]{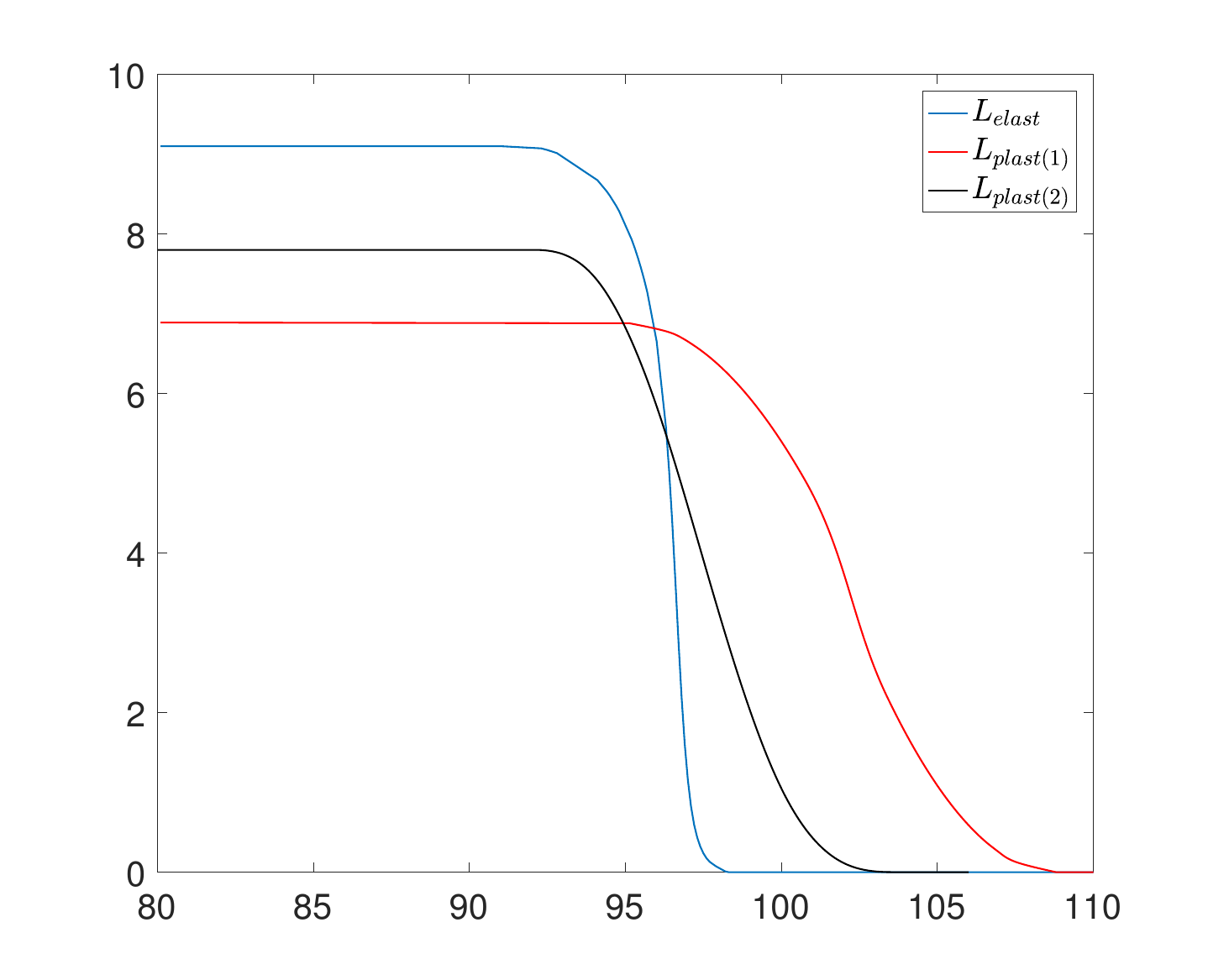}
\hspace{0mm}
\includegraphics[scale=0.30]{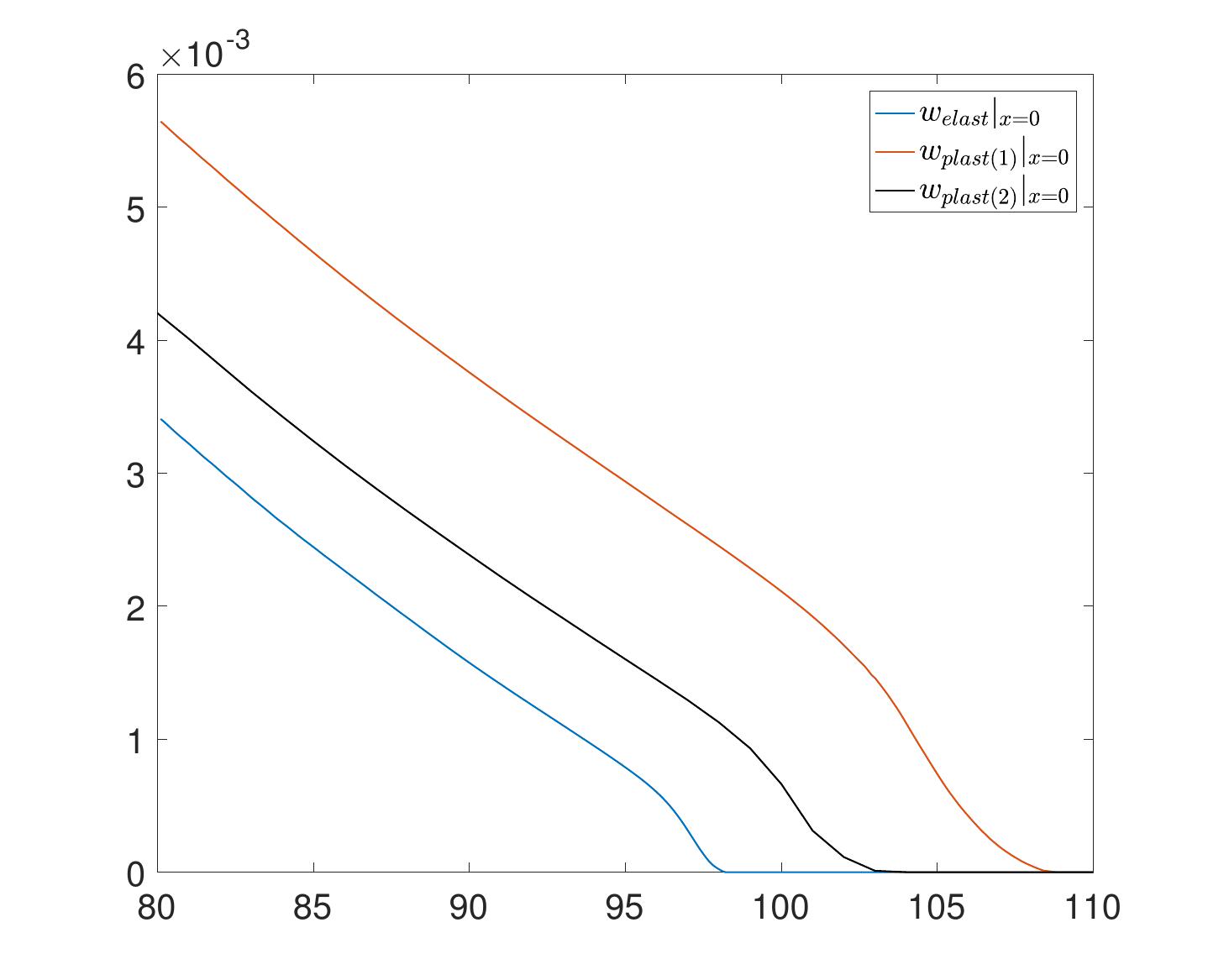}
\put(-320,0){$t$ [s]}
\put(-105,0){$t$ [s]}
\put(-440,155){$\textbf{a)}$}
\put(-215,155){$\textbf{b)}$}
\put(-420,70){\rotatebox{90}{$L(t)$ [m]}}
\put(-205,70){\rotatebox{90}{$w(0,t)$ [m]}}
\begin{center}
\includegraphics[scale=0.30]{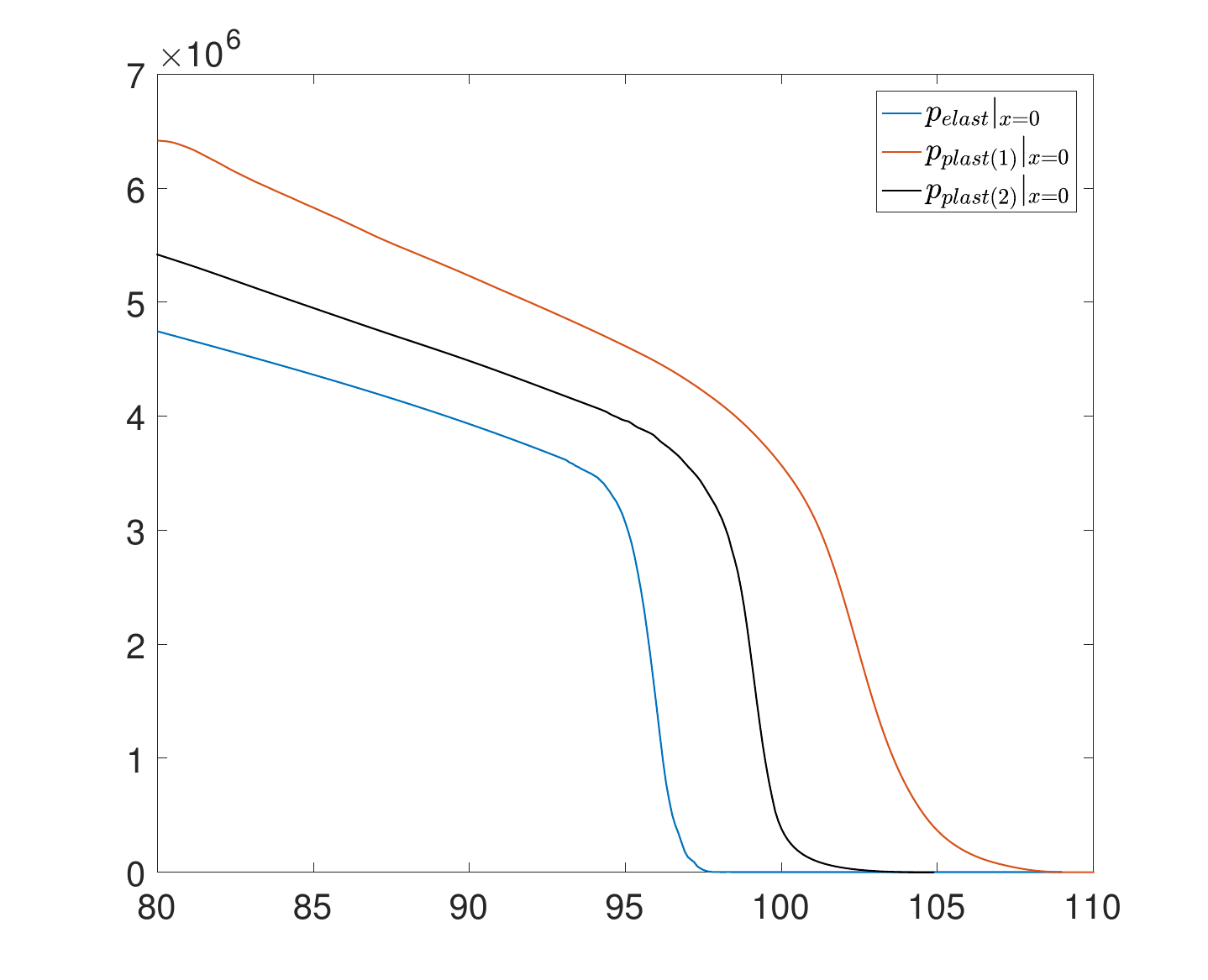}
\put(-210,155){$\textbf{c)}$}
\put(-205,75){\rotatebox{90}{$p_\text{w}$ [Pa]}}
\put(-105,0){$t$ [s]}
\end{center}
\caption{Fracture evolution during the closure stage: a) the crack half-length, $L(t)$ [m], b) the crack opening at the fracture mouth, $w(0,t)$ [m], c) the wellbore pressure, $p_\text{w}$ [Pa]. }
\label{L_w_p}
\end{center}
\end{figure}

\begin{figure}[htb!]
\begin{center}
\includegraphics[scale=0.30]{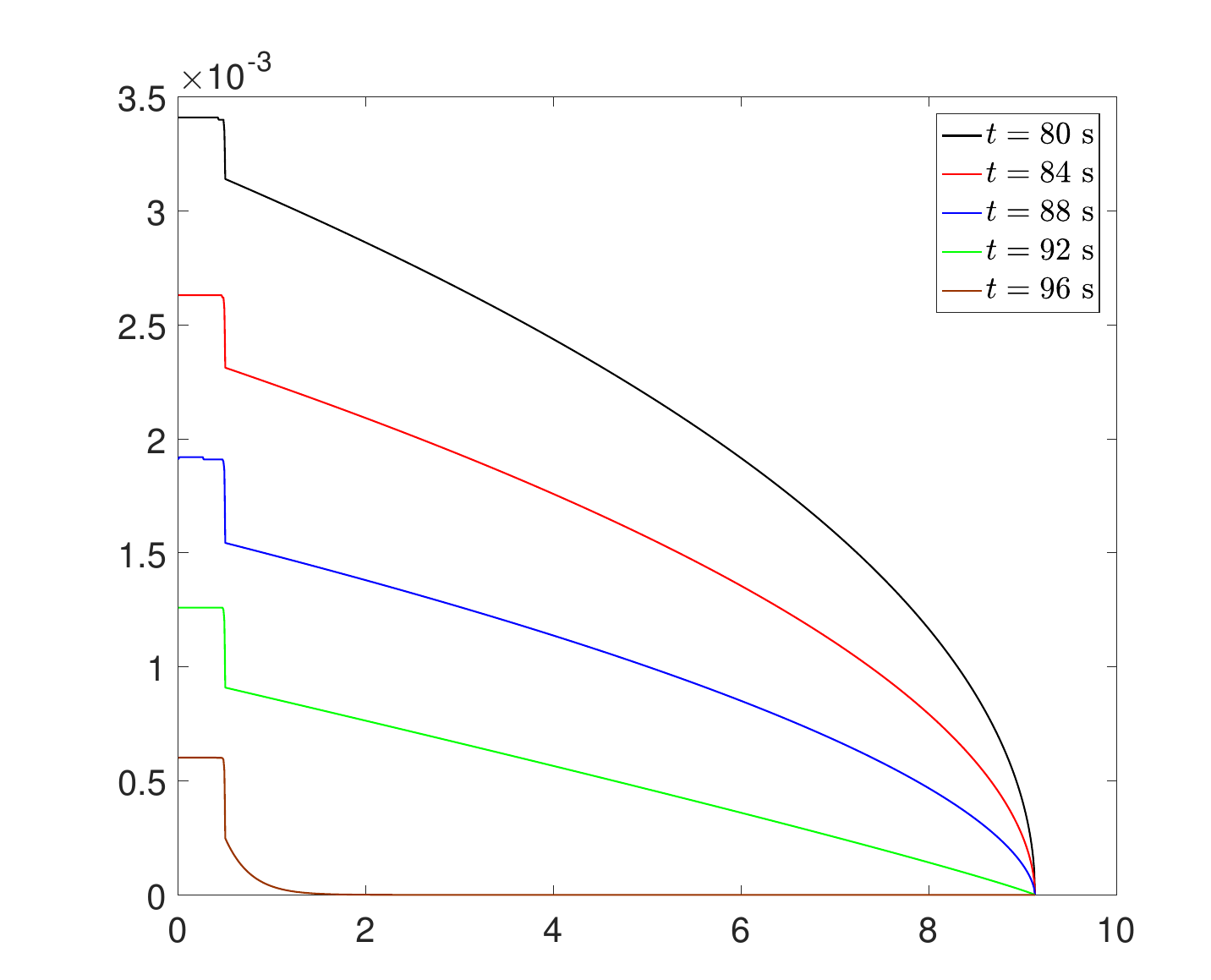}
\hspace{0mm}
\includegraphics[scale=0.30]{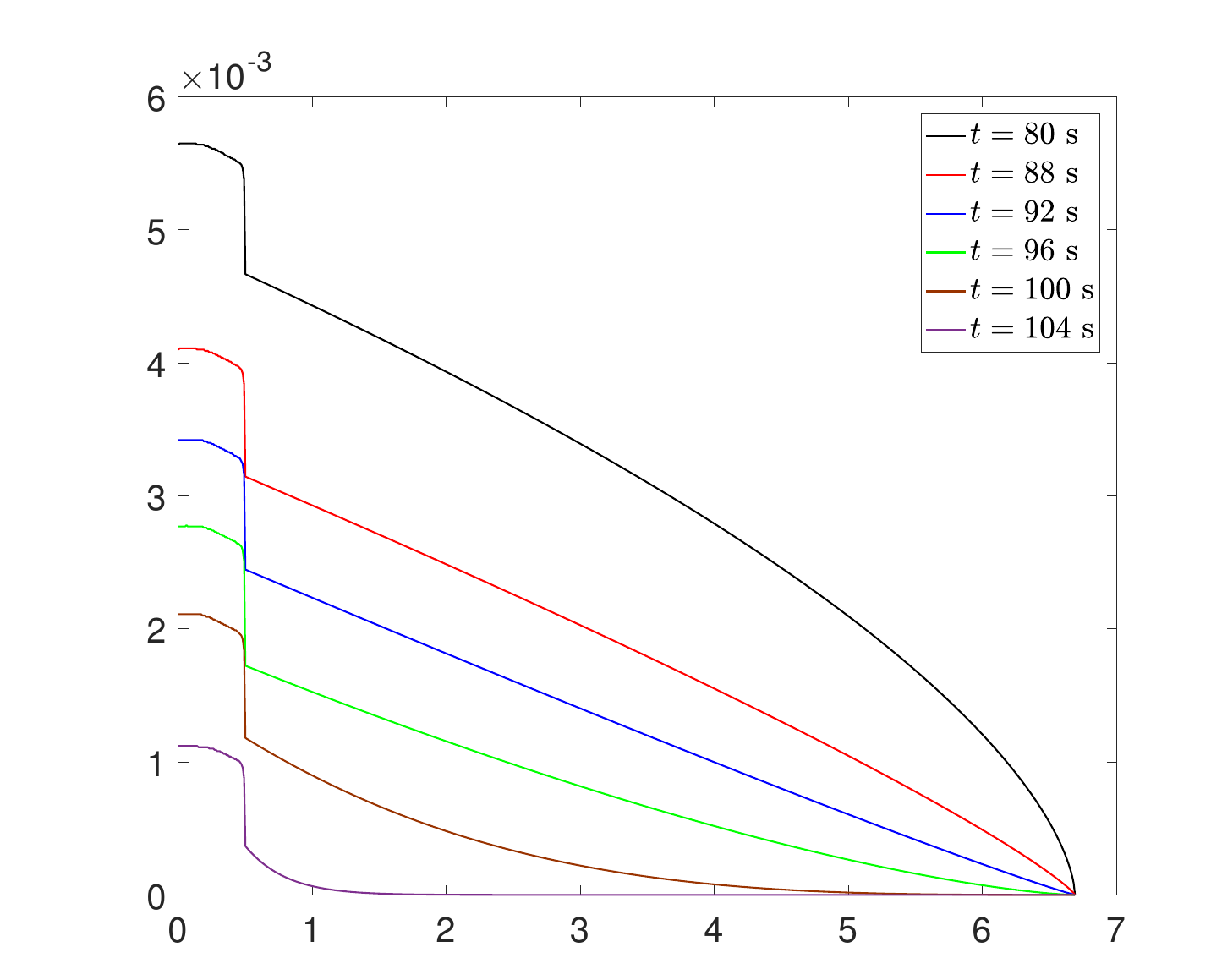}
\put(-320,-5){$x$ [m]}
\put(-110,-5){$x$ [m]}
\put(-440,155){$\textbf{a)}$}
\put(-215,155){$\textbf{b)}$}
\put(-420,65){\rotatebox{90}{$w(x)$ [m]}}
\put(-205,65){\rotatebox{90}{$w(x)$ [m]}}
\begin{center}
\includegraphics[scale=0.30]{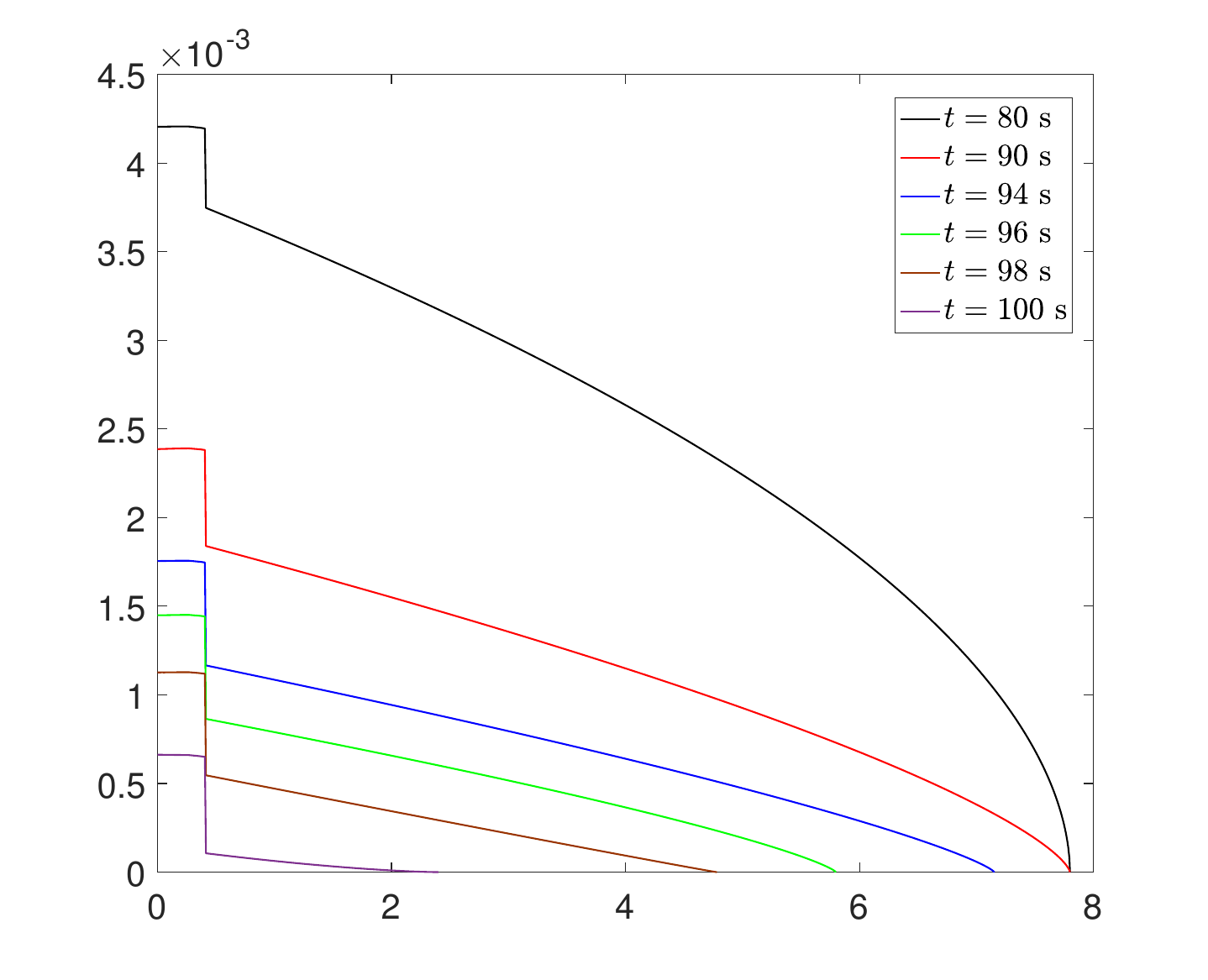}
\put(-210,155){$\textbf{c)}$}
\put(-210,70){\rotatebox{90}{$w(x)$ [m]}}
\put(-110,0){$x$ [m]}
\end{center}
\caption{Crack opening evolution, $w$ [m], during the fracture closure for: a) the elastic case, b) the large yield case ($c=1.33$ MPa), c) the small yield case ($c=2$ MPa). }
\label{w_profile}
\end{center}
\end{figure}
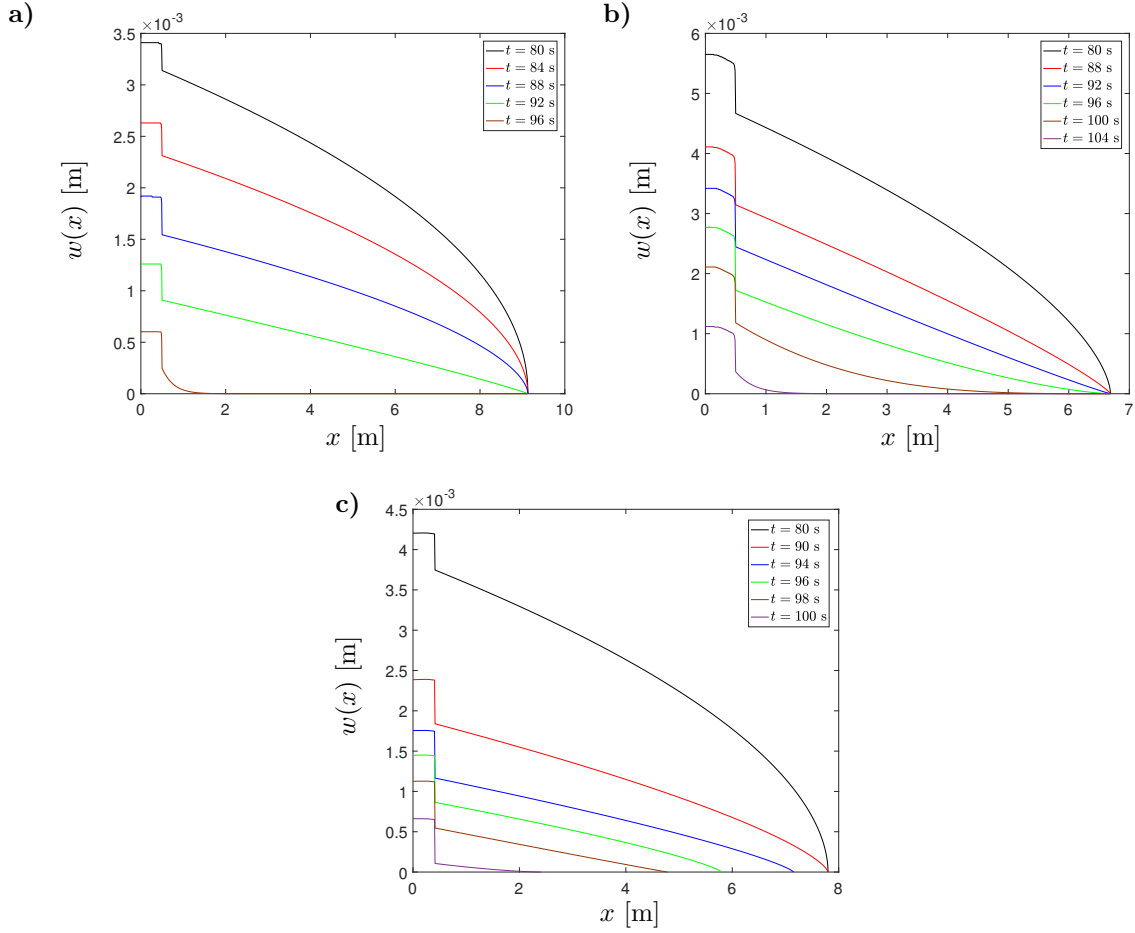

Let us consider the stage of fracture closure ($t>80$ s).  The evolution of crack geometry and the wellbore pressure for all variants of the problem (elastic and elasto-plastic) are depicted in Figures \ref{L_w_p} - \ref{w_profile}. One can observe that the crack closure process can be divided into two stages. In the first stage the crack length remains constant and the fracture volume decreases as a result of width reduction. In the second stage also the crack length declines. Notably, the transition between the respective stages can be easily identified in the graphs for $w(0,t)$ and $p_\text{n}(0,t)$ too. As can be seen in the figures, the second stage is much shorter for the elastic fracture. Thus, we can conclude that in the case of elastic fracture the hinge pattern of closure holds, while for the elasto-plastic fracture the zip closure mechanism prevails.

Let us now employ the elements of the classical $g$ function analysis \citep{Valko} to identify the values of the leak-off coefficient $C_\text{L}$ from the obtained results. We assume that the wellbore pressure is expressed by the following linear function of $g(\Delta t,\beta)$:
\begin{equation}
\label{p_lin}
p_\text{w}=p_\text{n}(0,t)=a-b \cdot g(\Delta t,\beta),
\end{equation}
where the dimensionless $g$ function is defined as:
\[
g(\Delta t,\beta)=\frac{4\beta \sqrt{\Delta t}+2\sqrt{1+\Delta t} \cdot {_2F_1}\left( 1/2,\beta;1+\beta;\frac{1}{1+\Delta t}\right)}{1+2\beta}.
\]
In the above relations $\Delta t$ stands for the normalized time:
\[
\Delta t=\frac{t-t_\text{e}}{t_\text{e}},
\]
with $t_\text{e}$ being the shut-in time, ${_2F_1}$ is the hypergeometric Gauss function, while $\beta$ is a dimensionless parameter depending on the HF model (for the KGD geometry $\beta=2/3$). Coefficients $a$ and $b$ from \eqref{p_lin} are defined as:
\begin{equation}
\label{a_def}
a=p_\text{c}+\frac{c_\text{f}V_i}{A_\text{e}}-2c_\text{f}S_p,
\end{equation}
\begin{equation}
\label{b_def}
b=2c_\text{f}C_\text{L}\sqrt{t_e},
\end{equation}
where: $p_\text{c}$ - closure pressure, $c_\text{f}=c_\text{f(KGD)}=E'/(\pi L)$ - proportionality coefficient between the net pressure and the average width, $V_i$ - volume of the injected fluid, $A_\text{e}$ - fracture surface at the end of pumping (one wing), $S_p$ - spurt loss coefficient.

In the classical $g$ function analysis the coefficients $a$ and $b$ from the representation \eqref{p_lin} are approximated from the given characteristics of $p_\text{w}$. Then, the leak-off coefficient, $C_\text{L}$, and the spurt loss coefficient, $S_p$, are computed from \eqref{a_def}--\eqref{b_def}.  Under non-ideal conditions the wellbore pressure characteristics can diverge from the linear relation which entails some modifications of the basic technique \citep{Valko,Economides}. In our study we aim at approximating the values  of $C_\text{L}$ and the closure stress, $p_\text{c}$, from the given characteristics of $p_\text{w}(g)$.

\begin{figure}[htb!]
\begin{center}
\includegraphics[scale=0.30]{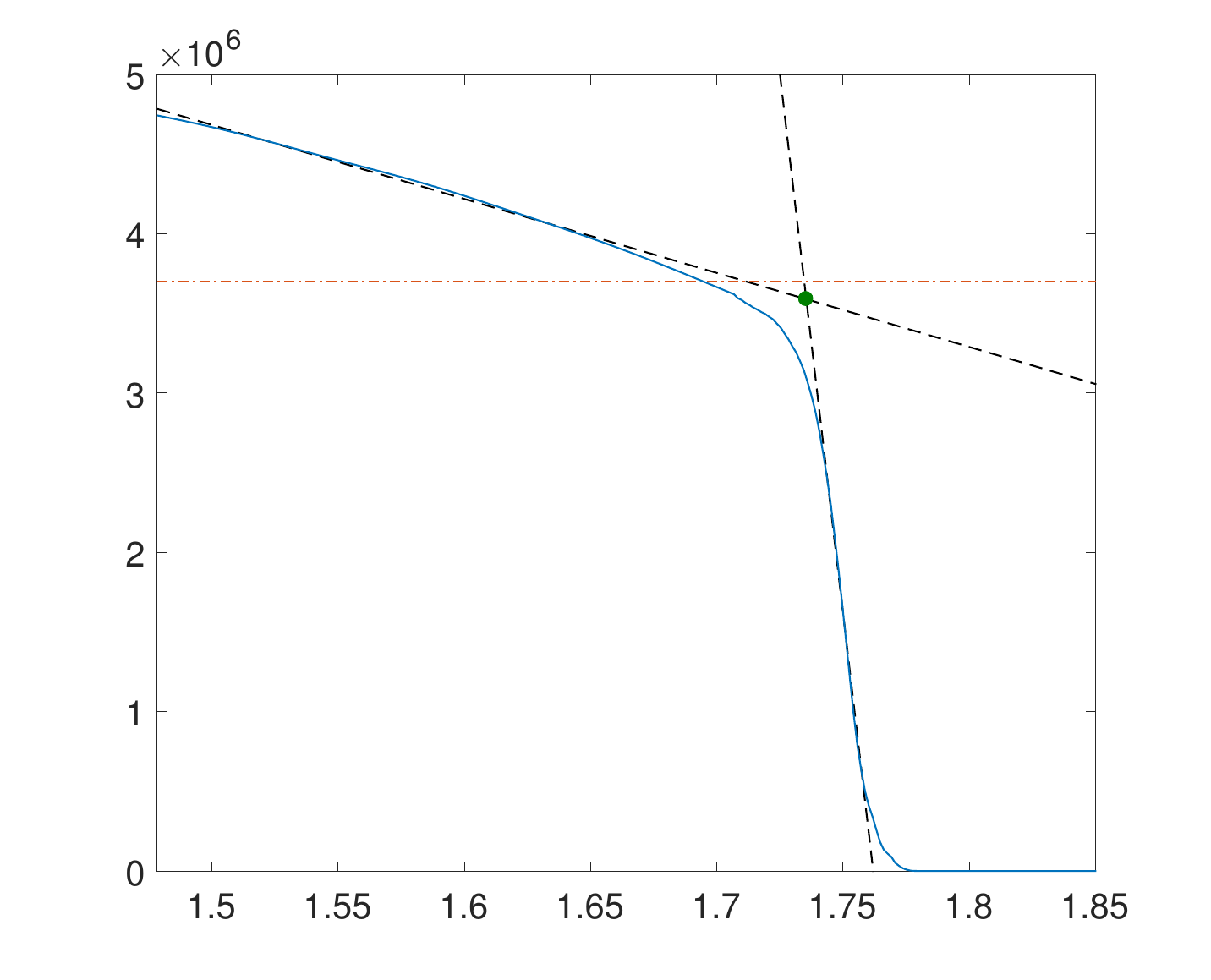}
\hspace{0mm}
\includegraphics[scale=0.30]{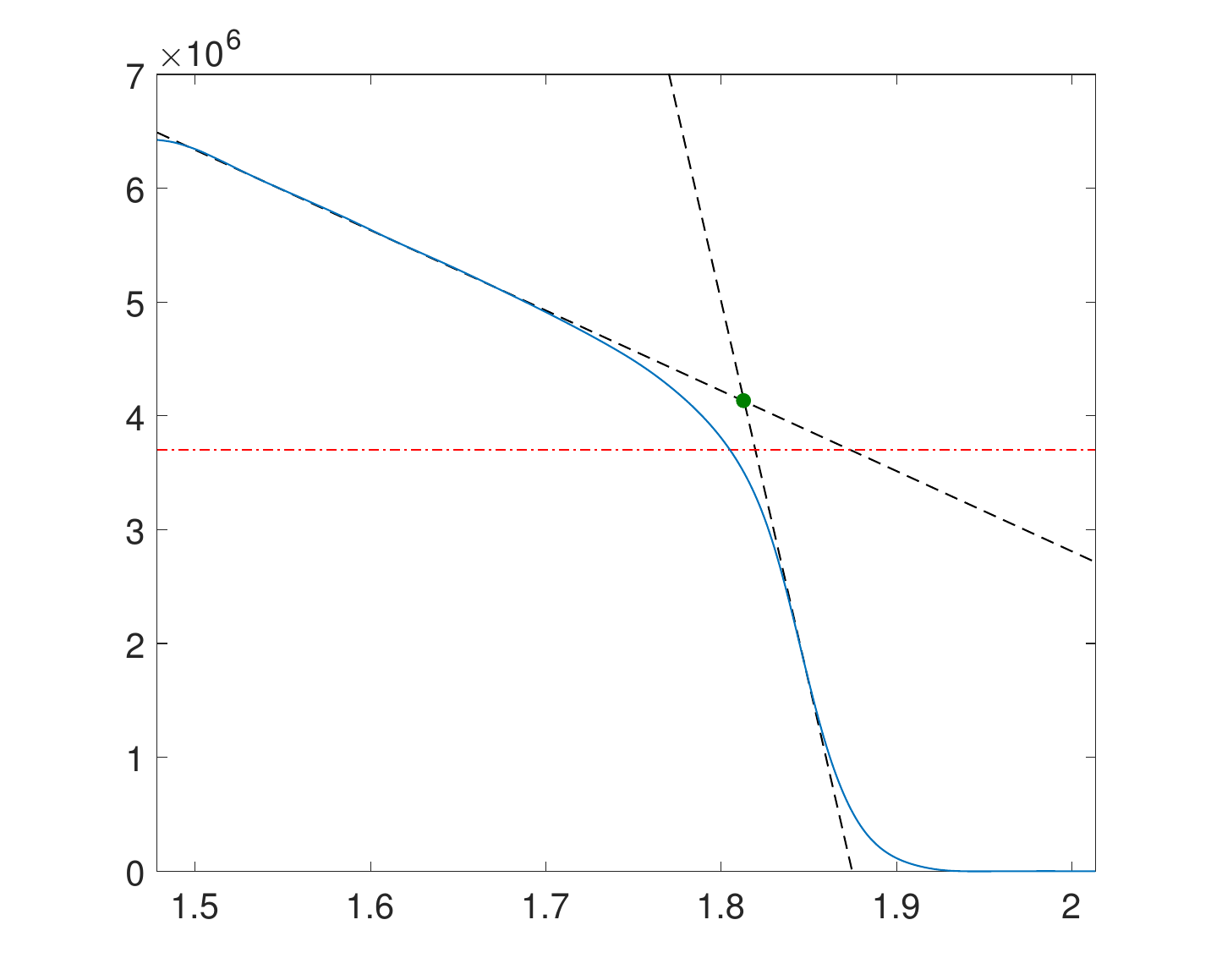}
\put(-325,0){$g(\Delta t)$}
\put(-115,0){$g(\Delta t)$}
\put(-440,155){$\textbf{a)}$}
\put(-215,155){$\textbf{b)}$}
\put(-420,70){\rotatebox{90}{$p_\text{w}$ [Pa]}}
\put(-202,70){\rotatebox{90}{$p_\text{w}$ [Pa]}}
\begin{center}
\includegraphics[scale=0.30]{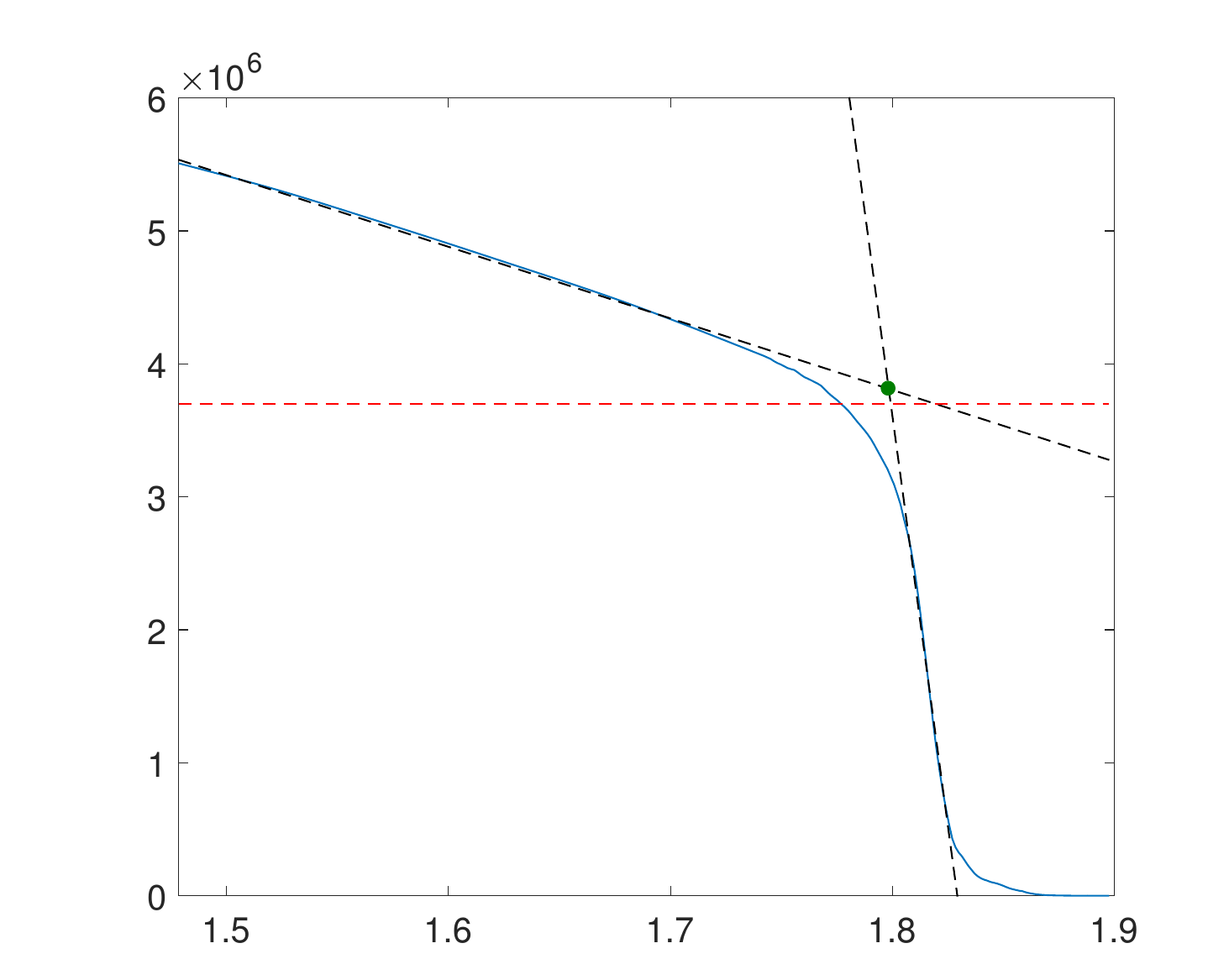}
\put(-210,155){$\textbf{c)}$}
\put(-200,70){\rotatebox{90}{$p_\text{w}$ [Pa]}}
\put(-105,-5){$g(\Delta t)$}
\end{center}
\caption{The $g$ function analysis for: a) the elastic case, b) the large yield case ($c=1.33$ MPa), c) the small yield case ($c=2$ MPa). The green dots mark the approximated values of closures pressure. The actual value of closure pressure is denoted by red dot-dash lines. }
\label{g_analysis}
\end{center}
\end{figure}
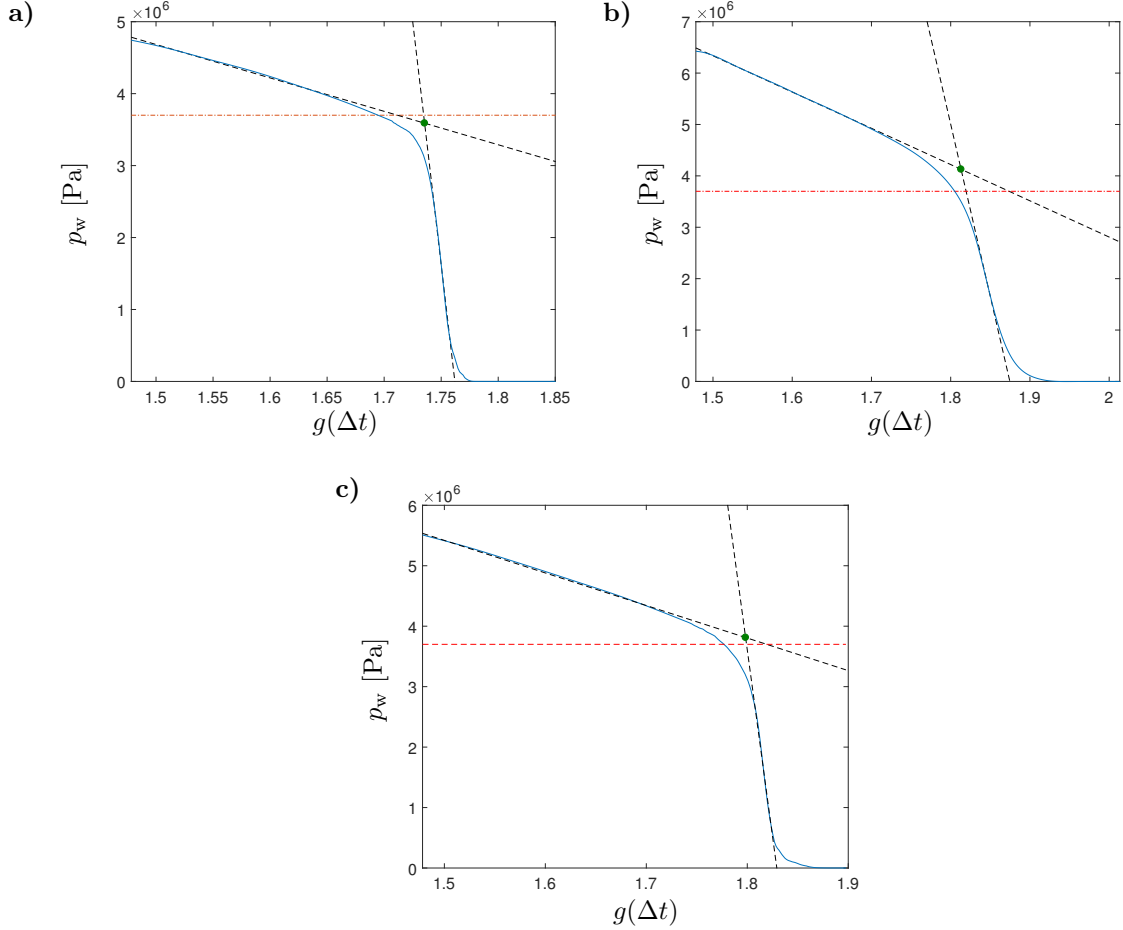

\begin{figure}[htb!]
\begin{center}
\includegraphics[scale=0.30]{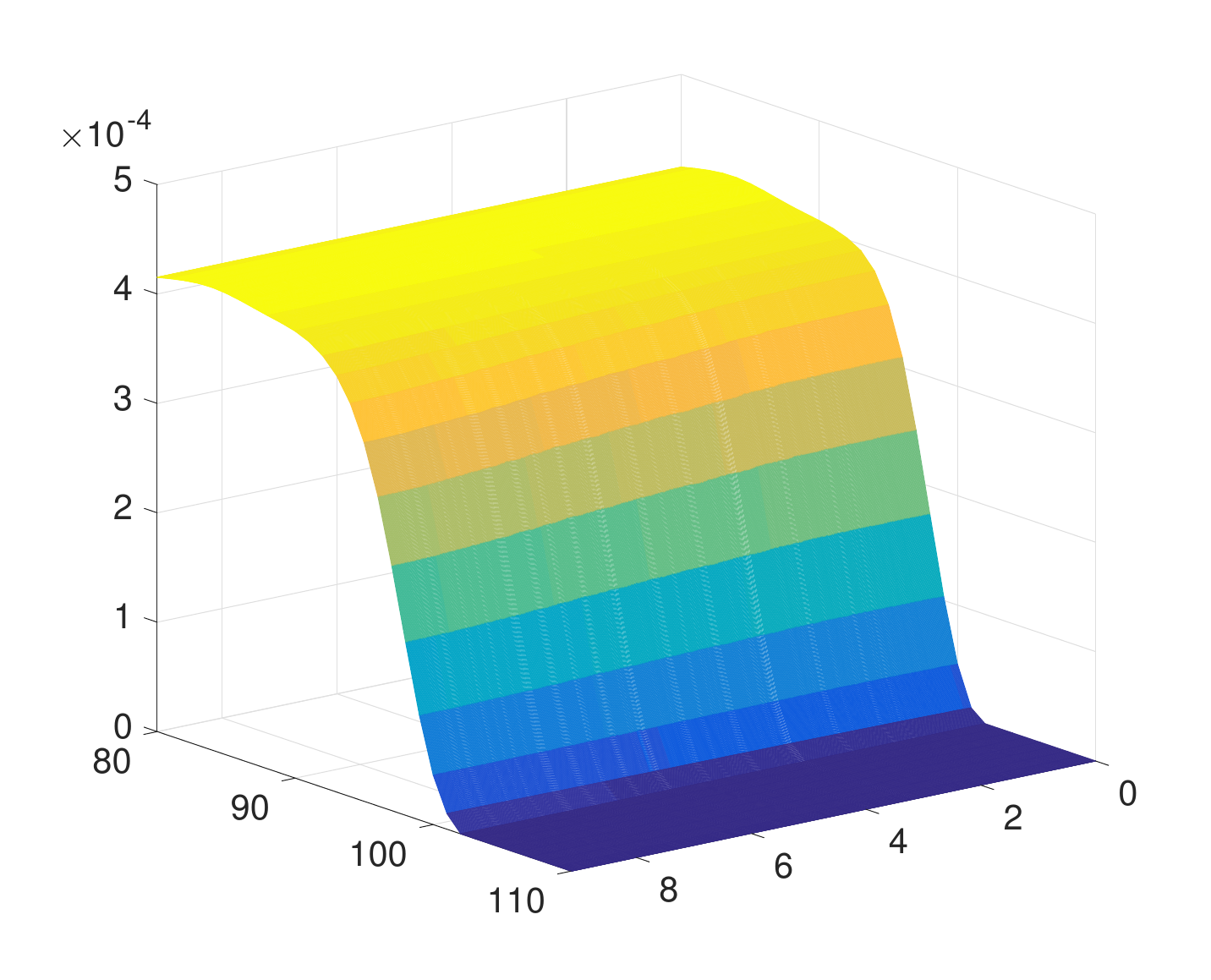}
\hspace{0mm}
\includegraphics[scale=0.30]{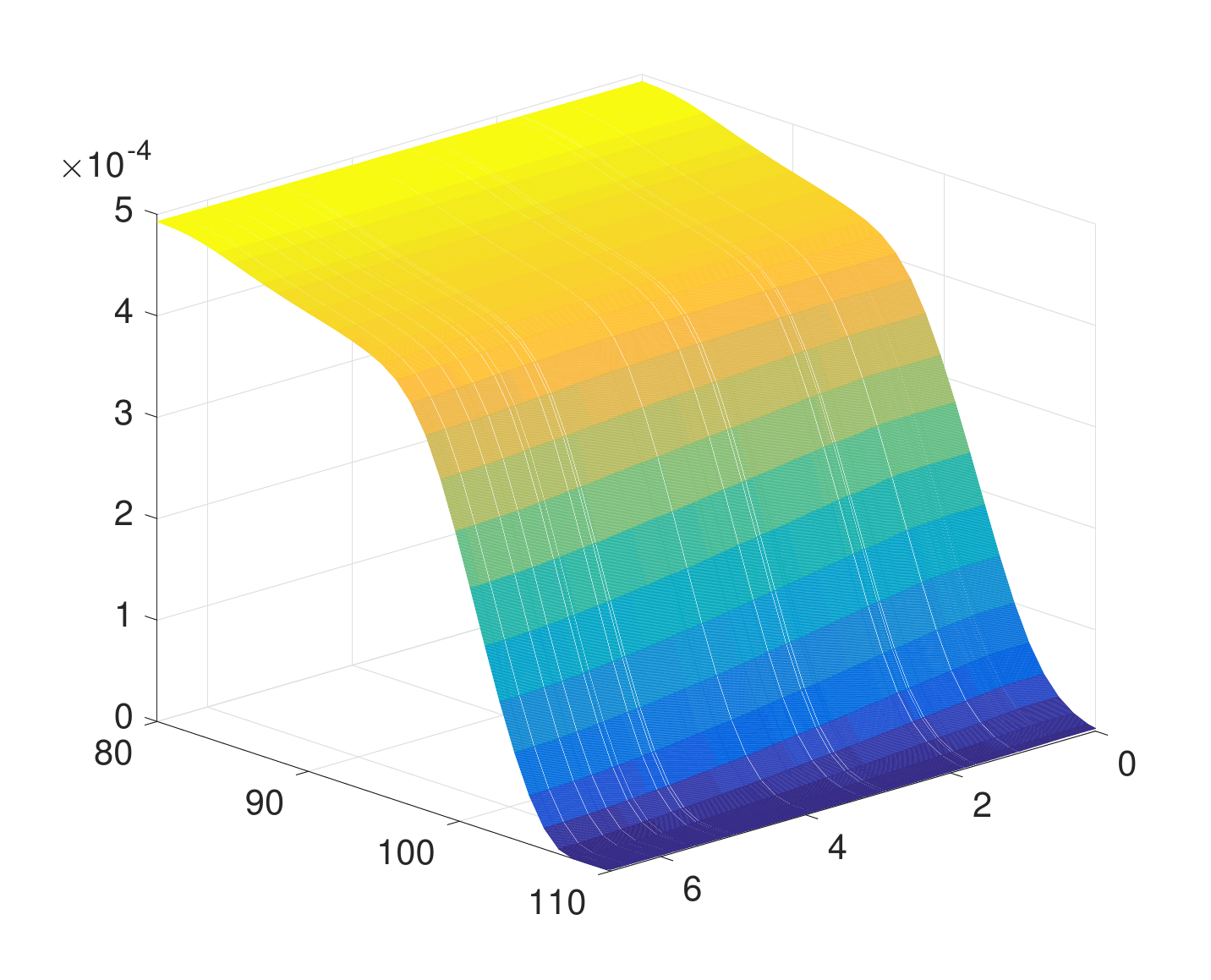}
\put(-270,5){$x$ [m]}
\put(-385,10){$t$ [s]}
\put(-55,10){$x$ [m]}
\put(-165,10){$t$ [s]}
\put(-440,155){$\textbf{a)}$}
\put(-215,155){$\textbf{b)}$}
\put(-430,75){\rotatebox{90}{$C_\text{L}$ $\left[\frac{\text{m}}{\sqrt{\text{s}}}\right]$}}
\put(-215,75){\rotatebox{90}{$C_\text{L}$ $\left[\frac{\text{m}}{\sqrt{\text{s}}}\right]$}}
\begin{center}
\includegraphics[scale=0.30]{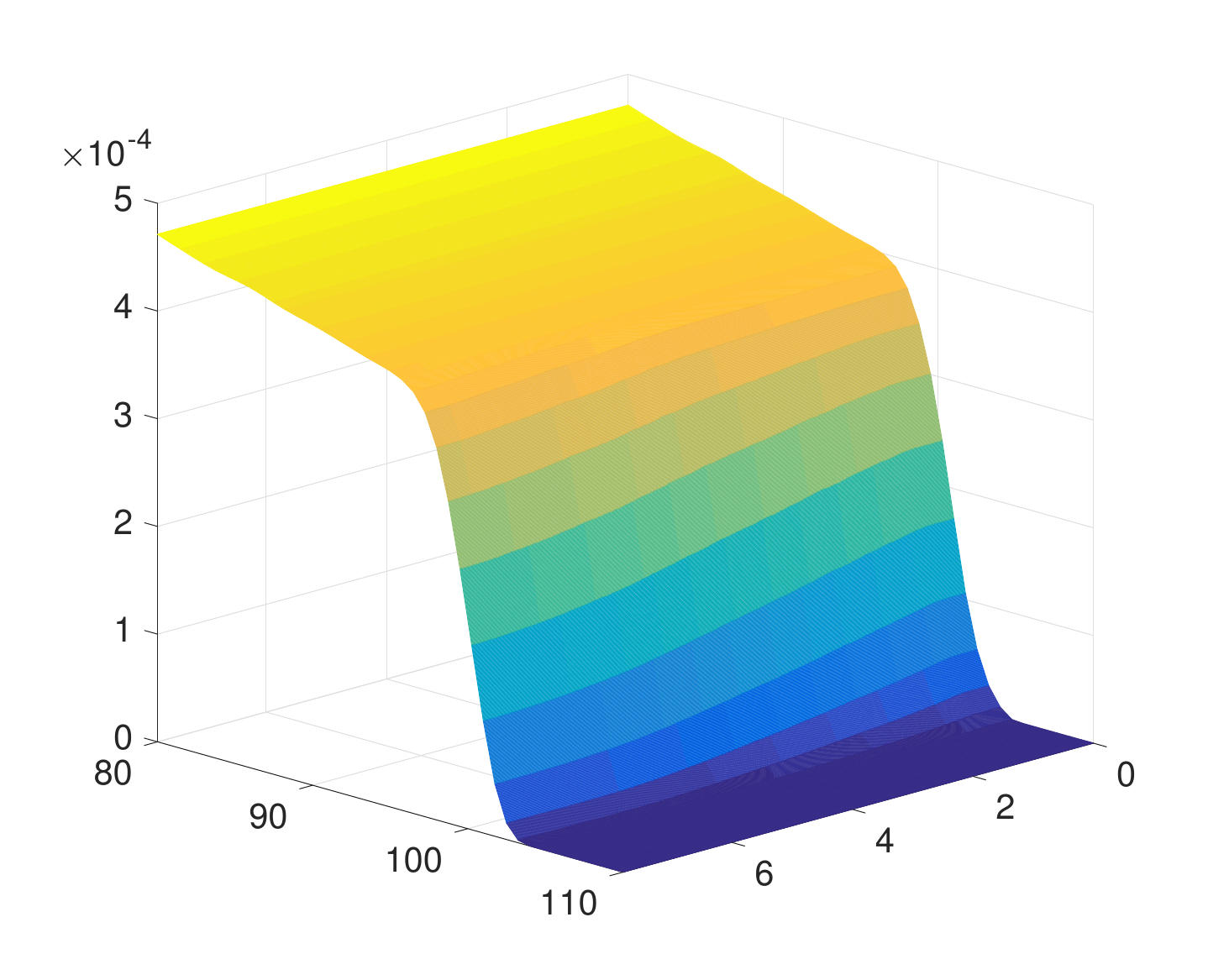}
\put(-210,155){$\textbf{c)}$}
\put(-215,75){\rotatebox{90}{$C_\text{L}$ $\left[\frac{\text{m}}{\sqrt{\text{s}}}\right]$}}
\put(-165,10){$t$ [s]}
\put(-55,10){$x$ [m]}
\end{center}
\caption{The pressure dependent leak-off coefficient, $C_\text{L}$ $\left[\frac{\text{m}}{\sqrt{\text{s}}}\right]$, for: a) the elastic case, b) the large yield case ($c=1.33$ MPa), c) the small yield case ($c=2$ MPa). }
\label{CL_graphs}
\end{center}
\end{figure}

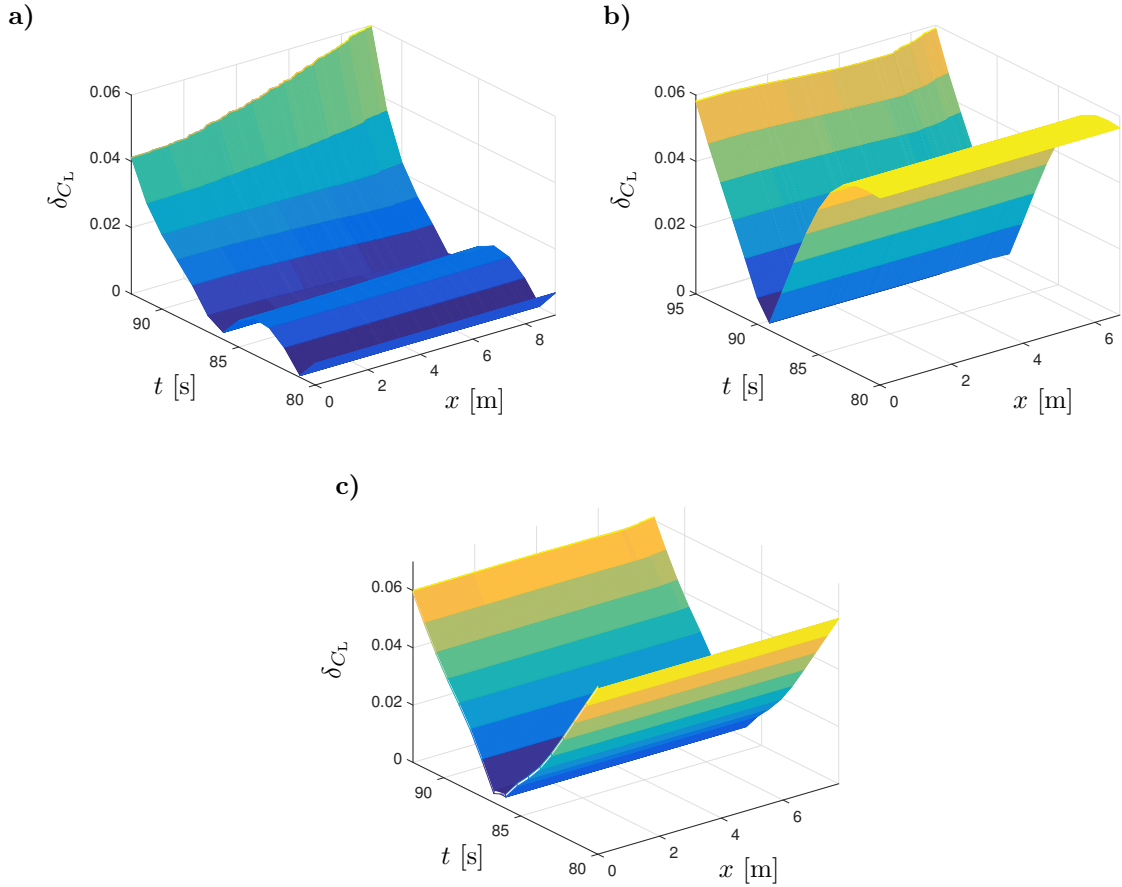
\begin{figure}[htb!]
\begin{center}
\includegraphics[scale=0.30]{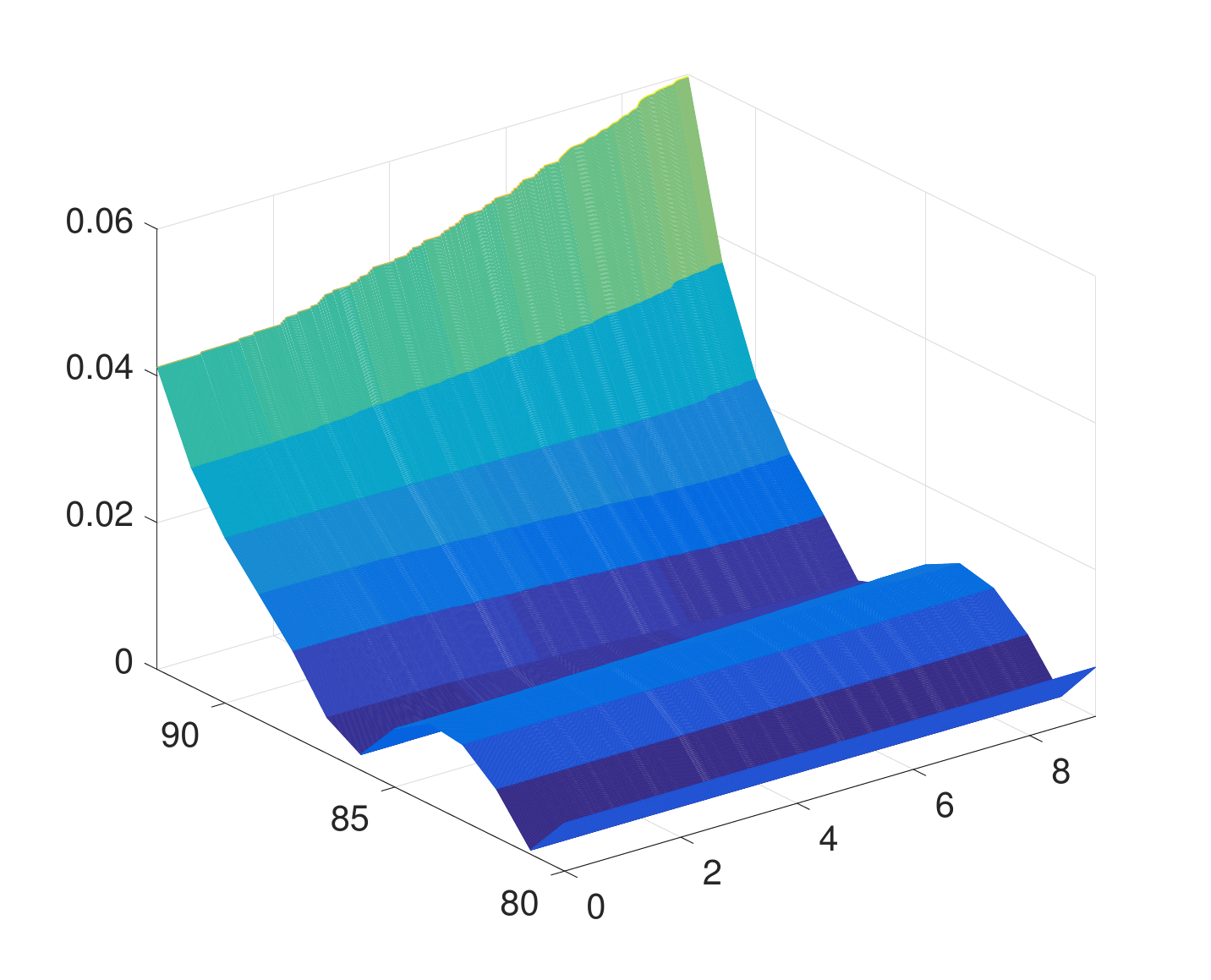}
\hspace{0mm}
\includegraphics[scale=0.30]{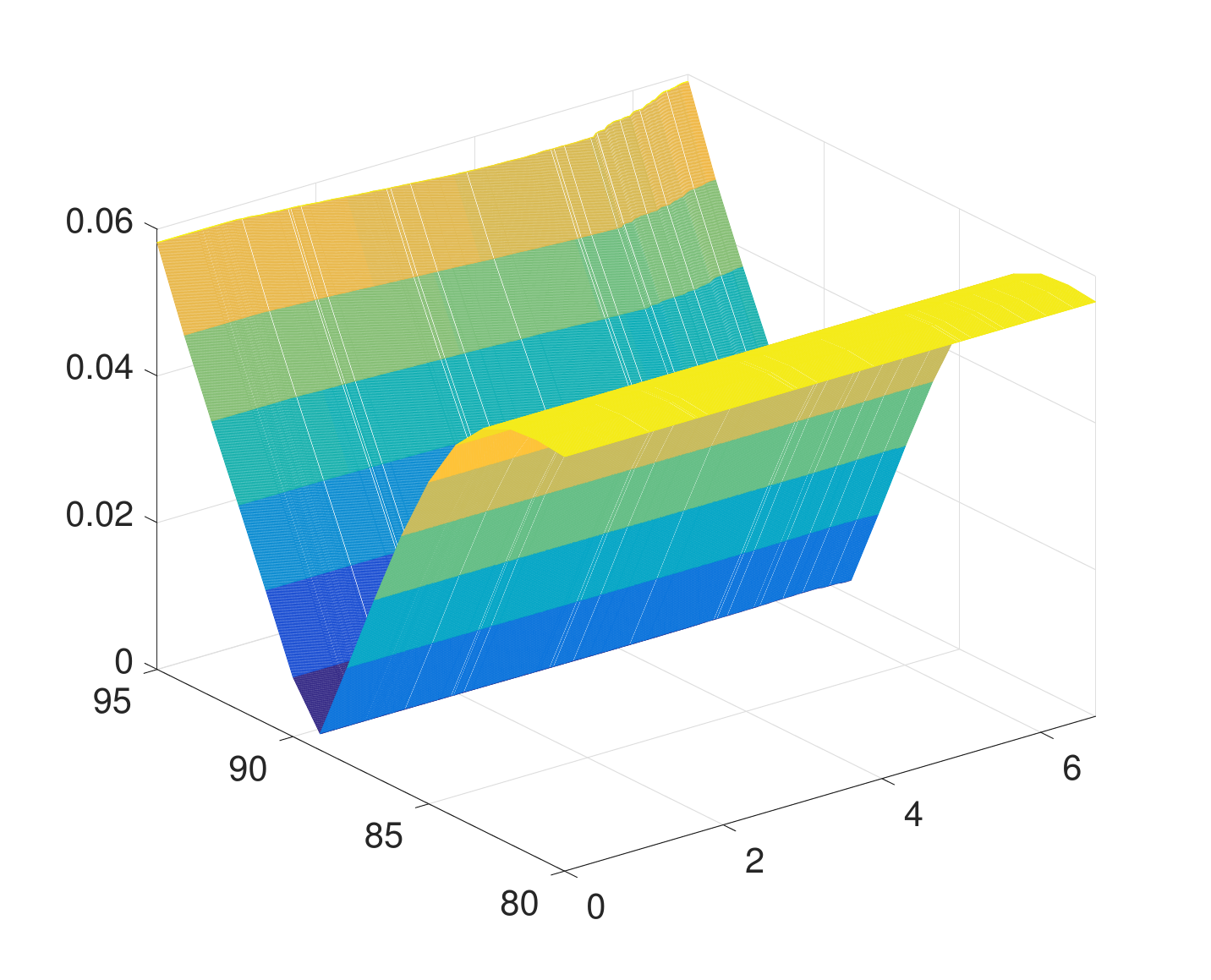}
\put(-275,10){$x$ [m]}
\put(-385,15){$t$ [s]}
\put(-60,10){$x$ [m]}
\put(-170,15){$t$ [s]}
\put(-440,155){$\textbf{a)}$}
\put(-215,155){$\textbf{b)}$}
\put(-425,85){\rotatebox{90}{$\delta_{C_\text{L}}$}}
\put(-212,85){\rotatebox{90}{$\delta_{C_\text{L}}$}}
\begin{center}
\includegraphics[scale=0.30]{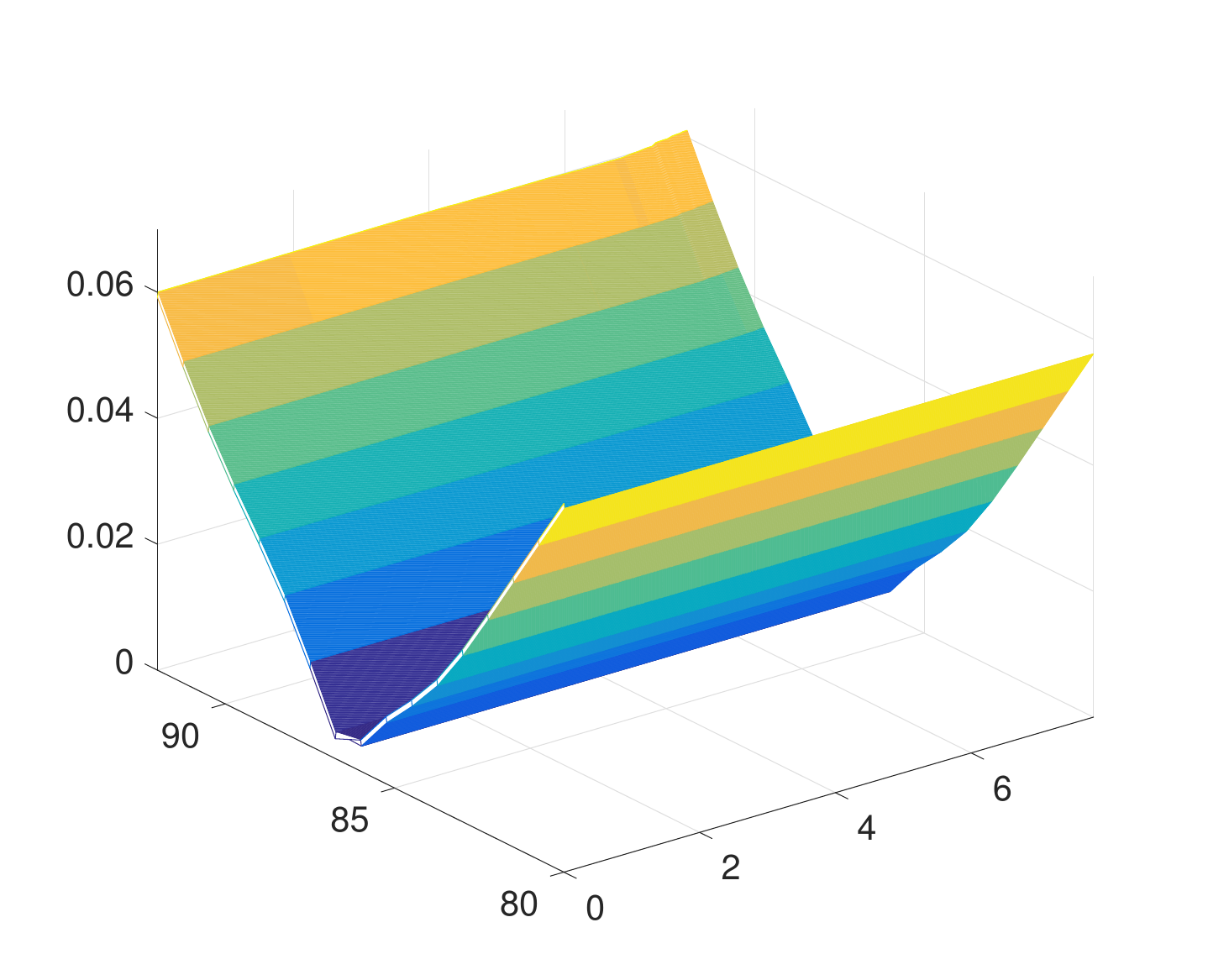}
\put(-210,155){$\textbf{c)}$}
\put(-215,85){\rotatebox{90}{$\delta_{C_\text{L}}$}}
\put(-170,15){$t$ [s]}
\put(-65,10){$x$ [m]}
\end{center}
\caption{The relative error of approximation of the leak-off coefficient,  $\delta_{C_\text{L}}$, for: a) the elastic case, b) the large yield case ($c=1.33$ MPa), c) the small yield case ($c=2$ MPa). }
\label{delta_CL_graphs}
\end{center}
\end{figure}

The graphs of $p_\text{w}(g)$ are shown in Figure \ref{g_analysis}. In each case the leak-off coefficient, $C_\text{L}$, is approximated from this part of the pressure characteristics that corresponds to the constant crack length (which is one of the basic assumptions of the $g$ function analysis). When employing the linear fitting in the respective cases we obtained the following values of leak-off coefficients: i) for the first variant of the problem (elastic fracture): $C_\text{L}=4.18\cdot 10^{-4}$ $\frac{\text{m}}{\sqrt{\text{s}}}$,  ii) for the large yield case ($c=1.33$ MPa): $C_\text{L}=4.65\cdot 10^{-4}$ $\frac{\text{m}}{\sqrt{\text{s}}}$ and iii) for the small yield case ($c=2$ MPa): $C_\text{L}=4.44\cdot 10^{-4}$ $\frac{\text{m}}{\sqrt{\text{s}}}$. The actual values of the pressure dependent (and thus solution dependent) leak-off coefficients are depicted in Figure \ref{CL_graphs}. It shows that the approximated constant values of $C_\text{L}$ reflect quite well the actual magnitudes up to the point where the crack length starts to decline. Slightly better agreement (in terms of error distribution) between the approximated and actual values is observed for the elastic fracture.
In Figure \ref{delta_CL_graphs} we present the relative errors of the leak-off coefficient approximation, $\delta_{C_\text{L}}$ (for the temporal interval that refers to the constant crack length). In all cases the maximal error does not exceed 6$\%$. The average errors amount to 1.3 $\%$ in the elastic case, 3.5 $\%$ for the large plastic yield variant and 2.3 $\%$ for the small plastic yield case. This constitutes a fairly good estimation of $C_\text{L}$ considering the fact the $g$ function methodology assumes neither the pressure dependence of the leak-off coefficient nor the plastic deformation effects.

In order to estimate the value of closure pressure, $p_\text{c}$, let us analyze again the $p_\text{w}(g)$ curves. As can be seen in Figure \ref{g_analysis} the distribution of $p_\text{w}(g)$ can be approximated quite well by a bi-linear characteristics (dashed lines in the figure). Better quality of approximation is produced for the elastic variant of the problem. The deflection point on the pressure graph (i. e. the point in which the linear approximations intersect) constitutes a reasonable estimation of the closure pressure. For the elastic variant of the problem the predicted closure pressure was 3.59 MPa. It differs from the actual closure stress of 3.7 MPa by less than 3 $\%$. For the large plastic yield case the respective value was 4.1 MPa  which is 10.8 $\%$ higher than the real magnitude of $p_\text{c}$. Finally, in the small plastic yield variant the approximation of $p_\text{c}$ produces 3.81 MPa which provides the relative error below 3 $\%$. The aforementioned mechanism has been computationally verified also against other examples. As the above results confirm, the quality of estimation of the closure pressure turned out to depend on the extent of plastic deformation (the smaller the extent of plastic yield the better approximation of $p_\text{c}$ is obtained). We expect this trend to hold also for other plasticity models. Note that the method employed to determine the closure stress is a counterpart of the tangent intersection method according to classification given by \cite{Hayashi_1989}.

\section{Final conclusions}
\label{concl}

In this paper we analyzed the problem of propagation and closure of a short hydraulic fracture. Two models of rock deformation were considered: i) the linear elastic model, ii) the elasto-plastic model based on a combination of  Mohr-Coulomb and Rankine yield criteria. The fluid leak-off to the rock formation was described by the Carter type relation with the leak-off coefficient depending on fluid pressure. The pressure decline analysis was performed for three computational examples, two of which were meant to constitute the limiting variants of the HF problem i. e.: i) the case of only elastic deformation of the fractured material, ii) the case of large plastic yield of solid. Methodology for determination of the closure stress and average leak-off coefficient was proposed which holds also in the case of inelastic deformations of rock. The basic trends of solution behaviour were reported and the accuracy of estimation of the aforementioned parameters was established.

The following conclusions can be drawn from the conducted analysis:
\begin{itemize}
\item{If the analyzed problem assumes inelastic deformation of the fractured material it is essential that the crack propagation stage is properly modelled (i. e. the plastic deformation zone develops fully around the fracture surfaces).  It comes from the fact that the character of plastic deformation of rock affects substantially the closure stage.}
\item{The crack closure pattern depends essentially on the character of rock deformation. For the elastic deformation the hinge-like mechanism prevails, while for the elasto-plastic case the zip pattern holds.}
\item{The $g$ function analysis can be employed also in the case of elasto-plastic deformation of the fractured material, however the quality of obtained results is worse than for the elastic variant of the problem.}
\item{When employing the proposed methodology to estimate the values of closure stress and the average leak-off coefficient, there holds a general trend of accuracy improvement with reduction of the extent of plastic yield.}
\item{The $g$ function analysis can be also used with the Crater type leak-off model for the pressure dependent leak-off coefficient. The conducted analysis shows that, in such a case, the approximated constant value of leak-off coefficient constitutes   quite accurate estimation of the actual leak-off coefficient in an average sense, up to the point where the crack length starts to decline.}
\item{The tangent intersection method yields good approximation of the closure stress. Better accuracy of this estimation is obtained in the case of elastic deformation of rock.}
\end{itemize}

\pagebreak
\noindent
{\bf Funding:} 
This work was funded by European Regional Development Fund and the Republic of Cyprus
through the Research Promotion Foundation (RESTART 2016 - 2020 PROGRAMMES, Excellence Hubs, Project EXCELLENCE/0421/0456). 

\vspace{5mm}
\noindent
{\bf Conflicts of interest/Competing interests:} 
The authors have no conflicts of interest to declare that are relevant to the content of this article.

\vspace{5mm}
\noindent
{\bf Acknowledgments:}
The authors are thankful to Professor Gennady Mishuris for his useful comments and discussions. MD would like to acknowledge the support provided by Welsh Government via Ser Cymru Future Generations Industrial Fellowship grant AU224 and by Royal Academy of Engineering for RAE Industrial Fellowship Award IF2122$\ $183.

\end{document}